\documentclass[twocolumn]{aastex63}
\usepackage{amsmath}
\usepackage{amssymb}
\usepackage{graphics}
\usepackage{enumerate}
\usepackage{soul}

\usepackage{savesym}
\savesymbol{tablenum}
\usepackage{siunitx}
\restoresymbol{SIX}{tablenum}
\usepackage{natbib}

\usepackage{hyperref}

\newcommand{\del}{\partial}

\DeclareRobustCommand{\delsf}{\bgroup\markoverwith{\textcolor{red}{\rule[.5ex]{2pt}{0.4pt}}}\ULon}
\newcommand{\isotope}[2]{{\ensuremath{{}^{#1}\mathrm{#2}}}}

\newcommand{\amu}{m_\mathrm{u}}

\shorttitle{Collapses and explosions of rotating SMSs}

\shortauthors{Fujibayashi et al.}

\begin{document}

%\preprint{APS/123-QED}

\title{Powerful Explosions From The Collapse of Rotating Supermassive Stars}% Force line breaks with \\

\author[0000-0001-6467-4969]{Sho Fujibayashi}
\affiliation{Frontier Research Institute for Interdisciplinary Sciences, Tohoku University, Sendai 980-8578, Japan}
\affiliation{Astronomical Institute, Graduate School of Science, Tohoku University, Sendai 980-8578, Japan}
\affiliation{Max-Planck-Institut f\"ur Gravitationsphysik (Albert-Einstein-Institut), Am M\"uhlenberg 1, D-14476 Potsdam-Golm, Germany}

\author[0009-0007-7617-7178]{C\'edric Jockel}
\affiliation{Max-Planck-Institut f\"ur Gravitationsphysik (Albert-Einstein-Institut), Am M\"uhlenberg 1, D-14476 Potsdam-Golm, Germany}

\author[0000-0003-4443-6984]{Kyohei Kawaguchi}
\affiliation{Max-Planck-Institut f\"ur Gravitationsphysik (Albert-Einstein-Institut), Am M\"uhlenberg 1, D-14476 Potsdam-Golm, Germany}
\affiliation{Institute for Cosmic Ray Research, The University of Tokyo, 5-1-5 Kashiwanoha, Kashiwa, Chiba 277-8582, Japan}
\affiliation{Center for Gravitational Physics and Quantum Information, Yukawa Institute for Theoretical Physics, Kyoto University, Kyoto, 606-8502, Japan}

\author[0000-0002-2648-3835]{Yuichiro Sekiguchi}
\affiliation{Center for Gravitational Physics and Quantum Information, Yukawa Institute for Theoretical Physics, Kyoto University, Kyoto, 606-8502, Japan}
\affiliation{Department of Physics, Toho University, Funabashi, Chiba 274-8510, Japan}

\author[0000-0002-4979-5671]{Masaru Shibata}
\affiliation{Max-Planck-Institut f\"ur Gravitationsphysik (Albert-Einstein-Institut), Am M\"uhlenberg 1, D-14476 Potsdam-Golm, Germany}
\affiliation{Center for Gravitational Physics and Quantum Information, Yukawa Institute for Theoretical Physics, Kyoto University, Kyoto, 606-8502, Japan}

% \collaboration{MUSO Collaboration}%\noaffiliation

\date{\today}

\correspondingauthor{Sho Fujibayashi}
\email{sho.fujibayashi@astr.tohoku.ac.jp}

\begin{abstract}
We perform new general relativistic hydrodynamics simulations for collapses of rotating supermassive star cores with an approximate nuclear burning up to carbon and a detailed equation of state. For all the models we investigate, the energy generation by nuclear burning plays only a minor role, leading to the formation of a black hole without a nuclear-powered explosion. For rotating models, however, the stellar explosion associated with shock heating is driven from a torus, which forms after the black hole formation. The explosion energy is up to $10^{-4}$ of the mass energy of the supermassive star cores ($\sim 10^{55}$--$10^{56}$\,erg). We find that, even if we increase the rotational angular momentum of the progenitor, the ejecta mass saturates at $\sim 1$\% of the total mass of the initial stellar core. The average ejecta velocity also saturates at $\approx 20\%$ of the speed of light. As a result, the ejecta kinetic energy is approximately proportional to the initial mass of the supermassive star core for the rapidly rotating case. We also perform viscous hydrodynamics simulations for exploring the evolution of the remnant torus. Although the viscous heating drives an outflow from the torus, we find that its effect is subdominant in terms of the kinetic energy because of the small velocity ($\approx 0.07c$) of the ejecta component.
%We focus on the properties of the ejecta driven by strong bounce of the torus formed around the black hole. 
\end{abstract} 


% \maketitle

\section{Introduction}
% Massive BH in early Universe.
The presence of supermassive black holes with estimated high masses of $\sim 10^7$--$10^{10}M_\odot$ in the early universe is an intriguing puzzle. The recent extensive searches for high-redshift galaxies indicate that a number of supermassive black holes of mass $\gtrsim10^9M_\odot$ were already present in the first billion years ($z\gtrsim6$) after the big bang (e.g., \citealt{Fan2023aug,Bogdan2024jan,Goulding2023sep,Kovacs2024apr}). This suggests that a rapid growth of the black holes from their massive or very massive seeds are required in the early universe (e.g., \citealt{Inayoshi2020aug,Volonteri2021sep} for reviews).

% formation of massive seed
%  ($\dot{M}_\mathrm{Edd} \propto M_\mathrm{BH}$, where $M_\mathrm{BH}$ is the black hole mass)
% ; it determines the Eddington limit, to which the mass accretion onto the black hole is likely limited.
The growth of the black hole mass is typically limited by the Eddington rate. In this respect, a high mass of the seed black holes is preferred for the swift formation of supermassive black holes. For example, the seed black holes may originate from collapses of $\sim100$--$1000M_\odot$ Population III stars, as indicated in numerical simulations~\citep{Hirano2014feb}. To reach a $10^9M_\odot$ black hole from one with $10^3M_\odot$ by $z=6$, a mass accretion rate with nearly Eddington rate is required throughout its growing history of about 1 billion years. However, several feedback effects, which can stem from, e.g., viscous heating and resultant enhanced radiation pressure during the mass accretion onto the seed black hole and nearby supernova explosions, could make such a high duty cycle unlikely.
%\footnote{The Eddington limit is mitigated if the mass accretion rate of the black hole is so high that photons are trapped in the accretion flow. To achieve this, it is however necessary to supply the mass to the region close to the black hole horizon, against radiative and mechanical feedback effects (e.g., \citealt{Inayoshi2020aug} and references therein). } !! MS thinks that this descrption is not so accurate that it wound be better to delete it. 

Several scenarios have been proposed for the formation of a high-mass seed black hole (e.g., \citealt{Rees1978oct,Inayoshi2020aug,Volonteri2021sep}). One of the scenarios is the so-called direct collapse scenario (e.g., \citealt{Bromm2003oct}). In this scenario, a supermassive star with mass of $\sim10^4$--$10^6M_\odot$ is formed in a rapidly accreting (with a mass accretion rate of $\sim 0.1M_\odot$/yr) high-temperature primordial gas cloud, which is cooled primarily by atomic hydrogen line emissions.\footnote{The gas cloud that hosts a supermassive star should lack molecular hydrogen due to, e.g., strong irradiation of far-ultraviolet photons~\citep{Omukai2001jan}, collisional dissociation in a dense and hot region that experienced a strong shock~\citep{Inayoshi2012may}, or mechanical heating by frequent merger of the host halos~\citep{Chon2016dec,Hirano2017sep,Wise2019jan}.} The supermassive star then collapses to a massive black hole due to the general relativistic instability~\citep{1964ApJ...140..417C}. The resulting large initial mass of the seed black hole helps to grow to a supermassive black hole in a shorter timescale. Although there may still be issues related to the duty cycle, i.e., the accretion rate should still be nearly the Eddington limit for sub-billion years~\citep{Tanaka2009may}, the collapse of supermassive stars is one of the most promising scenarios for the formation of the supermassive black hole seeds in the early universe.
%They may originate from stellar mergers in dense metal-poor clusters (e.g., \citealt{Omukai2008oct,Devecchi2009mar}).

% Explosion of SMS
Supermassive stars may not just collapse to a black hole, but show some astrophysical transients. It was shown in \cite{1973ApJ...183..941F}, \cite{1986ApJ...307..675F}, and \cite{Montero2012apr} that an explosion due to the energy generation from hydrogen burning via the carbon-nitrogen-oxygen (CNO) cycle is possible if a supermassive star is metal-rich. More recently, \cite{2014ApJ...792...44C}, \cite{ Nagele2020aug}, \cite{Nagele2022dec}, \cite{Nagele2023aug}, and \cite{nagele2024arxiv}, showed, based on the initial data generated by a stellar evolution code, that a thermonuclear explosion is also possible for some special cases during the helium burning phase.

% rotation
The rotation of the star may also play an important role for generating a bright transient. As the molecular clouds observed in the present-day universe, the primordial gas clouds hosting supermassive stars are likely to have a significant amount of angular momentum. 
% Cosmological simulations indeed showed that a cosmological object with mass $M$ and radius $R$ has an average angular momentum of $\approx 0.035\sqrt{2GM^{3/2}R}$ at the time of the virialization~\cite{Barnes1987aug,Bullock2001jul}. For an object with $R\sim\SI{1}{pc}$~\cite{Patrick2023jul}, this corresponds to the centrifugal radius of $\sim\SI{e16}{cm}$, which is much larger than the radius of the supermassive star $\sim\SI{e14}{cm}$. 
Supermassive stars may thus rotate rapidly (see, however, a  discussion in \citealt{Lee_2016, 2018ApJ...853L...3H}). If this is the case, the rotating supermassive stars not only collapse into black holes, but also form a torus surrounding the black holes. This scenario suggests further activities, potentially with astrophysical transients, because the torus formation process can be accompanied by strong shock-wave formation which subsequently drives a powerful outflow of unbound matter~\citep{2007PhRvD..76h4017L,Uchida2017oct,Lee2006apr}. In addition, the effective viscosity induced by the magnetohydrodynamical processes in the torus could drive a post-collapse outflow from it. The black hole-torus system may also drive a relativistic outflow~\citep{Matsumoto2015sep}, if a magnetosphere is developed along the rotational axis of the formed black hole. 

% this paper
Motivated by the above considerations, and also by rapidly progressing observations of the early universe, especially with the James Webb Space Telescope (JWST), we revisit the collapse of rotating supermassive star cores in this paper. We focus in particular on the properties of the matter ejected as a result of the explosive shock heating at the formation of a torus. For this purpose, we first perform a set of new axisymmetric fully general relativistic hydrodynamics simulations starting from equilibrium configurations of supermassive star cores that are subject to the general relativistic instability. In these simulations, we take into account the effect of hydrogen burning and triple-alpha reactions in a simple, but consistent way. We further perform a viscous hydrodynamics simulation for the remnant torus to investigate the effect of the post-collapse mass ejection, assuming a hypothetical enhancement of the effective viscosity that can be developed in the presence of a magnetohydrodynamical turbulence \citep{Balbus:1998ja}. 

% content
This paper is organized as follows: In Sec.~\ref{sec:method}, we present the numerical methods used in the present simulation. Then in Sec.~\ref{sec:results}, the results are described, focusing on the properties of the ejecta. We discuss a possible astrophysical transient based on our results in Sec.~\ref{sec:discussion}. Finally, Sec.~\ref{sec:summary} is devoted to a summary. Throughout this paper, $G$, $c$, and $k_\mathrm{B}$ denote the gravitational constant, speed of light, and Boltzmann's constant, respectively.

\section{Method} \label{sec:method}
A general relativistic neutrino radiation hydrodynamics code is employed for the simulations. Einstein's equations are solved with a version of the puncture-Baumgarte-Shapiro-Shibata-Nakamura formalism~\citep{shibata1995a,baumgarte1998a,Campanelli:2005dd} with a Z4c constraint propagation scheme~\citep{Hilditch2013a}. The so-called cartoon method~\citep{Alcubierre2001a,Shibata2000a,Shibata2012mar} is used to impose the axial symmetry.  The basic method of radiation hydrodynamics is the same as that in our previous studies (e.g., \citealt{Sekiguchi2010a, fujibayashi2017a,Fujibayashi2020c}). To account for the energy generation by the nuclear burning, the mass fractions of several nuclear species are evolved. Neutrino radiation transport is taken into account with an energy-integrated truncated moment formalism~\citep{Thorne1981a, shibata2011a}. In the supermassive star collapse, the optical depth of the matter to neutrinos is always low, and hence, the neutrino process simply acts as cooling.

\subsection{Nuclear burning}
In addition to the usual hydrodynamical variables, mass fractions of several nuclear species are solved. Here, we summarize the basic equations that govern their evolution.

\subsubsection{Basic equation}
The equation for the number density of a nuclear species, $n_I$, is written as
\begin{align}
\del_\mu( \sqrt{-g} n_I u^\mu) = \sqrt{-g} \dot{n}_I|_\mathrm{reac},\label{eq1}
\end{align}
where $g$ is the determinant of the spacetime metric, $u^\mu$ is the common four velocity of the fluid, and $\dot{n}_I|_\mathrm{reac}$ is the change rate in the number density of the $I$-th nuclear species in the fluid rest frame.\footnote{Throughout this paper, Greek and Latin indices (except for $I$) run over spacetime and space, respectively. The subscript $I$ denotes a species of particles.} We define a mass fraction $X_I= A_I n_I/n_\mathrm{b}$, where $A_I$ is the mass number of the $I$-th species, and $n_\mathrm{b}$ is the baryon number density. Then Eq.~\eqref{eq1} is rewritten to the evolution equation for the mass fraction as
\begin{align}
\del_\mu( \rho_* X_I u^\mu) = A_I\amu \sqrt{-g} \dot{n}_I|_\mathrm{reac}, \label{eq:mass-fraction}
\end{align}
where $\amu$ is the atomic mass unit, and $\rho_* = \rho u^t \sqrt{-g}$ is the conserved rest-mass density with $\rho=\amu n_\mathrm{b}$.

Following \cite{Uchida2017oct} and \cite{Montero2012apr}, we consider the CNO-cycle of the hydrogen burning and triple-alpha reaction of the helium burning for the relevant energy-generation processes. We thus consider only three nuclear species, \isotope{1}{H}, \isotope{4}{He}, and ``CNO species", which are denoted by subscripts $p$, $\alpha$, and CNO, respectively. The ``CNO species" denotes the total species that catalyse the CNO cycle (isotopes of carbon, nitrogen, and oxygen). The right-hand side of Eq.~\eqref{eq:mass-fraction} is calculated so that the energy-generation rate of each reaction becomes consistent with that in \cite{Montero2012apr}.

\subsubsection{CNO-cycle}
For temperatures of $T\gtrsim\SI{1e7}{K}$, the CNO cycle dominates the hydrogen burning. The CNO cycle in a low temperature region is called cold CNO cycle, which is what we find in the context of hydrostatic stellar nucleosynthesis. The number in the cold CNO cycles that occur per unit time per unit volume is described as
\begin{align}
\dot{n}_\mathrm{cCNO} &= \frac{\rho \epsilon_\mathrm{cCNO}}{Q_\mathrm{cCNO}} \notag \\
&= \SI{1.1e30}{cm^{-3}.s^{-1}}\ \rho^2 X_p X_\mathrm{CNO}\notag\\
&\times \bigg[{T_9}^{-2/3}\exp\big(-15.231{T_9}^{-1/3}\big)\notag\\
&+\SI{8.3e-5}{} T_9^{-3/2}\exp\big(-3.0057{T_9}^{-1}\big)\bigg], \label{eq:ccno}
\end{align}
where $T_9 := T / (10^{9}K)$, $X_p$ and $X_\mathrm{CNO}$ are mass fractions of \isotope{1}{H} and CNO species. Equation~\eqref{eq:ccno} is derived from the expression in \cite{Shen2007may,Wiescher1999jun} with the liberated energy $Q_\mathrm{cCNO} = (26.73-0.71-1.00)\SI{}{MeV}$, which is the difference of the masses of four \isotope{1}{H} and one \isotope{4}{He} with the subtraction of the average energies of two neutrinos emitted by beta decays of $\isotope{13}{N}$ and $\isotope{15}{O}$ (the values are taken from the Evaluated Nuclear Data File (ENDF) database \footnote{\url{https://www-nds.iaea.org/public/download-endf/ENDF-B-VIII.0/decay/}}).

The \textit{hot} CNO cycle is the dominant process if the timescale of the reaction $\isotope{13}{N}(p,\gamma)\isotope{14}{O}$ is shorter than that of $\isotope{13}{N}(e^+\nu_e)\isotope{13}{C}$ (the half life is approximately $\SI{10}{min}$). The bottleneck reactions of the cycle are the $\isotope{14}{O}(e^+\nu_e)\isotope{14}{N}$ and $\isotope{15}{O}(e^+\nu_e)\isotope{15}{N}$ beta decays. Therefore, the hot CNO sycle has a temperature-independent rate
\begin{align}
\dot{q}_\mathrm{hCNO} = \SI{4.6e15}{erg.g^{-1}.s^{-1}}\ X_\mathrm{CNO}. \label{eq:hcno}
\end{align}
Considering the average energies of neutrinos for the two decays, \SI{1.05}{MeV} and \SI{1.00}{MeV}, respectively, dividing the energy generation rate by $Q_\mathrm{hCNO} = (26.73 - 1.05 - 1.00)\SI{}{MeV}$ yields
\begin{align}
\dot{n}_\mathrm{hCNO} = \SI{1.16e20}{cm^{-3}.s^{-1}} \ \rho X_\mathrm{CNO}.
\end{align}
Since we do not solve all nuclear species relevant for the cold and hot CNO cycles, we do not know the abundance of the bottleneck species, which are \isotope{14}{N} for cold, and \isotope{14}{O} and \isotope{15}{O} for hot CNO cycles, respectively. Therefore, we simply assume that the mass fraction of these species is the same as that of the CNO species.

When the hot CNO cycle works in a high-temperature environment, the cold CNO cycle does not work any more. Because the expression of the cold CNO cycle (Eq.~\eqref{eq:hcno}) is lower than that of the hot CNO cycle (Eq.~\eqref{eq:ccno}) when the latter works for the hydrogen burning, to account for the transition from cold to hot CNO cycles, we take the smaller rate for the total process as
\begin{align}
\dot{n}_\mathrm{CNO} = \min(\dot{n}_\mathrm{cCNO}, \dot{n}_\mathrm{hCNO}).
\end{align}
Then the source terms for the mass fractions of \isotope{1}{H} and \isotope{4}{He} for this process are, respectively,
\begin{align}
\dot{n}_p|_\mathrm{CNO} &= -4\dot{n}_\mathrm{CNO},\\
\dot{n}_\alpha|_\mathrm{CNO} &= \dot{n}_\mathrm{CNO}.
\end{align}

\subsubsection{Triple-alpha reaction}
The energy generation rate of the triple-alpha reaction is (e.g., \citealt{1990sse..book.....K})
\begin{align}
\dot{q}_{3\alpha} = \SI{5.1e8}{erg.g^{-1}.s^{-1}}\ \rho^2 X_\alpha^3 T_9^{-3} \exp\big(-4.4T_9^{-1}\big),
\end{align}
where $X_\alpha$ is the mass fraction of \isotope{4}{He}.
We can derive the number rate of the reaction, $\dot{n}_{3\alpha}$, simply by multiplying $\rho$ and dividing $3m(\isotope{4}{He})-m(\isotope{12}{C}) = 3\cdot \SI{2.425}{MeV} = \SI{1.166e-5}{erg}$ (no neutrino loss) as
\begin{align}
\dot{n}_{3\alpha} = \SI{4.37e13}{cm^{-3}.s^{-1}}\ \rho^3X_\alpha^3 T_9^{-3} \exp\big(-4.4T_9^{-1}\big).
\end{align}
Using this rate, the source terms for the mass fractions of \isotope{4}{He} and CNO species are written as
\begin{align}
\dot{n}_\alpha|_{3\alpha} &= -3\dot{n}_\mathrm{3\alpha},\\
\dot{n}_\mathrm{CNO}|_{3\alpha} &= \dot{n}_\mathrm{3\alpha}.
\end{align}
In summary, the evolution equations of the mass fractions are written as
\begin{align}
\del_\mu( \rho_* X_p u^\mu) &= \amu \sqrt{-g}(-4 \dot{n}_\mathrm{CNO}),\\
\del_\mu( \rho_* X_\alpha u^\mu) &= 4 \amu \sqrt{-g} ( \dot{n}_\mathrm{CNO} - 3\dot{n}_\mathrm{3\alpha}),\\
\del_\mu( \rho_* X_\mathrm{CNO} u^\mu) &= 12 \amu \sqrt{-g} ( \dot{n}_\mathrm{3\alpha}).
\end{align}

% \subsection{Summary of the system equation}
% The evolution equations of mass fractions of the free proton, alpha particle, and $\isotope{12}{C}$ are
% \begin{align}
% \frac{D}{Dt} X_p &= 1\cdot\frac{\amu}{\rho u^t} (-4 \dot{n}_\mathrm{CNO}),\label{eq:dXp/dt}\\
% \frac{D}{Dt} X_\alpha &= 4\cdot\frac{\amu}{\rho u^t} (\dot{n}_\mathrm{CNO} - 3\dot{n}_{3\alpha}),\\
% \frac{D}{Dt} X_\mathrm{C} &= 12\cdot\frac{\amu}{\rho u^t} \dot{n}_{3\alpha},
% \end{align}
% where $D/Dt = \del_t + v^k \del_k$. They satisfy $D/Dt (X_p+X_\alpha+X_\mathrm{C}) = 0$.

\subsection{Equation of state}
In the present work, we assume that the matter consists of ions with a mixture of \isotope{1}{H}, \isotope{4}{He}, and CNO species, photons, electrons, and thermally generated electron-positron ($e^-e^+$) pairs. We further assume that the atoms are fully ionized, because this is a good approximation inside the star, and the photons are thermalized with the same temperature as the ions and electrons. Then, the specific internal energy, $\varepsilon$, is divided into the three components as
\begin{align}
\varepsilon = \varepsilon_\mathrm{ion} + \varepsilon_\gamma + \varepsilon_\mathrm{e}, \label{eq:eps-tot}
\end{align}
where the contribution of the ions, photons, electrons and positrons can be written as
\begin{align}
\varepsilon_\mathrm{ion} &= \frac{3}{2}\frac{k_\mathrm{B}T}{\langle A \rangle \amu} + \frac{\langle \Delta m\rangle c^2}{\amu},\\
\varepsilon_\gamma &= \frac{a_\mathrm{rad}T^4}{\rho},\\
\varepsilon_\mathrm{e} &= \frac{e_\mathrm{e} (n_\mathrm{e}, T)}{\rho}.
\end{align}
Here, $a_\mathrm{rad}$ is the radiation constant, $e_\mathrm{e}$ is the internal energy density of electrons (including the rest mass of $e^-e^+$ pairs), $\langle A\rangle$ is the average mass number of the ions, and $\langle \Delta m\rangle$ is the average mass excess per baryon (which also includes the mass of electrons). The definitions of the latter two are
\begin{align}
\frac{1}{\langle A \rangle} &:=\sum_I \frac{X_I}{A_I},\\
\langle \Delta m \rangle &:= \sum_I (m_I-A_I\amu) \frac{X_I}{A_I} = \sum_I \Delta m_I \frac{X_I}{A_I},
\end{align}
with $\Delta m_I = m_I - A_I \amu$, where $m_I$ is the atomic mass, which contains the mass of electrons $Z_I m_\mathrm{e}$. For the CNO species, we assumed $A_\mathrm{CNO}=12$, $Z_\mathrm{CNO}=6$, and $\Delta m_\mathrm{CNO}=0$.

In the same manner, the pressure $P$ is divided into three components as
\begin{align}
P = P_\mathrm{ion} + P_\gamma + P_\mathrm{e}\, ,
\end{align}
where
\begin{align}
P_\mathrm{ion} &= \frac{\rho k_\mathrm{B}T}{\langle A \rangle \amu},\\
P_\gamma &= \frac{a_\mathrm{rad}}{3}T^4,\\
P_\mathrm{e} &= P_\mathrm{e} (n_\mathrm{e}, T).
\end{align}
The internal energy density and the partial pressure of electrons and thermal $e^-e^+$ pairs, $e_e$ and $P_e$, are functions of the net number density of electrons $n_\mathrm{e}$ and temperature $T$, and are tabulated using the Timmes equation of state~\citep{timmes2000a}. The detailed implementation is described in Appendix~\ref{app:eeos}. The electron fraction $Y_\mathrm{e} = n_\mathrm{e}/n_\mathrm{b}$ is defined, by using the charge neutrality, as
\begin{align}
Y_\mathrm{e} = \sum_I \frac{Z_I}{A_I} X_I.
\end{align}

% \addsf{The stress-energy tensor of the system is
% \begin{align}
% T_{\mu\nu} = \rho h u_\mu u_\nu + P g_{\mu\nu}\label{eq:tmunu}
% \end{align}
% where $h=1+\varepsilon/c^2 + P/\rho c^2$ is the specific enthalpy. It is worth mentioning that Eq.~\eqref{eq:tmunu} properly takes into account the difference of the rest masses of nuclear species by the term
% \begin{align}
% \rho \bigg(1+\frac{\langle \Delta m\rangle}{\amu}\bigg) = \sum_I m_I n_I.
% \end{align}
% When we perform viscous hydrodynamics simulations as in Sec.~\ref{subsec:vis}, we add terms accounting for the viscosity, and evolve the viscous tensor following \cite{shibata2017apr}.
% }

\subsection{Neutrino transfer}
Neutrinos are emitted as a result of the hydrogen burning through beta-decays. We have to implement this process to take into account the energy generation rate consistently. We solve the neutrino transfer equations with a truncated moment formalism~\citep{shibata2011a} in the energy-integrated way. In this formalism, the energy and flux density of neutrinos, which are denoted by $E$ and $F_i$, respectively, are evolved. In Cartesian coordinates, their evolution equations are written as
\begin{align}
\del_t(\sqrt{\gamma}E) &+ \del_k[\sqrt{\gamma}(\alpha F^k - \beta^k E)] \notag\\
&= \sqrt{\gamma} [\alpha P^{ij} K_{ij} - F^k\del_k \alpha] \notag\\
&\ \ - \alpha \sqrt{\gamma} S^\alpha n_\alpha,\\
\del_t(\sqrt{\gamma}F_i) &+ \del_k[\sqrt{\gamma}(\alpha P^k{}_i - \beta^k F_i)]\notag\\
&= \sqrt{\gamma} \bigg[-E\del_i \alpha + F_k \del_i\beta^k +\frac{\alpha}{2} P^{kl} \del_i \gamma_{kl}\bigg] \notag\\
&\ \ + \alpha \sqrt{\gamma} S^\alpha \gamma_{i\alpha},
\end{align}
where $\gamma_{ij}$ and $\gamma$ are the induced three-metric and its determinant, $\alpha$ and $\beta^i$ are the lapse function and shift vector, $K_{ij}$ is the extrinsic curvature, and $n_\mu = (-\alpha,0,0,0)$ is the time-like unit vector orthogonal to spatial hypersurfaces of the constant time coordinate. $S^\alpha$ is the source term due to the reaction. To close the system, the second moment $P_{ij}$ is approximated by the M1-closure as in~\cite{fujibayashi2017a}.

The matter in the present simulations has only a tiny opacity to neutrinos, and thus, the neutrinos propagate essentially freely after being generated. Therefore, we do not take any absorption and scattering processes into account in the neutrino transfer.

% In cylindrical coordinates, there is geometrical source term for the direction of the cylindrical radius:
% \begin{align}
% \del_t(\sqrt{\gamma}F_x) &+ \del_k[\sqrt{\gamma}(\alpha P^k{}_x - \beta^k F_x)] \notag\\
% &= -\frac{1}{x}\bigg[\alpha\sqrt{\gamma} (P_x{}^x-P_y{}^y)-\sqrt{\gamma}(\beta^xF_x - \beta^yF_y)\bigg]\notag\\
% &\ \ +(\text{the other terms})
% \end{align}

% \subsubsection{Source term}
We only consider the electron-type neutrinos emitted from the CNO cycle. In this  process, the extracted energy is $\approx \SI{2}{MeV}$ per cycle. Thus, energy emission rate per unit volume per unit time in the fluid rest frame is 
\begin{align}
Q_{\nu,\mathrm{CNO}} \approx \SI{2}{MeV}\ \dot{n}_\mathrm{CNO},
\end{align}
which implies that the source term can be written as
\begin{align}
S^\alpha &=Q_{\nu,\mathrm{CNO}} u^\alpha,\\
\alpha \sqrt{\gamma} S^\alpha n_\alpha &= \alpha \sqrt{\gamma} Q_{\nu,\mathrm{CNO}} w,\\
\alpha \sqrt{\gamma} S^\alpha \gamma_{\alpha i} &= \alpha \sqrt{\gamma} Q_{\nu,\mathrm{CNO}} u_i,
\end{align}
where $w=\sqrt{1+\gamma^{ij} u_i u_j}$ is the Lorentz factor. In this work, the cooling by the thermal production of neutrinos (e.g., those presented in \citealt{Itoh1996feb}) are not taken into account because such an effect is not important in low-density and low-temperature environment in collapse of supermassive stars (see Sec.~\ref{subsec:neutrino}; \citealt{Uchida2017oct}).
% To be consistent with the actual change in the proton mass fraction, it is better to rewrite the term as
% \begin{align}
% \alpha \sqrt{\gamma} Q_{\nu,\mathrm{CNO}} &= \SI{2}{MeV} \cdot \alpha \sqrt{\gamma}\bigg(-\frac{\dot{n}_p}{4}\bigg)\notag\\
% &= \SI{2}{MeV} \cdot \frac{\rho_*}{1\cdot \amu} \bigg(-\frac{1}{4}\frac{DX_p}{Dt}\bigg),
% \end{align}
% where $DX_p/Dt := (\del_t + v^k\del_k) X_p = \amu\dot{n}_p|_\mathrm{CNO}/\rho$. When the change in proton mass fraction $\Delta X_p = \Delta t \cdot DX_p/Dt$ is obtained, the corresponding change in the energy and momentum density of neutrino and matter is
% \begin{align}
% \Delta (\sqrt{\gamma}E_{\nue}) &= \SI{2}{MeV}\cdot \frac{\rho_*}{\amu} \bigg(-\frac{1}{4}\Delta X_p\bigg) w,\\
% \Delta (\sqrt{\gamma}F_{i,\nue}) &= \SI{2}{MeV}\cdot \frac{\rho_*}{\amu} \bigg(-\frac{1}{4}\Delta X_p\bigg) u_i,\\
% \Delta S_0 &= -\Delta (\sqrt{\gamma}E_{\nue}),\\
% \Delta S_i &= -\Delta (\sqrt{\gamma}F_{i,\nue}).
% \end{align}

\begin{table*}[]
    \centering
    \caption{List of initial data and their key properties. From left to right, the model name, gravitational mass, equatorial radius, the ratio of kinetic to gravitational potential energy, central lapse, central adiabatic index minus 4/3, the parameter that indicates the degree of differential rotation $\hat A$, and entropy per baryon. Note that the total baryon rest mass is approximately equal to the gravitational mass.}
    \begin{tabular}{cccccccc}
    \hline
    \hline
    model  & $M_0$ $(M_\odot)$ & $R_{\mathrm{e0}}$ (cm) & $T_\mathrm{kin}/|W|$ & $\alpha_{\mathrm{c,0}}$ & $\gamma_\mathrm{c,0}-4/3$ & $\hat A$ & $s/k_\mathrm{B}$\\
    \hline
        H1 &\SI{2.1e5}{} & \SI{1.7e13}{} &0.002& 0.992 & 0.0026 & $\infty$ & 450\\
        H2 &\SI{3.2e5}{} & \SI{2.3e13}{} &0.004& 0.990 & 0.0021 & $\infty$ & 550\\
        H3 &\SI{4.3e5}{} & \SI{2.7e13}{} &0.006& 0.988 & 0.0018 & $\infty$ & 630\\
        H4 &\SI{6.9e5}{} & \SI{4.4e13}{} &0.009& 0.985 & 0.0014 & $\infty$ & 800 \\
        Hdif1 &\SI{9.2e5}{} & \SI{5.0e13}{} &0.011& 0.983 & 0.0012 & 2     & 920 \\
        Hdif2 &\SI{1.1e6}{} & \SI{5.3e13}{} &0.013& 0.981 & 0.0012 & 1.5   & 1000\\
        Hdif3 &\SI{1.9e6}{} & \SI{7.4e13}{} &0.018& 0.976 & 0.0009 & 1.0   & 1300 \\
\hline
        He1 &\SI{5.0e4}{} & \SI{4.3e12}{} &0.002& 0.992 & 0.0023 & $\infty$ & 210 \\
        He2 &\SI{7.1e4}{} & \SI{5.1e12}{} &0.004& 0.990 & 0.0019 & $\infty$ & 250 \\
        He3 &\SI{9.6e4}{} & \SI{6.1e12}{} &0.006& 0.988 & 0.0016 & $\infty$ & 300 \\
        He4 &\SI{1.6e5}{} & \SI{1.0e13}{} &0.009& 0.985 & 0.0013 & $\infty$ & 380 \\
\hline
    \end{tabular}
    \label{tab:ID-list}
\end{table*}

\subsection{Initial profiles of supermassive stars}
As the initial conditions of the simulations, we employ marginally stable general relativistic equilibrium states of supermassive star cores, which are constructed in the same way as in \cite{Uchida2017oct}. In the construction, we assume uniform radiation entropy per baryon and uniform composition. This is a good approximation for the supermassive star cores fully mixed by convection. 
%(but be careful for the outer region of accreting supermassive stars, for which this is not the case in general; e.g., \cite{Hosokawa2013dec}). 
With this assumption, together with the fact that the system is highly radiation-pressure-dominated, a polytropic equation of state with the polytropic index close to $3$ can be used to construct the equilibrium states. 

The stability of supermassive star cores against the general relativistic instability is identified in terms of a fitting formula derived in \cite{Shibata2016feb} (i.e., Eq.~(28) of that paper). Although this formula is valid only for rigidly rotating supermassive star cores, we use it for approximately identifying the stability of the differentially rotating case. Our present numerical simulations show that the formula works well for identifying the stability at least for moderately differentially rotating cases with $\hat A \geq 1$ (see below).

A word of caution is appropriate here. Supermassive stars in reality are likely to increase their mass with a very high accretion rate $\gtrsim0.1M_\odot$/yr until the onset of the general relativistic instability. As a result, they are likely to have an inflated envelope with radius $\sim\SI{e15}{cm}$~\citep{Hosokawa2013dec,Umeda2016oct,Saio2024arxiv}. Such a supermassive star has a convective region only in its core. Therefore, our initial data focus on the core of accreting supermassive stars. As the envelope is very dilute, it is not likely that the envelope has a significant effect on the core collapse and subsequent black hole plus torus formation.

The initial data are listed in Table~\ref{tab:ID-list}. For the ``H"-series, we assume that the general relativistic instability sets in during the early hydrogen burning phase and thus the stellar composition is assumed to be primordial with $X_p=0.75$ and $X_\alpha=0.25$ together with a low metallicity of $X_\mathrm{C}=\SI{5e-9}{}$~\citep{Bond:1984sn}. 
%This corresponds to the case in which the general relativistic instability sets in before the significant change in its composition. 
Assuming that the energy generation rate is equal to the Eddington luminosity, we find the central temperature as $T_\mathrm{c}\approx\SI{1.5e8}{K}$. For the ``He"-series, on the other hand, we assume that the instability sets in at the beginning of the helium burning phase and thus we initialize the star with $X_\alpha=1$ and $T_\mathrm{c}\approx \SI{3e8}{K}$. %This corresponds to the case, instead, in which the star becomes unstable in the beginning of the helium burning phase. 
We note that the models H1, He1, H4, and He4 are essentially the same as models A1, A2, A3, and A4 in \cite{Uchida2017oct}.

For most of the models, we assume rigid rotation for the angular velocity. The H4 and He4 models are at mass-shedding limit, i.e., the rotation velocity at the surface of the core in the equatorial plane is that of a Keplerian orbit. The other rigid rotation models (H1--H3 and He1--He3) have smaller surface velocity (see the column of $T_\mathrm{kin}/|W|$ of Table~\ref{tab:ID-list}, which indicates how fast the star rotates). Here, the kinetic energy and proper mass of the system are defined by
\begin{align}
T_\mathrm{kin} &= \int \rho u^t \sqrt{-g} \big(1+\langle\Delta m\rangle/\amu\big) \frac{1}{2} (w^2-1) d^3x, \label{eq:tkin}\\
M_\mathrm{p} &= \int \rho u^t \sqrt{-g} \big(1+\varepsilon/c^2) d^3x,
\end{align}
where the factor $(1+\langle\Delta m\rangle/\amu)$ in Eq.~\eqref{eq:tkin} corrects the difference of the mass per baryon from $\amu$. We then define the gravitational potential energy of the system as $W= M_0c^2 - M_\mathrm{p}c^2 - T_\mathrm{kin}$, where $M_0$ is the gravitational (Arnowitt-Deser-Misner) mass of the system~\citep{arnowitt1960a}. To explore the effect of more rapid rotations, we prepare ``Hdif"-series, for which we assume a differential rotation using the so-called $j$-constant law with varying the degree of differential rotation as $\hat A=1$--2 (\citealt{Baumgarte:1999cq}; see Table~\ref{tab:ID-list} for the value of $\hat{A}$). Note that $\hat{A}\rightarrow\infty$ corresponds to the rigidly rotating configuration. For these models, the axial ratio (polar radius to equatorial radius) is set to be the same as that of H4, which is $\approx 2/3$. 

In Table~\ref{tab:ID-list}, we find that the central adiabatic index is closer to $4/3$ and that the central lapse is smaller for supermassive star cores with higher values of $T_\mathrm{kin}/|W|$. This indicates that the rotation stabilises the supermassive star cores against the general relativistic instability, and hence, the rotating stars have to be more compact and radiation-dominated to become unstable (details are discussed in our accompanying paper~\citealt{Shibata2024prep}).

The mass of the marginally stable supermassive star cores which are studied in this paper is in the range between $\approx 2\times 10^5$ and $\approx 2\times 10^6M_\odot$ for the hydrogen-burning models and between $\approx 5 \times 10^4$ and $\approx 2\times 10^5M_\odot$ for the helium-burning models. For a given value of $T_\mathrm{kin}/|W|$, the mass of the marginally stable supermassive star becomes smaller for more evolved one. For example, for an evolved supermassive star core in the helium burning phase, the fractions of carbon and oxigen increase. For such a supermassive star, the threshold mass for the collapse associated with the general relativistic instability is by a factor of 2--3 lower than $5\times 10^4M_\odot$ \citep{Shibata2024prep}. In this paper we do not pay attention to such relatively low-mass supermassive star cores but only to high-mass ones. In the follow-up work, we plan to explore the fate of the collapse for the low-mass supermassive star cores.

% \begin{table}[]
%     \centering
%     \caption{List of the grid parameters.}
%     \begin{tabular}{cccccc}
%     \hline
%     Res.  & $dx_0/r_\mathrm{g}$ & $r_\mathrm{uniform}/r_\mathrm{g}$ & $L/r_\mathrm{g}$ & $N$ & $\eta$\\
%     \hline
%         L & 0.5$\rightarrow$0.040 & 1 & 1800 & 236 $\rightarrow$ 423 & 0.0117$\rightarrow$0.0117 \\
%         H & 0.5$\rightarrow$0.025 & 1 & 1800 & 236 $\rightarrow$ 487 & 0.0117$\rightarrow$0.0117\\
%     \end{tabular}
%     \label{tab:coor}
% \end{table}

\subsection{Grid setup}
Following our previous works~(e.g., \citealt{Fujibayashi2020a, Fujibayashi2020b, Fujibayashi2020c}), we employ cylindrical coordinates denoted by $(x,z)$ with mirror symmetry with respect to the $z=0$ plane. For both directions, the grid is assigned in the following manner: $x_i = x_{i-1} + dx_i$ with the innermost grid located at $x_0=0$ and $dx_i = dx_0$ for $x_{i-1} < r_\mathrm{uniform}$. Otherwise, $dx_i = dx_{i-1} (1+\eta)$ with a small number $\eta>0$. Here $i=0$--$N$ with $N$ denoting the grid size. 

In this work, the grid spacing is determined so that the number of grid points for a large radius is suppressed, while keeping enough angular resolution in that region. For a given innermost grid spacing $dx_0$, the size of the uniform-grid region $r_\mathrm{uniform}$, the location of the outer boundary $L = x_{N-1/2}:= (x_{N-1} + x_{N})/2$ ($i=N$ corresponds to the first ghost cell), and approximate angular resolution $dx_{N-1}/x_{N-1/2}$, we determine $N$ and $\eta$. In this study, we always take $L=1800\,r_\mathrm{g}$, $r_\mathrm{uniform} = r_\mathrm{g}$, and $dx_{N-1}/L=(\pi/2)/96$, where $r_\mathrm{g}=GM_0/c^2$. %with the initial gravitational mass of the system $M_0$.

\begin{figure*}[t]
    \centering
    \includegraphics[width=0.49\textwidth]{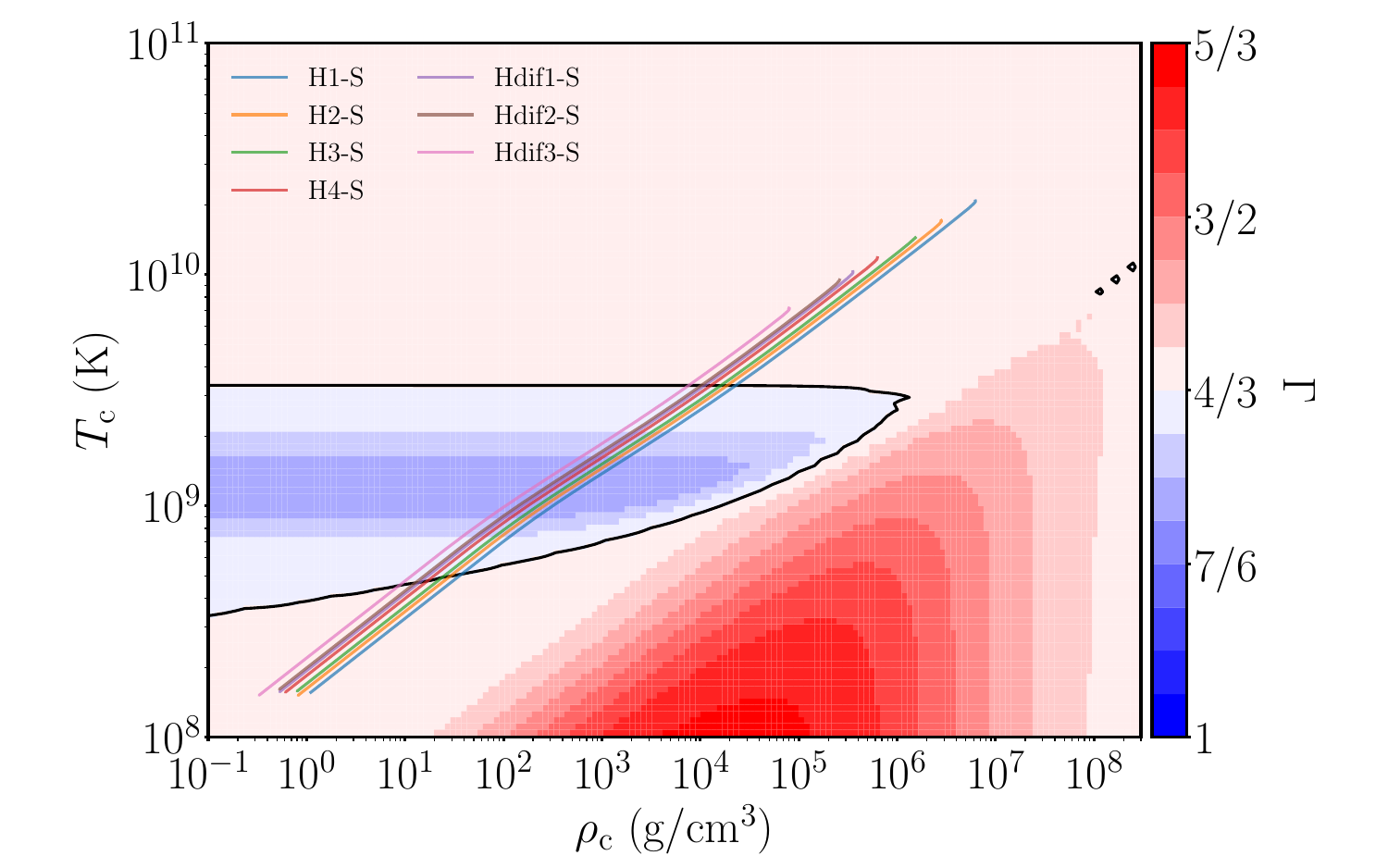}
    \includegraphics[width=0.49\textwidth]{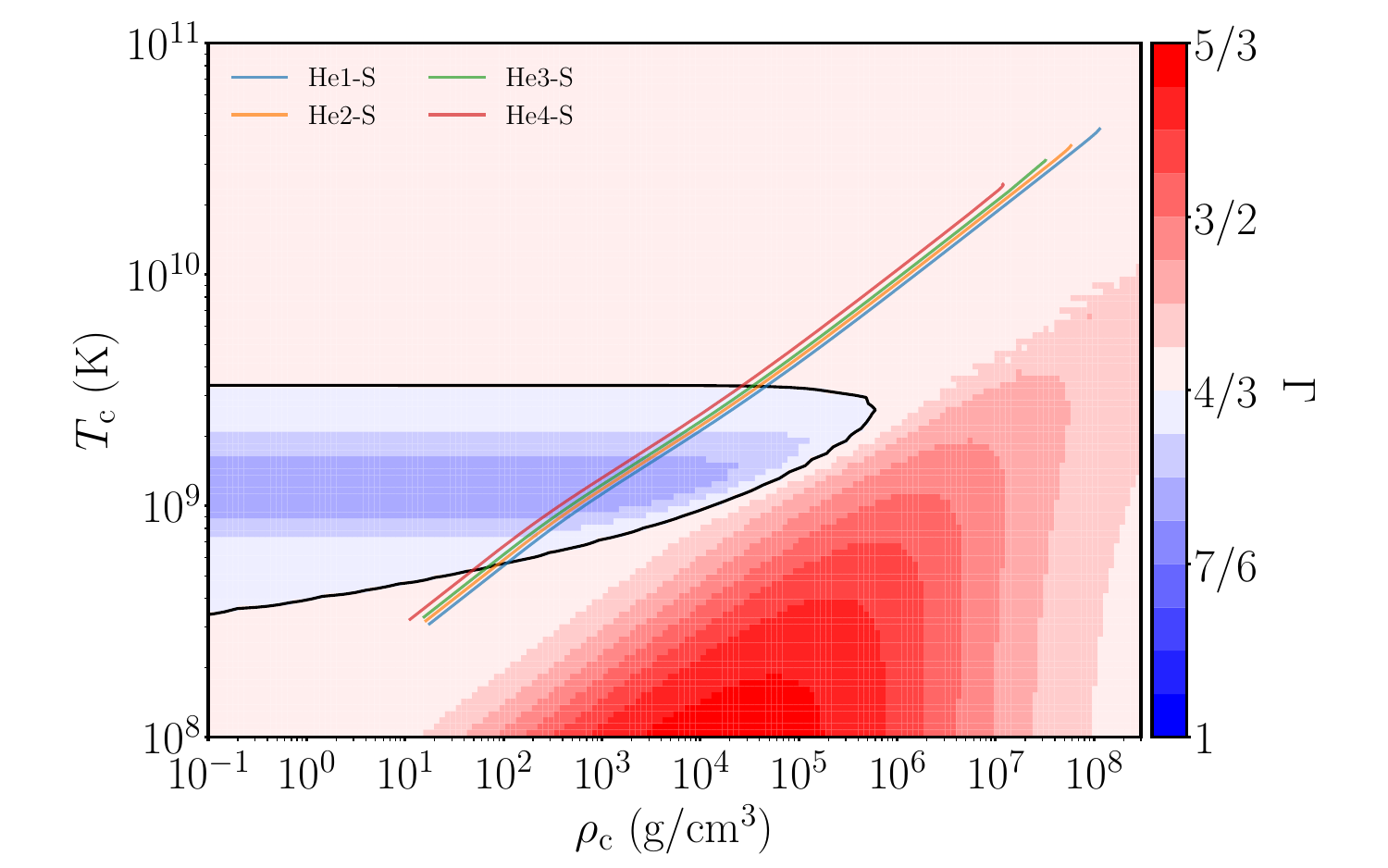}
    \caption{Evolution of the central density and central temperature of supermassive star cores up to the black hole formation (solid curves). The adiabatic index is shown in color. The black curves denote the contour of $\Gamma=4/3$.}
    \label{fig:gamma-c}
\end{figure*}

\subsection{Regridding} \label{subsec:regridding}
The collapsing supermassive stars are becoming more and more compact with time, and thus, a grid which resolves the curvature scale of the collapsing matter at each moment is required. In addition, to numerically evolve the formed black hole accurately, a sufficient grid resolution with $dx_0 \ll r_\mathrm{g}$ is necessary, although we need such high grid resolutions only after a black hole is formed. To save computational resources, thus, we employ a \textit{regridding} algorithm (e.g., \citealt{Shibata:2002br}) in this work.

For the first run of each simulation, we take $dx_0 = 0.5r_\mathrm{g}$ and decrease it successively during the collapse. For each run with a given value of $dx_0$, we determine the time of regridding using the central lapse $\alpha_\mathrm{c}$, because it indicates the effective compactness of the star, $GM/c^2R \sim 1-\alpha_\mathrm{c}$. When $1-\alpha_\mathrm{c}$ of a run becomes half of that at its start time, we stop the run and restart the simulation with finer grid resolutions. We set a new value of $dx_0$ which is typically half of that in the previous run. As $dx_0$ decreases, the total number of grid points and the parameter $\eta$ are set to keep the same values of $L$ and $dx_{N-1}/L$. By doing this, the collapsing stellar radius is always resolved with a similar number of grids. 

In the regridding process, we map the quantities to be evolved onto the new, finer grid points with a third-order Lagrange interpolation scheme. At the beginning of each run, assuming a conformally flat space, the constraint equations of general relativity are solved to obtain the initial condition for the metric variables based on a predetermined energy-momentum distribution. We note that assuming the spatial conformal flatness is a good approximation because the maximum non-diagonal component of the spatial metric $\tilde{\gamma}_{ij}$ is always in the order of $10^{-4}$ when the criterion of the regridding is satisfied. In other words, the regridding has to be performed for the case that the collapsing star is not very compact. 
%We then solve the constraint equations assuming a conformal-flat metric. 
In the present work, the final regridding is performed when $\alpha_\mathrm{c}=0.85$. For the final run,  we choose $dx_0=0.04r_\mathrm{g}$ and $0.025r_\mathrm{g}$ for standard- and high- resolution runs, respectively (they are labeled with the letters ``L" and ``H", respectively). 

The coordinate parameters for the first run are calculated as $(N,\eta)=(252, 0.01668)$. After the final regridding, on the other hand, they are $(N,\eta)=(424, 0.01682)$ and $(488, 0.01693)$ for standard- and high-resolution runs, respectively. 

%We note that from the viewpoint of the numerical stability, such a large value of $\eta$ is not suitable for longterm simulations with the duration of $\sim 10^5M_0$. However, in the present work, the value of $\eta ¥approx 0.017$ is acceptable. 

\begin{table*}[]
    \centering
    \caption{Key results. From left to right, the model name, ejecta mass, asymptotic kinetic energy and average velocity of the ejecta, mass and dimensionless spin of the formed black hole. In the third and fifth columns, the ratios of the ejecta to total mass and kinetic to total mass energy are also shown in the parentheses. The mass and dimensionless spin of the black hole are measured at $t-t_\mathrm{BH}=1000GM_0/c^3$ for each simulation. The black hole mass and dimensionless spin remain blank for the viscous models see those for H4-S.}
    \begin{tabular}{lccccccc}
    \hline
    \hline
    model & $M_\mathrm{ej}$ & ($M_\mathrm{ej}/M_0$) & $K_\mathrm{ej}$ & ($K_\mathrm{ej}/M_0c^2$) & $V_\mathrm{ej}$ & $M_\mathrm{BH}$ & $\chi_\mathrm{BH}$\\
          & ($M_\odot$)  & (\%)  & (erg) & (\%) & $(c)$ & $(M_\odot)$ &\\
    \hline
      H1-S  & \SI{2.8e+02}{} &(0.14) & \SI{9.5e+54}{} &(0.003) & 0.19 & \SI{2.09e+05}{} & 0.47\\
%       H1-S  & \SI{4.2e+02}{} &(0.20) & \SI{3.0e+54}{} &(0.001) & 0.09 & \SI{2.14e+05}{} & 0.47\\
      H2-S  & \SI{1.7e+03}{} &(0.52) & \SI{5.1e+55}{} &(0.009) & 0.18 & \SI{3.17e+05}{} & 0.58\\
      H3-S  & \SI{3.1e+03}{} &(0.74) & \SI{9.9e+55}{} &(0.013) & 0.19 & \SI{4.20e+05}{} & 0.63\\
      H4-S  & \SI{5.8e+03}{} &(0.85) & \SI{1.9e+56}{} &(0.016) & 0.19 & \SI{6.66e+05}{} & 0.67\\
   Hdif1-S  & \SI{8.1e+03}{} &(0.88) & \SI{2.8e+56}{} &(0.017) & 0.19 & \SI{8.84e+05}{} & 0.69\\
   Hdif2-S  & \SI{9.8e+03}{} &(0.90) & \SI{3.4e+56}{} &(0.017) & 0.19 & \SI{1.04e+06}{} & 0.71\\
   Hdif3-S  & \SI{1.9e+04}{} &(0.97) & \SI{5.9e+56}{} &(0.017) & 0.19 & \SI{1.85e+06}{} & 0.74\\
     He1-S  & \SI{5.5e+01}{} &(0.11) & \SI{1.7e+54}{} &(0.002) & 0.18 & \SI{5.00e+04}{} & 0.47 \\
%      He1-S  & \SI{1.2e+01}{} &(0.02) & \SI{8.2e+52}{} &(0.0001) & 0.09 & \SI{5.03e+04}{} & 0.47\\
     He2-S  & \SI{3.1e+02}{} &(0.44) & \SI{8.6e+54}{} &(0.007) & 0.17 & \SI{7.02e+04}{} & 0.58\\
     He3-S  & \SI{9.0e+02}{} &(0.94) & \SI{1.9e+55}{} &(0.011) & 0.15 & \SI{9.42e+04}{} & 0.63\\
     He4-S  & \SI{1.6e+03}{} &(0.99) & \SI{4.2e+55}{} &(0.015) & 0.17 & \SI{1.53e+05}{} & 0.67\\
 H4-v0.03M  & \SI{5.9e+03}{} &(0.86) & \SI{1.9e+56}{} &(0.016) & 0.19 & -- & -- \\
 H4-v0.10M  & \SI{1.0e+04}{} &(1.46) & \SI{2.1e+56}{} &(0.017) & 0.15 & -- & -- \\
H4-v0.03SS  & \SI{7.8e+03}{} &(1.13) & \SI{2.0e+56}{} &(0.016) & 0.17 & -- & -- \\
H4-v0.10SS  & \SI{1.3e+04}{} &(1.86) & \SI{2.2e+56}{} &(0.018) & 0.14 & -- & -- \\
\hline
      H1-H  & \SI{3.0e+02}{} &(0.14) & \SI{9.9e+54}{} &(0.003) & 0.19 & \SI{2.08e+05}{} & 0.48\\
%      H1-H  & \SI{4.7e+02}{} &(0.22) & \SI{3.8e+54}{} &(0.001) & 0.09 & \SI{2.12e+05}{} & 0.48\\
      H4-H  & \SI{6.0e+03}{} &(0.87) & \SI{2.0e+56}{} &(0.017) & 0.19 & \SI{6.58e+05}{} & 0.68\\
     He1-H  & \SI{4.9e+01}{} &(0.10) & \SI{1.5e+54}{} &(0.002) & 0.18 & \SI{4.98e+04}{} & 0.48\\
%     He1-H  & \SI{1.5e+01}{} &(0.03) & \SI{8.4e+52}{} &(0.0001) & 0.08 & \SI{4.99e+04}{} & 0.48\\
     He4-H  & \SI{1.6e+03}{} &(1.02) & \SI{4.4e+55}{} &(0.016) & 0.17 & \SI{1.51e+05}{} & 0.68\\
\hline
    \end{tabular}
    \label{tab:key-result}
\end{table*}

\subsection{Ejecta diagnostics}
We define the unbound matter (ejecta) as the component that has a positive value of the specific binding energy as
\begin{align}
e_\mathrm{bind} = \frac{-T^t{}_t}{\rho u^t} - (c^2+\varepsilon_\mathrm{min}).\label{eq:ebind}
\end{align}
This is conserved along flow lines in stationary spacetime~\citep{Uchida2017oct,fujibayashi2021oct}. 
Here, $T_{\mu\nu}$ is the energy-momentum tensor of the fluid, and $\varepsilon_\mathrm{min} := \langle \Delta m\rangle c^2/\amu$ is the minimum specific internal energy. In this study, we evolve the composition dynamically. Thus, $\varepsilon_\mathrm{min}$ is defined more precisely than those defined using an equation of state for which nuclear statistical equilibrium is assumed in their construction.

With the definition of the specific binding energy by Eq.~\eqref{eq:ebind}, the mass and asymptotic kinetic energy of the ejecta at a given time are, respectively, defined as
\begin{align}
M_\mathrm{ej} &= \int \rho u^t \sqrt{-g} \Theta(e_\mathrm{bind}) d^3x \notag\\
&\ \ + \int^t \int \rho u^k \sqrt{-g} \Theta(e_\mathrm{bind}) ds_k\, dt,\\
K_\mathrm{ej} &= \int e_\mathrm{bind} \rho u^t \sqrt{-g} \Theta(e_\mathrm{bind}) d^3x \notag\\
&\ \ + \int^t \int f_\mathrm{bind}^k \sqrt{-g} \Theta(e_\mathrm{bind}) ds_k\, dt,
\end{align}
where the first and second terms in each expression are the contributions of the matter located inside the computational domain and that flown out from the domain. $ds_k$ is the area element at the outer boundary of the computational domain, $\Theta$ is the Heaviside function, and 
\begin{align}
f_\mathrm{bind}^k = -T^k{}_t - (c^2+\varepsilon_\mathrm{min})\rho u^k
\end{align}
is the flux density associated with the energy density. % conserved in stationary spacetime. 
The average asymptotic velocity of the ejecta is then defined by
\begin{align}
V_\infty/c = \sqrt{1-\Gamma_\infty^{-2}}
\end{align}
with the average asymptotic Lorentz factor,
\begin{align}
\Gamma_\infty = 1+K_\mathrm{ej}/M_\mathrm{ej} c^2.
\end{align}

\begin{figure*}[t]
    \centering
    \includegraphics[width=0.49\textwidth]{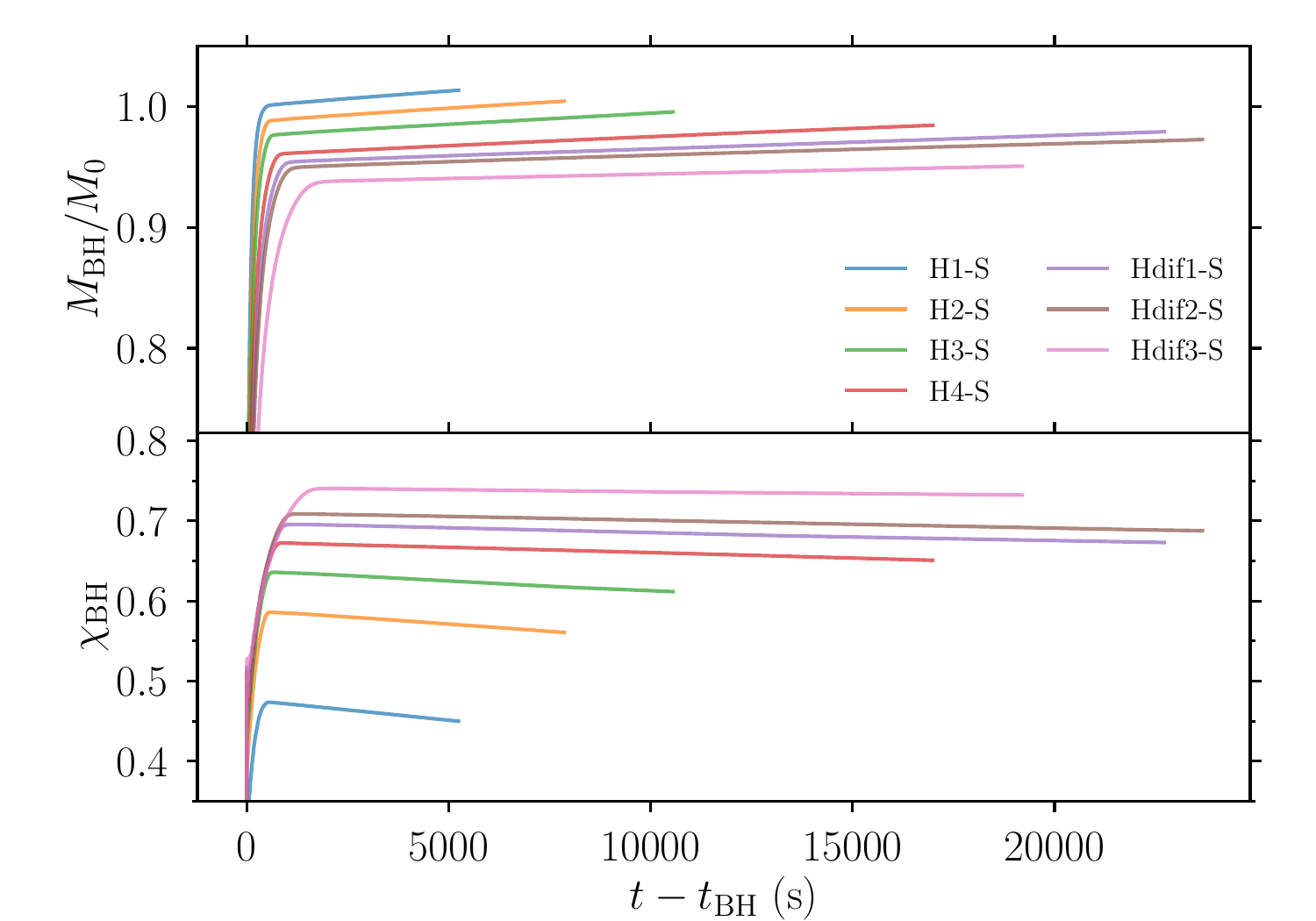}
    \includegraphics[width=0.49\textwidth]{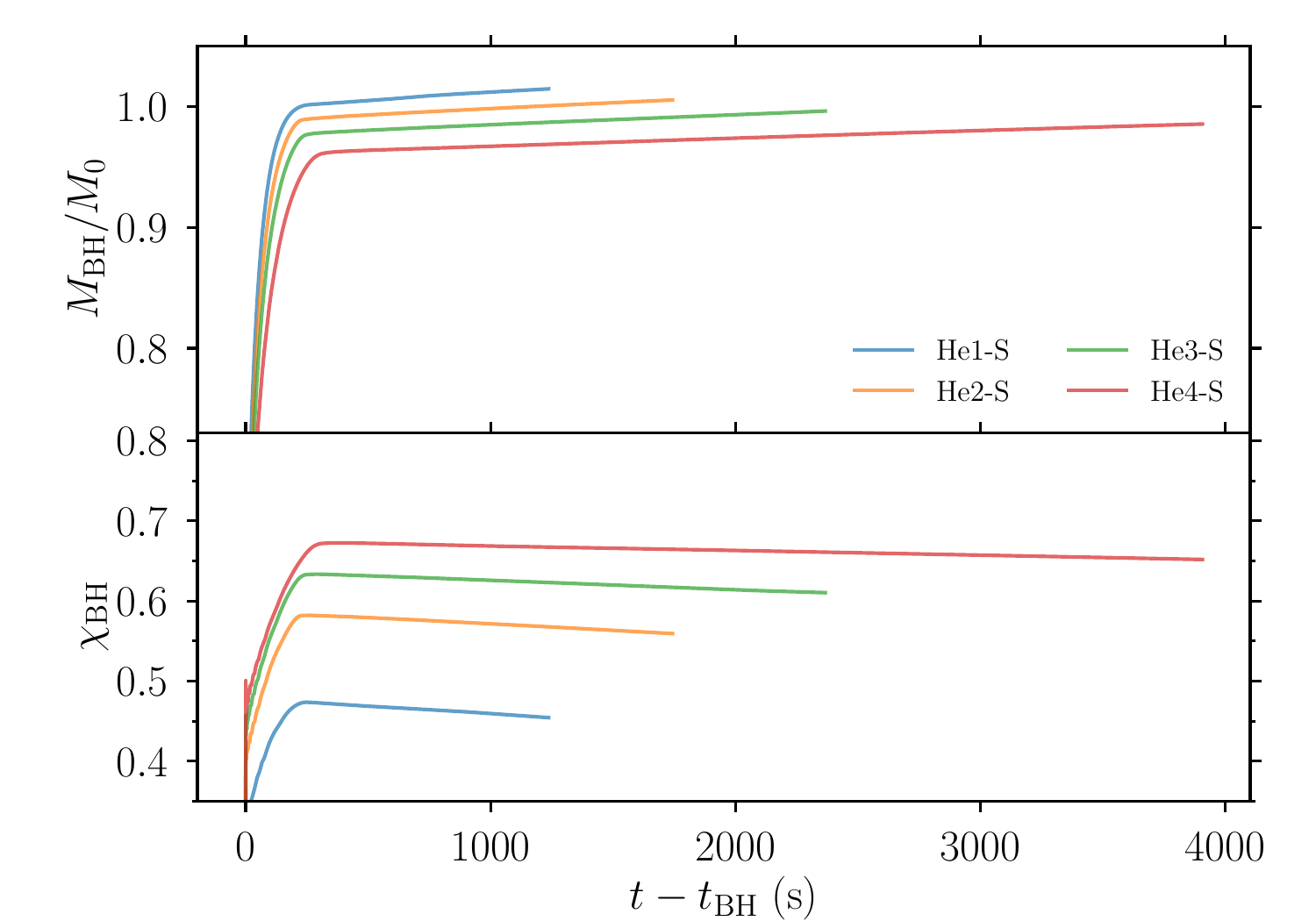}
    \caption{Time evolution of the black hole mass and dimensionless spin for all the models with standard grid resolutions studied in this paper. %Each curve is terminated at $t-t_\mathrm{BH} = 2000 GM_0/c^3$. }
    }
    \label{fig:bh}
\end{figure*}
\section{Results} \label{sec:results}
\subsection{Evolution outline}
\begin{figure*}[t]
    \centering
    \includegraphics[width=0.49\textwidth]{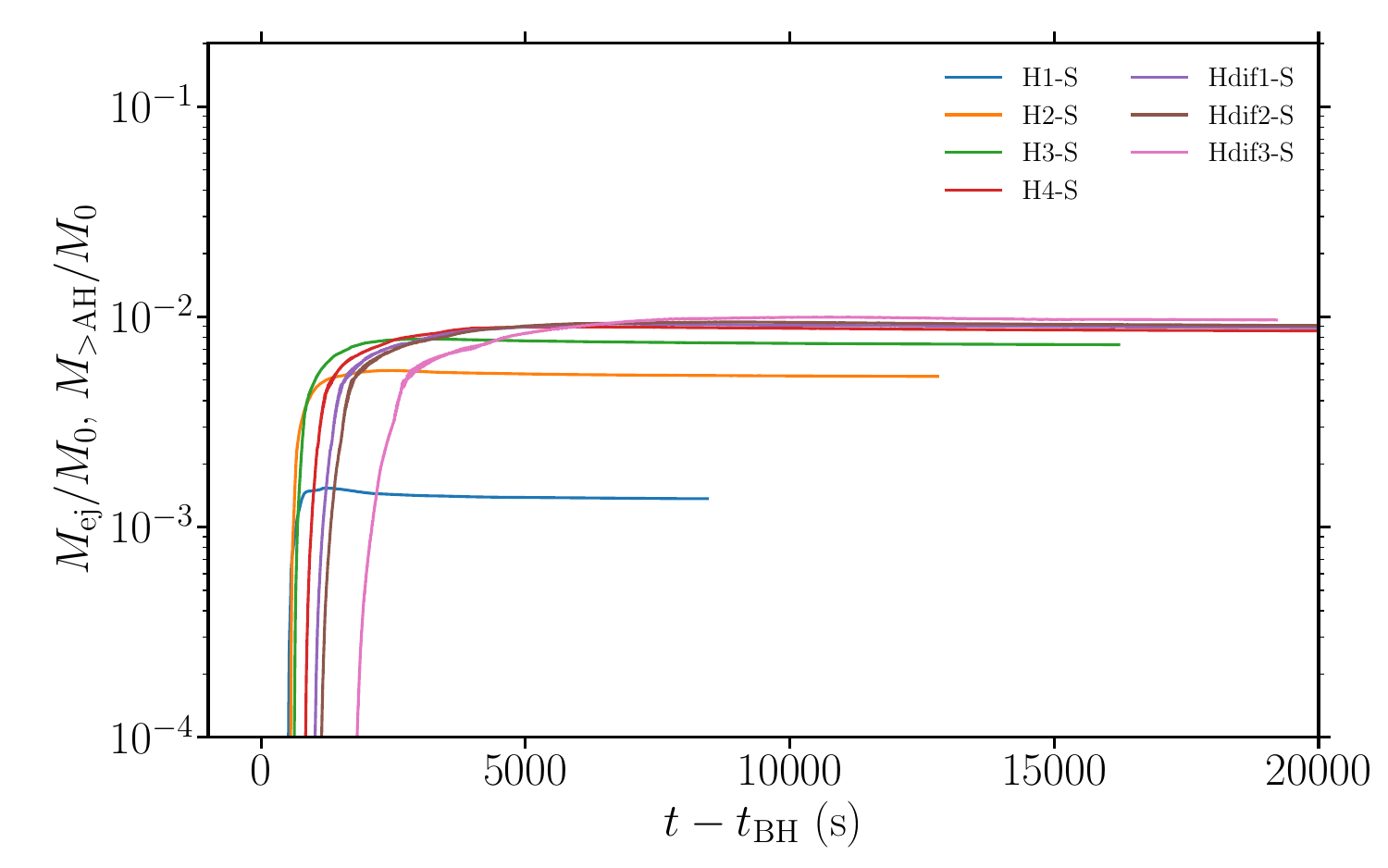}
    \includegraphics[width=0.49\textwidth]{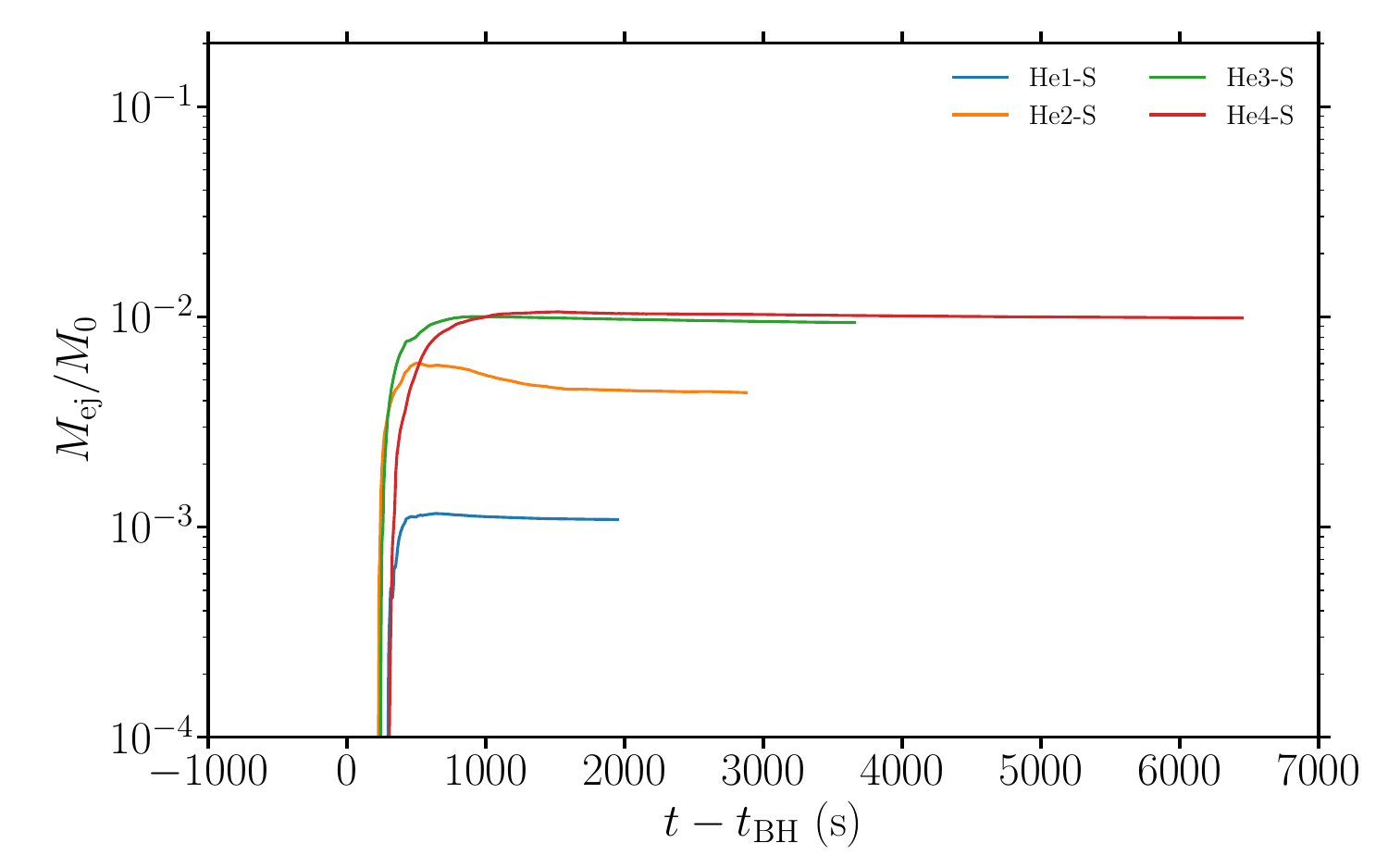}
    \includegraphics[width=0.49\textwidth]{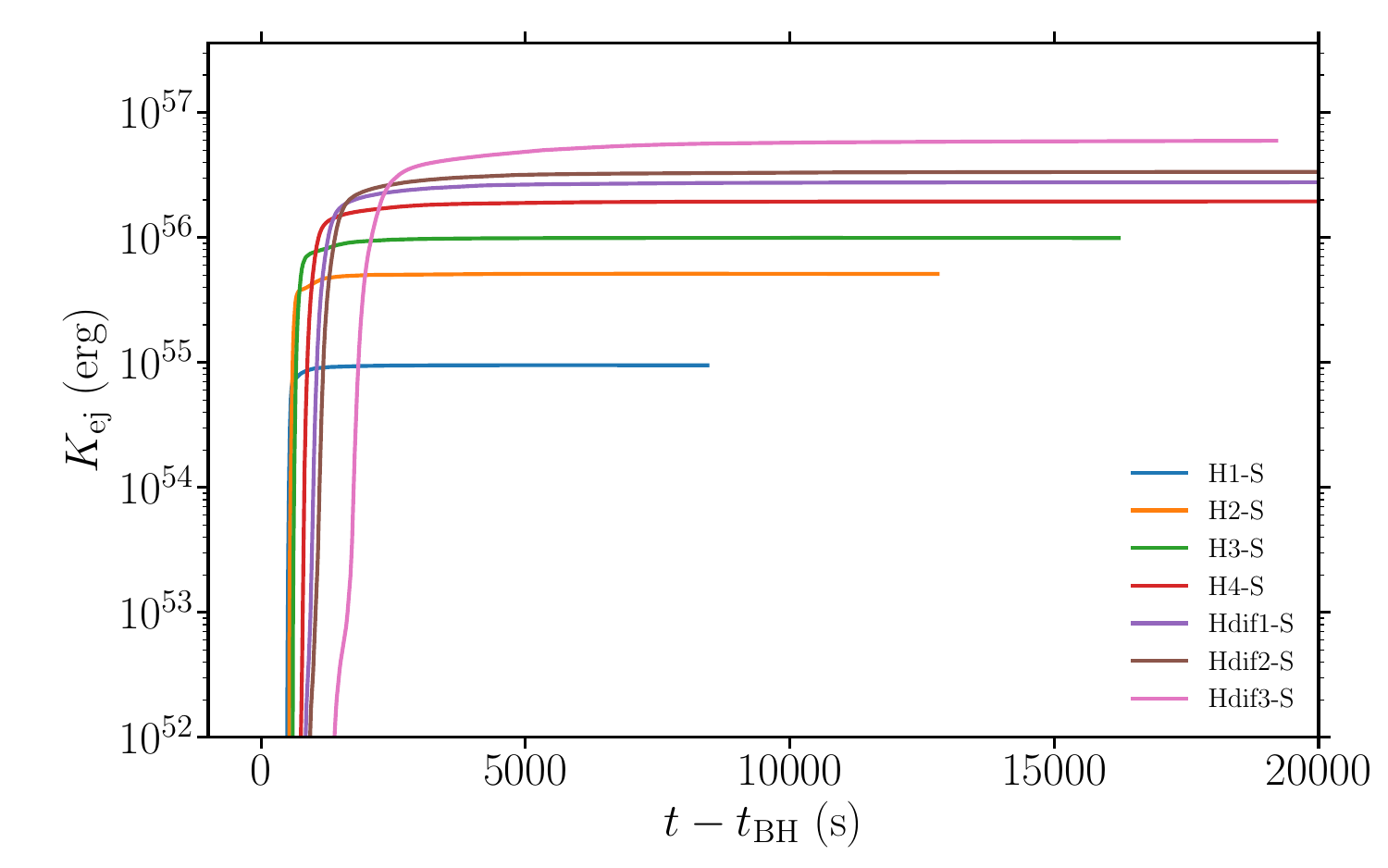}
    \includegraphics[width=0.49\textwidth]{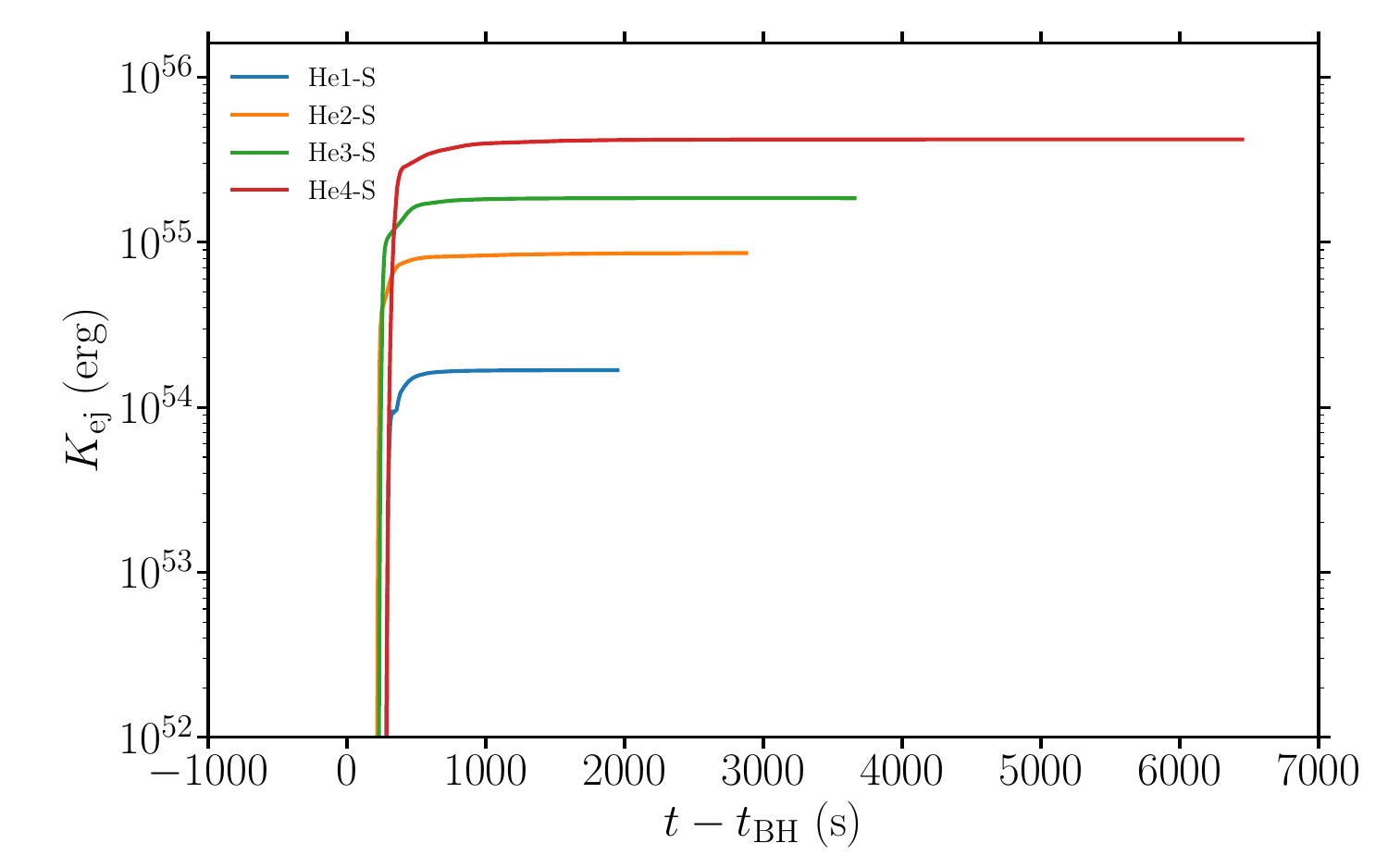}
    \includegraphics[width=0.49\textwidth]{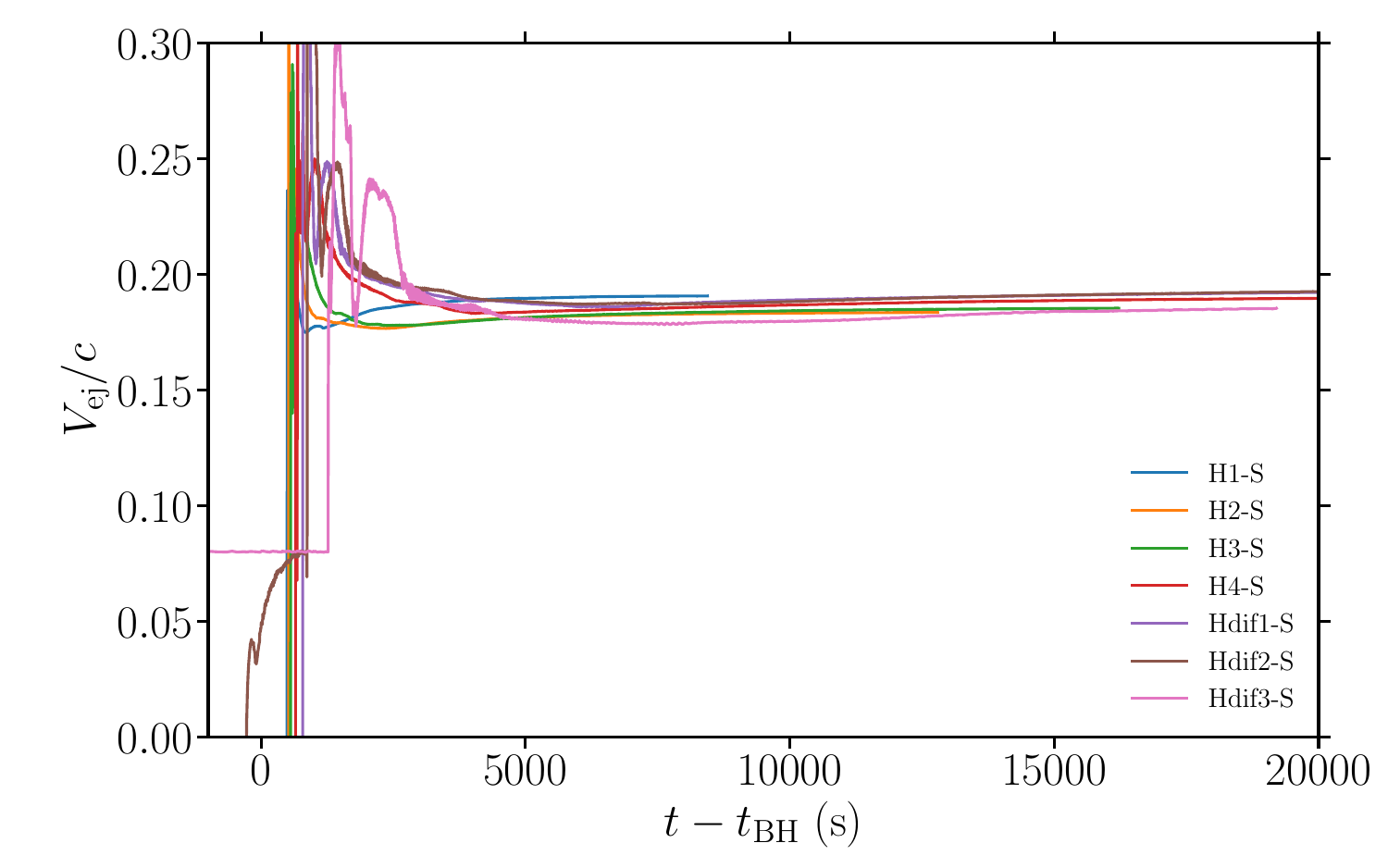}
    \includegraphics[width=0.49\textwidth]{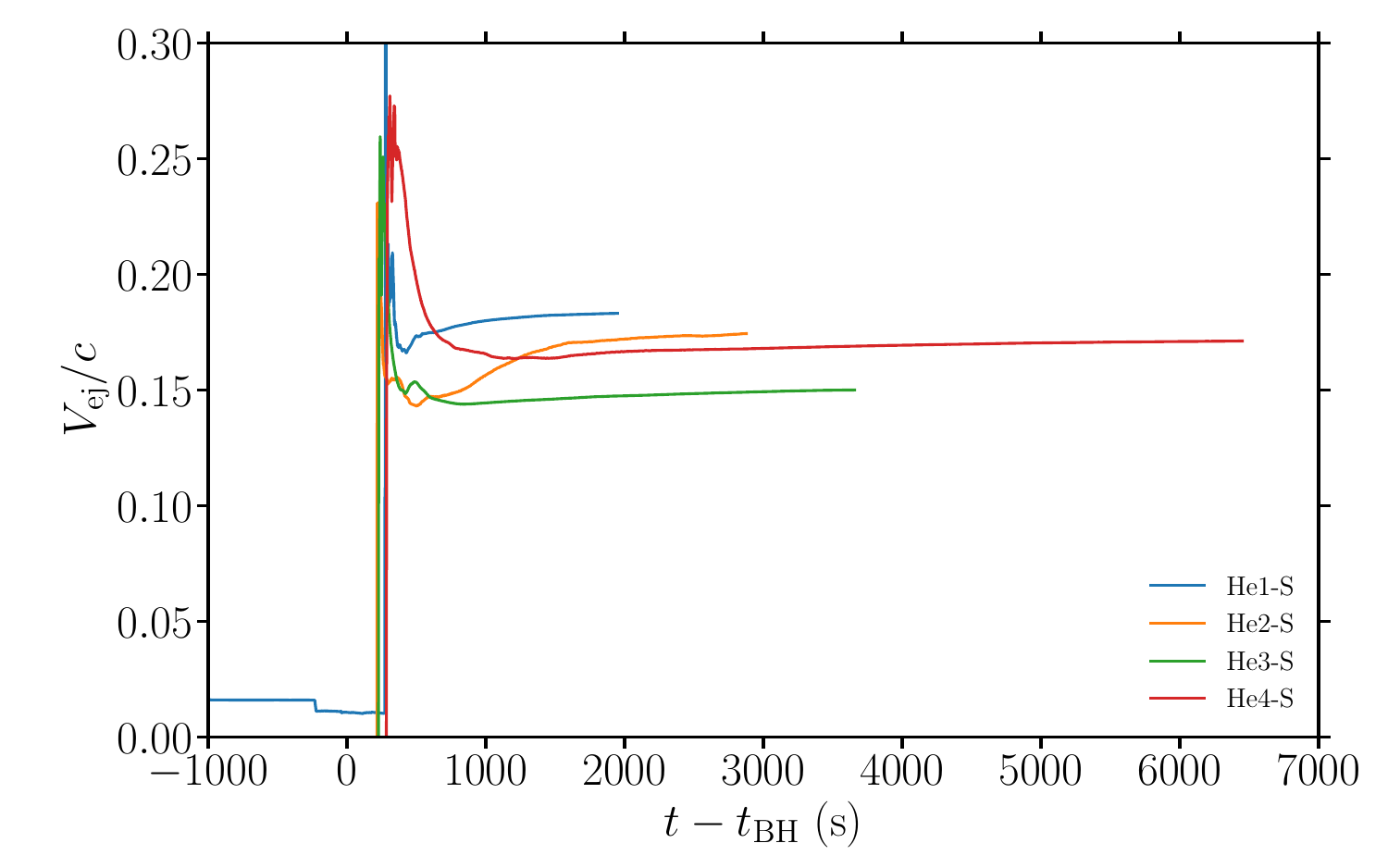}
    \caption{Time evolution of the mass (top panels), asymptotic kinetic energy (middle), and average asymptotic velocity (bottom) of the ejecta. The left and right panels are the results of H- and He-series, respectively. }
    \label{fig:ejecta}
\end{figure*}
For all the models, effects of nuclear burning play only a minor role during the collapse, and as a result, the collapse proceeds monotonically until a black hole is formed in a dynamical timescale. This result is consistent with the previous findings by \cite{Uchida2017oct}.
% (adiabatic index) initial condition: gamma>4/3 -> collapse is due to GR instability.

Figure~\ref{fig:gamma-c} displays the evolutionary paths of the central density and central temperature until the formation of the black hole in the $\rho$-$T$ plane with the adiabatic index shown in color. For all the models, the initial position of the central density and central temperature (left-bottom edge of each line; see also Table~\ref{tab:ID-list}) lies outside the domain for the pair-production instability, i.e., $\gamma_\mathrm{c}>4/3$. This indicates that the collapses are triggered by the general relativistic instability.

In the early phase of the collapse, the density and temperature increase approximately adiabatically, i.e., $T\propto \rho^{1/3}$. After the evolutionary path goes through the pair-unstable region, the temperature gradient with respect to the density, $dT/d\rho$, becomes slightly shallower; a part of the internal energy gained by the compression is converted to the rest-mass energy of $e^-e^+$ pairs. The path goes outside the pair-unstable region eventually for $T\gtrsim\SI{3e9}{K}$, but the collapse proceeds further without bounce in our models, and finally, a black hole is formed.

The top panels of Fig.~\ref{fig:bh} show the time evolution of the black hole mass estimated from the equatorial circumference length of the apparent horizon $C_\mathrm{e}$ \citep[e.g.,][]{Shibata2016a},
\begin{align}
M_\mathrm{BH} = \frac{c^2}{G}\frac{C_\mathrm{e}}{4\pi},
\end{align}
normalized by the initial ADM mass of the system. It shows that over 90\% of the stellar matter in mass becomes a black hole. The fraction is smaller for the higher $T_\mathrm{kin}/|W|$ cases because a more fraction goes into a torus formed. The bottom panels shows the dimensionless spin $\chi$ of the formed black hole. Assuming the relations for Kerr black holes, it is determined by solving
\begin{align}
\frac{C_\mathrm{p}}{C_\mathrm{e}} &= \frac{\sqrt{2\hat r_+(\chi)}}{\pi}\int_0^{\pi/2} \bigg(1-\frac{\chi^2}{2\hat r_+(\chi)}\sin\theta\bigg)^{1/2}d\theta
\end{align}
for $\chi$ (e.g., \citealt{Kiuchi2009sep}). Here, $C_\mathrm{p}$ is the polar circumference length of the apparent horizon and $\hat r_+(\chi) = 1+\sqrt{1-\chi^2}$ is the event-horizon radius normalized by $GM_\mathrm{BH}/c^2$. The dimensionless spins of the formed black holes are found to span from $\approx 0.45$ to $\approx 0.75$ depending on the initial values of $T_\mathrm{kin}/|W|$. 

The matter with a sufficiently high angular momentum, which is located at a large cylindrical radius initially, forms a centrifugally supported torus around the black hole. The torus formation starts at $t-t_\mathrm{BH}=1000$--\SI{2000}{s} ($t_\mathrm{BH}$ is the time when the BH forms) for H-series and 200--\SI{300}{s} for He-series. After the formation of the torus, the mass infall to the black hole is suppressed: The mass and dimensionless spin of the black hole saturate at the torus formation. The slow changes in the black hole mass and dimensionless spin after the saturation are caused mostly by numerical artifacts (see Sec.~\ref{subsec:resolution}).

The collapse triggered by the general relativistic instability proceeds rather coherently. In addition, there is no efficient cooling mechanism (the process is approximately adiabatic; see Sec.~\ref{subsec:neutrino}). As a result, the torus experiences a strong bounce soon after its formation due to the centrifugal barrier. The shock wave formed by the bounce drives the mass ejection (see \citealt{Uchida2017oct,Lee2006apr,Liu2007oct} for a detailed description of the ejecta formation process). After that, the torus relaxes to a quasi-stationary state around the black hole in the simulations that do not take into account viscous effects (cf. Fig.~\ref{fig:ejecta-density-temperature}).

\subsection{Properties of torus-shock-induced ejecta} \label{subsec:bounce-ejecta}

\begin{figure}[t]
    \centering
    \includegraphics[width=0.48\textwidth]{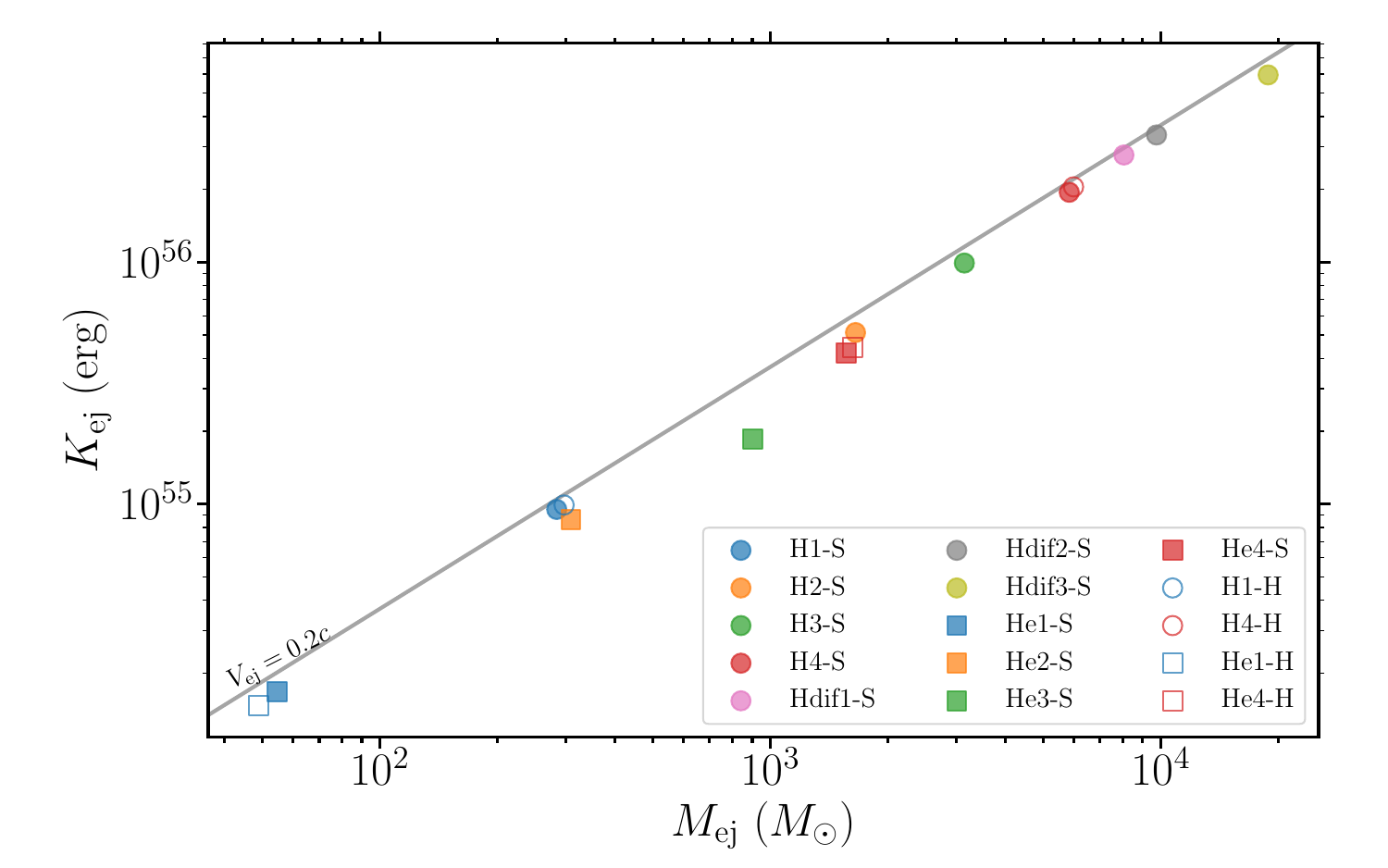}
    \caption{Correlation between the mass and kinetic energy of the ejecta driven by the shock formed in the bounce of the torus. The circle and square markers denote the results of H- and He-series, respectively, while the filled and open markers denote the standard- and high-resolution models, respectively. The grey line denotes the relation assuming a constant ejecta velocity of $0.2c$.}
    \label{fig:M-E}
\end{figure}

\begin{figure*}[t]
    \centering
    \includegraphics[width=0.48\textwidth]{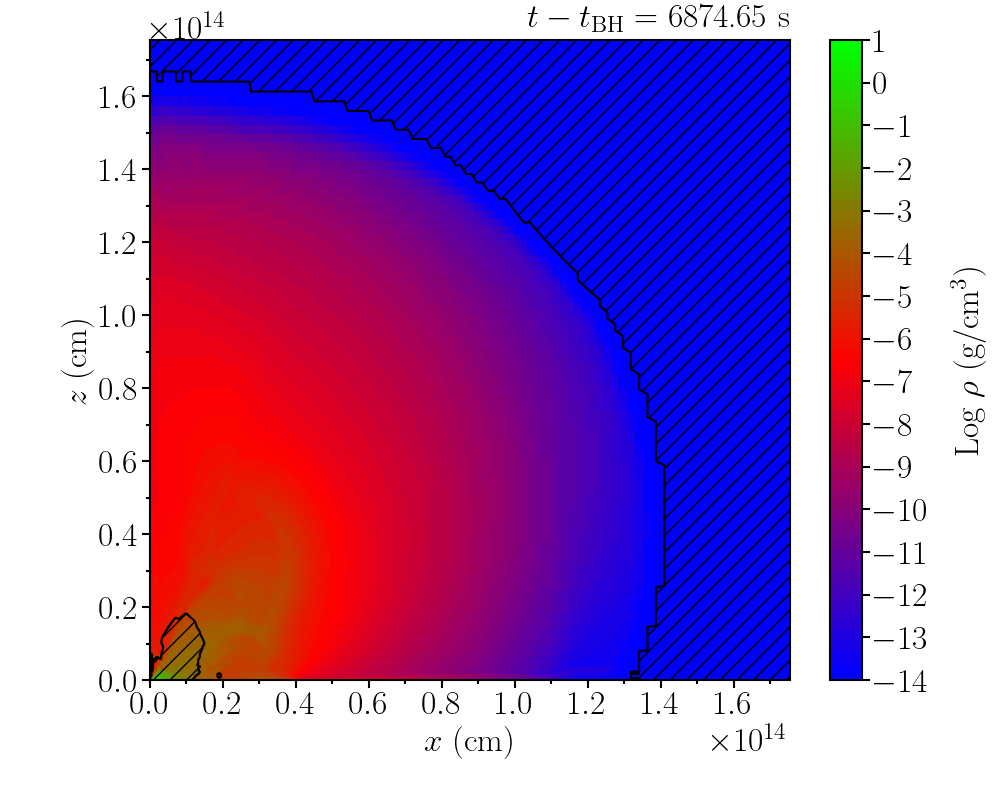}
    \includegraphics[width=0.48\textwidth]{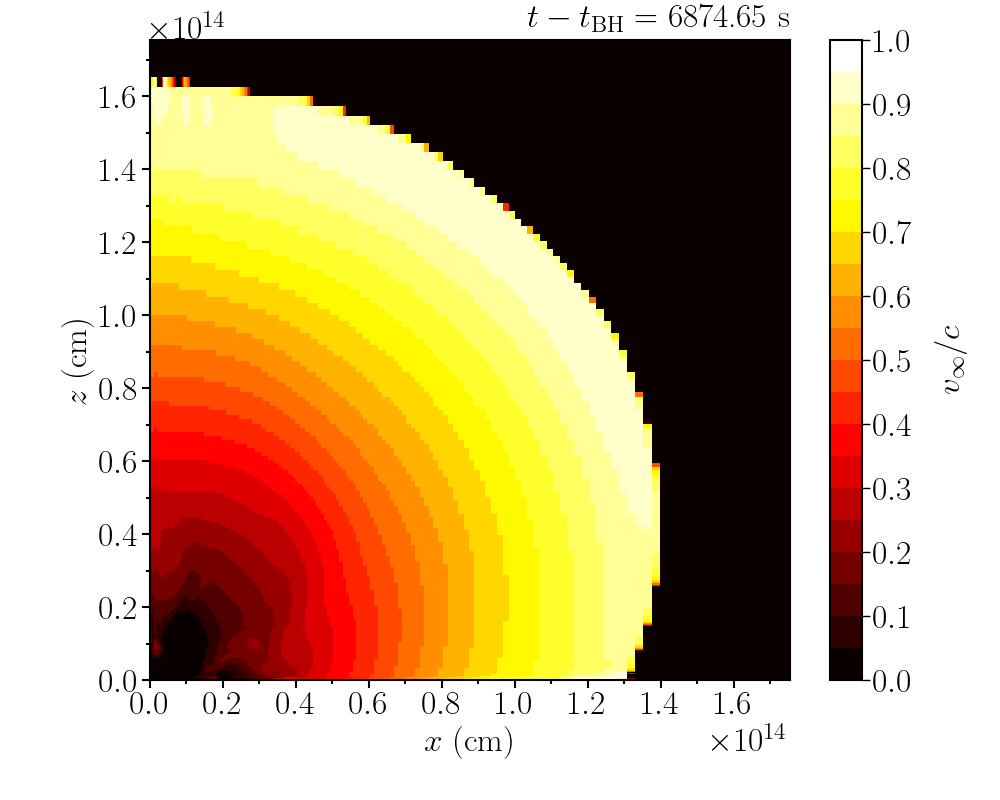}
    \caption{Density (left) and terminal velocity (right) distribution for model H4-H at $t-t_\mathrm{BH}\approx\SI{7000}{s}$. In the left panel, the domain with bound matter is marked by hatched areas.}
    \label{fig:2d-ejecta}
\end{figure*}
Figure~\ref{fig:ejecta} shows the evolution of the mass, kinetic energy, and average velocity of the ejecta generated by a shock formed at a inner surface of the torus for the H series (left) and He series (right). The ejecta mass of each model is normalized by the initial gravitational mass of the system (note that the relative difference between the gravitational mass and baryon rest mass is $\sim 10^{-5}$). For each curve, the time origin is shifted by the black hole formation time, $t_\mathrm{BH}$. As found in~\cite{Uchida2017oct}, $M_\mathrm{ej}/M_0$ increases with the increase of $T_\mathrm{kin}/|W|$ for relatively low values of $T_\mathrm{kin}/|W|$. However, the ejecta mass fraction saturates at $T_\mathrm{kin}/|W|\sim 0.01$ as $M_\mathrm{ej}/M_0 \sim 0.01$. This saturation level is similar for the differentially rotating cases (Hdif-series) and the more compact He models (see Table~\ref{tab:key-result}). 

Figure~\ref{fig:M-E} shows the relation between the kinetic energy and mass of the bounce-shock-driven ejecta for all the models studied in this paper. We find that the kinetic energy of the ejecta is approximately proportional to the ejecta mass, in particular for rapidly rotating models. This is reflected in the result that the average velocity of the ejecta is universal among the models as $\sim 0.2c$ (see the grey line in Fig.~\ref{fig:M-E}). This universally high velocity indicates that the mass ejection is driven in the vicinity of the black hole with the typical radius of $\sim 10$--$20\,r_\mathrm{g}$.%, which is close to that of the innermost stable circular orbit around the formed black hole.

% \delsf{The lowest-$T_\mathrm{kin}/|W|$ model for each of H- and He-series show a deviation from the linear relation between $M_\mathrm{ej}$ and $K_\mathrm{ej}$. It suggests that the generation of the bounce-driven ejecta requires a sufficient rotation of the supermassive star, and is suppressed under a certain threshold $T_\mathrm{kin}/|W|$. The threshold value is likely $\lesssim0.002$ based on our numerical results.}

For a variety of models, the kinetic energy of the ejecta exceeds $10^{55}$\,erg. Supermassive stars are believed to be formed in the center of proto-galaxies, likely surrounded by an accreting dense gas cloud. The large kinetic energy of the ejecta is likely to be injected into such a cloud, being dissipated and leading possibly to subsequent electromagnetic radiation. This process is similar to the stage prior to shock breakout of a supernova explosion in a massive star. The cloud is swept by the ejecta originating from the supermassive star, becoming a part of the ejecta. Since the kinetic energy and mass of the total ejecta are much larger than those of massive stars, the luminosity and duration for the subsequent radiation can be much larger than the typical supernovae (see, e.g., \citealt{Uchida2017oct}). This point will be discussed in Sec.~\ref{sec:discussion}. %\addms{COMMENT: Actually nothing is quantitatively discussed in Sec.4. We need to add quantitative numbers in Sec. 4.2}

% The properties of such a transient depend primarily on the mass and energy of the ejecta. In the current case, the total ejecta mass is likely dominated by that of the gas cloud swept up by the torus-shock-induced ejecta. 

Figure~\ref{fig:2d-ejecta} shows the spatial distribution of the rest-mass density and terminal velocity at $t-t_\mathrm{BH}\approx\SI{6900}{s}$ for model H4-H. The terminal velocity for unbound ($e_\mathrm{bind}>0$) matter is locally defined as
\begin{align}
\bigg[1-(v_\infty/c)^2\bigg]^{-1} = 1 + e_\mathrm{bind}/c^2.
\end{align}
The outer edge of the ejecta is found to have $v_\infty/c\alt 1$. This shows that a fraction of ejecta component is accelerated to such high velocity at the stellar surface, which has a steep density gradient~(e.g., \citealt{Waxman1993apr}).

It is also found that the ejecta has a quasi-spherical shape, although the bounce of the torus drives mass ejection toward the polar direction~\citep{Uchida2017oct}. After breaking out from the stellar surface, the ejecta expands laterally to become a spherical geometry. In addition, the shock wave associated with the bounce is diffracted in the stellar envelope toward the equatorial direction, and then, an outer part of the stellar envelope becomes unbound by being swept up by the shock wave. This effect also contributes to developing the quasi-spherical ejecta.

\begin{figure}[t]
    \centering
    \includegraphics[width=0.48\textwidth]{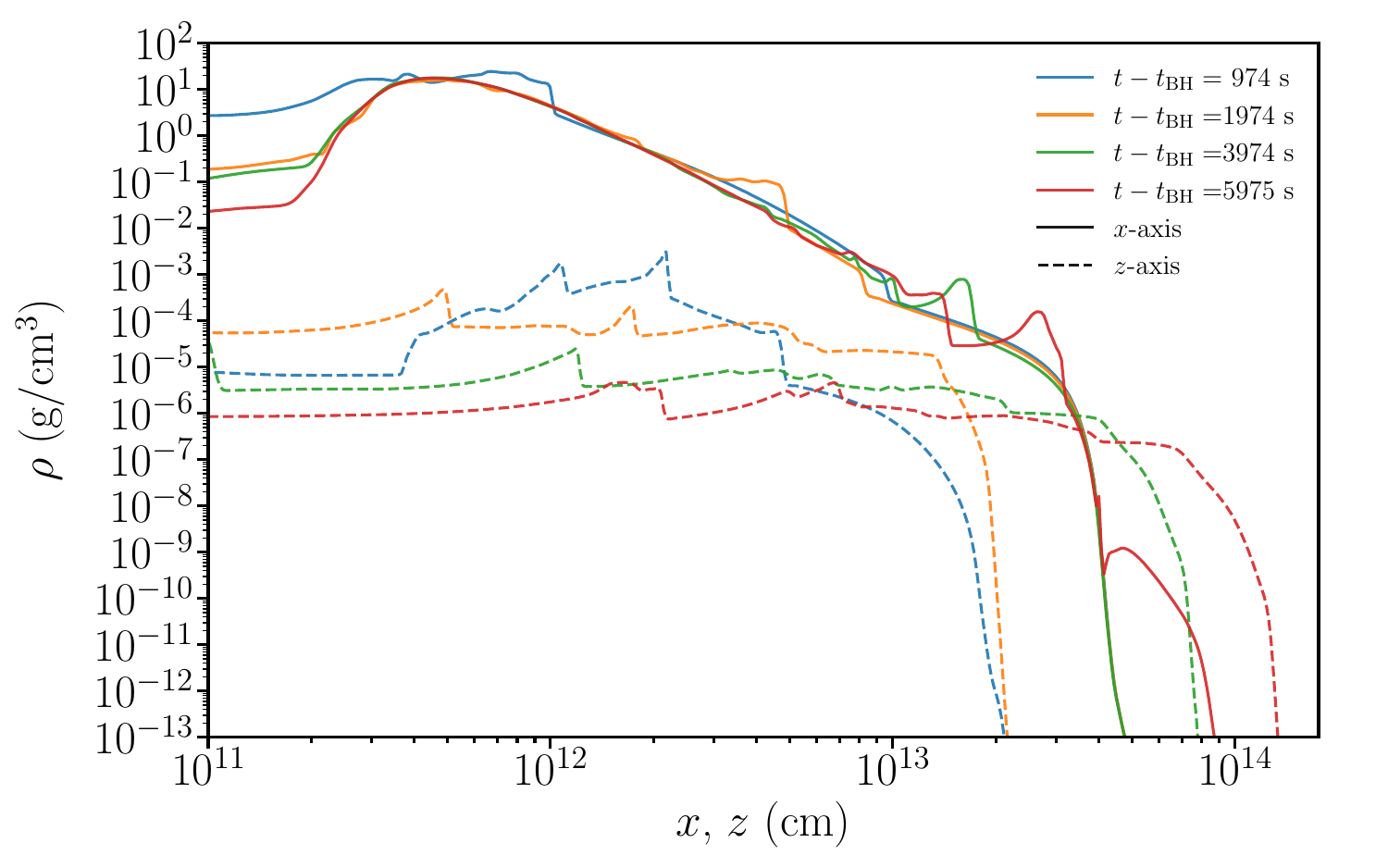}
    \includegraphics[width=0.48\textwidth]{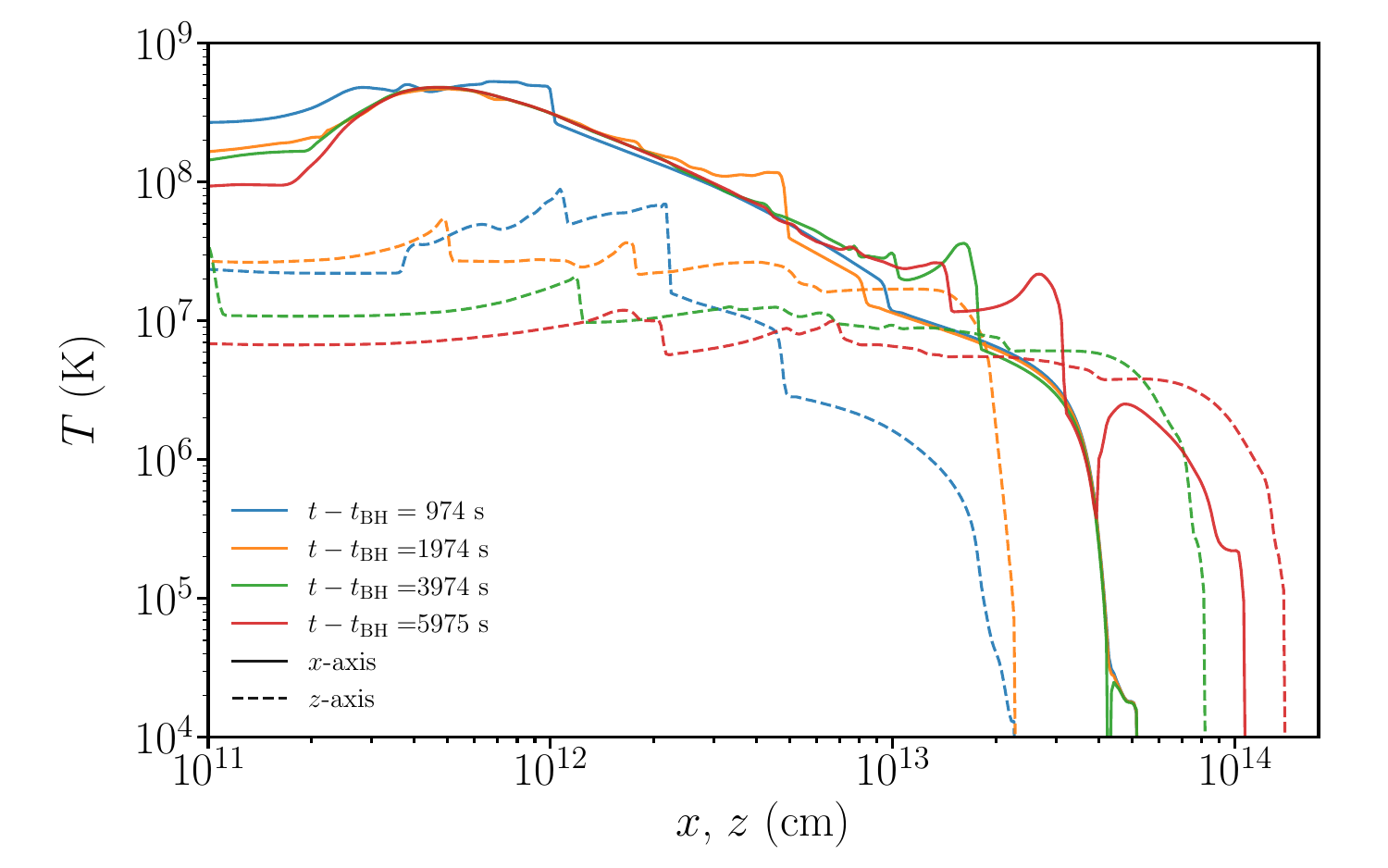}
    \caption{Distribution of density (upper panel) and temperature (lower panel) along the $x$- and $z$-axes for model H4-H, especially focusing on the times at which the ejecta is developed.}
    \label{fig:ejecta-density-temperature}
\end{figure}

\begin{figure}[t]
    \centering
    \includegraphics[width=0.48\textwidth]{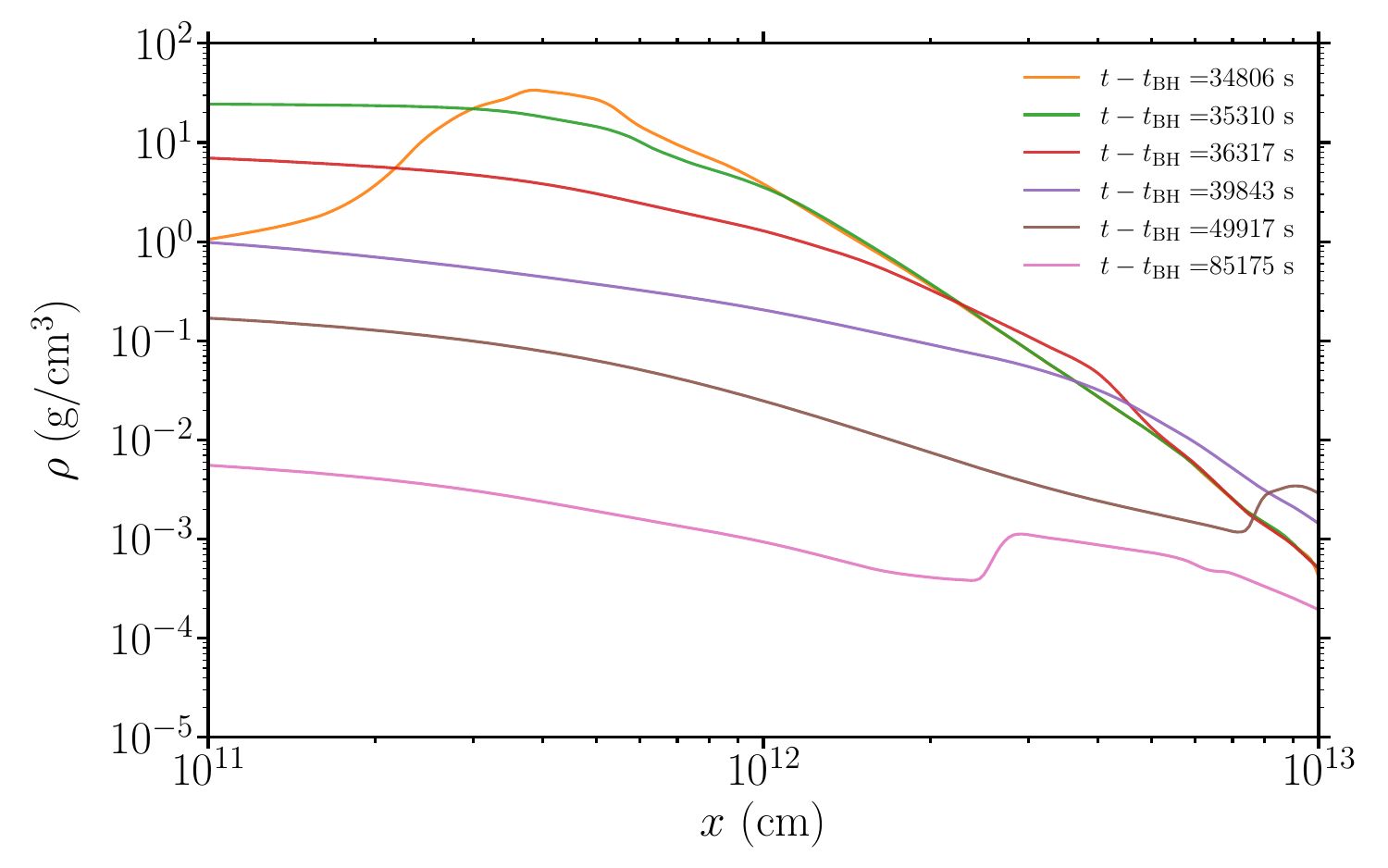}
    \includegraphics[width=0.48\textwidth]{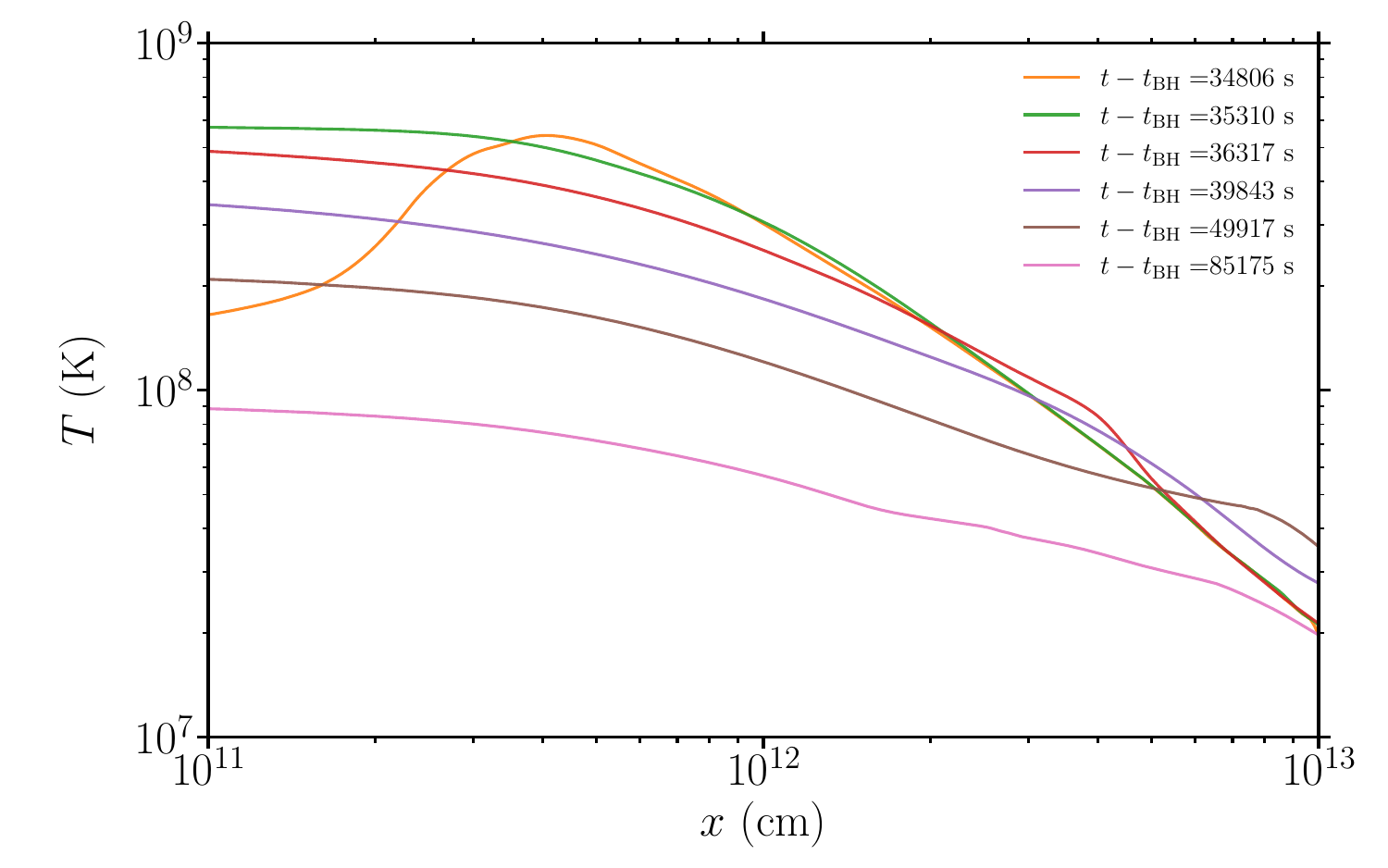}
    \caption{Density and temperature distribution along the $x$-axis for model H4-v0.10SS, after switching on the viscosity (at $t-t_\mathrm{BH}=\SI{3.4e4}{s}$).}
    \label{fig:torus-density-temperature}
\end{figure}

The nuclear composition of the ejecta is essentially the same as that in the initial condition because the matter in which nuclear burning proceeds efficiently is swallowed by the newly formed black hole. In addition, the density and temperature of the ejecta are too low for efficient nuclear reactions: Figure~\ref{fig:ejecta-density-temperature} shows the radial profiles of the rest-mass density and temperature along $x$- and $z$-axes for model H4-H. The density and temperature of the ejecta are, at highest, $\rho\sim\SI{10}{g/cm^3}$ and $T\approx\SI{5e8}{K}$ at the time of the shock formation (at $t-t_\mathrm{BH}\approx\SI{1000}{s}$). At such density and temperature, the timescales of the (hot) CNO cycle and triple-alpha reactions, defined by $n_\mathrm{b}/\dot{n}_\mathrm{CNO}$ and $n_\mathrm{b}/\dot{n}_{3\alpha}$, are on order of $\SI{e12}{s}$ at shortest. This timescale is much longer than the dynamical (expansion) timescale of the ejecta. Thus, the nuclear burning does not significantly proceed inside the torus and ejecta.

Figure~\ref{fig:ejecta-density-temperature} also shows that a stationary torus is developed on the equatorial plane in a timescale of $\sim 10^3$\,s. On the other hand, along the symmetric axis ($z$-axis), the density and temperature decrease gradually due to the mass accretion onto the central black hole.

\begin{figure*}[t]
    \centering
    \includegraphics[width=0.49\textwidth]{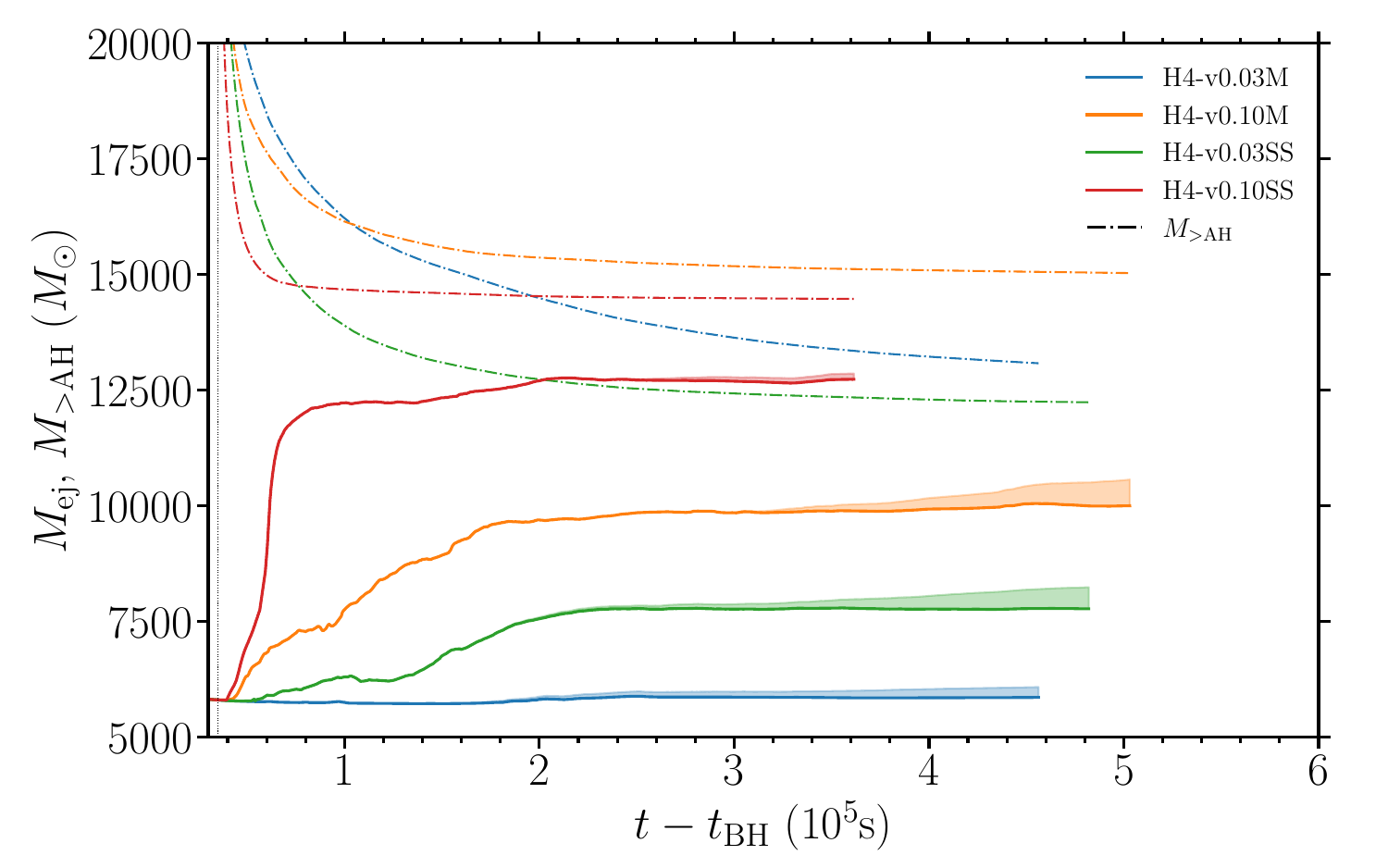}
    \includegraphics[width=0.49\textwidth]{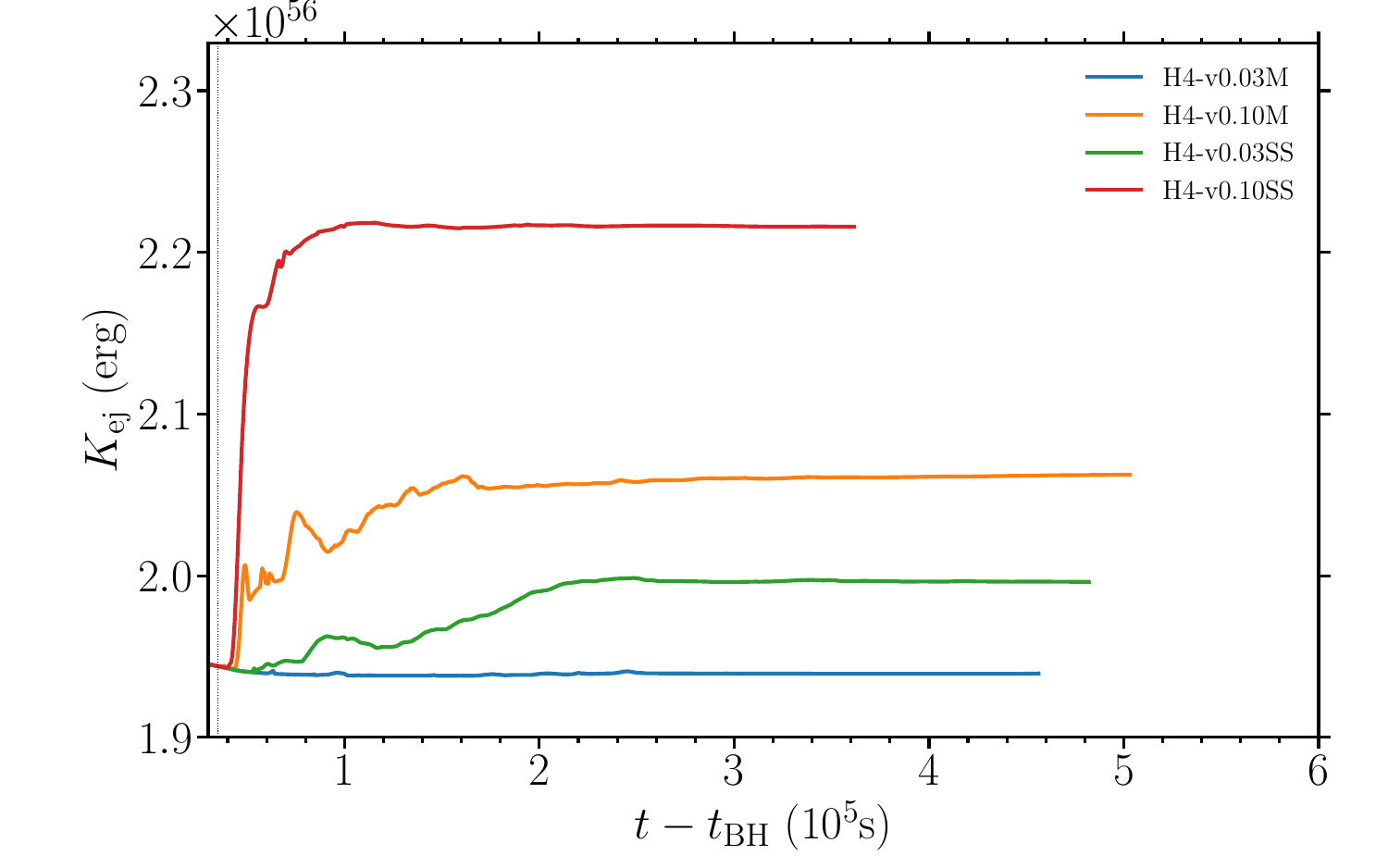}
    \caption{Mass (left) and asymptotic kinetic energy (right) of the total ejecta focusing on the development of the viscosity-driven ejecta. For each panel, the dashed vertical lines at $t-t_\mathrm{BH}\approx\SI{3.4e4}{s}$ mark the time at which the viscosity is switched on. In the left panel, the shaded bands denote the possible range of the ejecta, considering the still bound matter that already escaped from the computational domain. The dashed curves in the left panel shows the mass of the matter located outside the apparent horizon $M_\mathrm{>AH}$.}
    \label{fig:vis-ejecta}
\end{figure*}

\subsection{Viscous evolution of torus} \label{subsec:vis}
The torus formed around the black hole may evolve with magnetohydrodynamical processes through the magnetorotational instability \citep{Balbus1991a, Balbus:1998ja} in the presence of a seed magnetic field in the supermassive stars. The magnetorotational instability in the torus induces a turbulent state, which then acts as an effective viscosity and drives angular momentum transport and viscous heating. Because no efficient cooling mechanism is present in the present case, a part of the torus matter can become ejecta in the presence of the viscous effects (e.g., \citealt{Fujibayashi2020a}). To investigate the possible outcomes of the viscous evolution, we perform several viscous hydrodynamics simulations with the formalism used in \cite{shibata2017apr}. To assess the largest possible impact, we pick up the post-collapse data for the H4 model as the initial condition of the viscous hydrodynamics. We switch on the viscosity at $t-t_\mathrm{BH}\approx \SI{3.4e4}{s}$ at which the torus settles down to a stationary state as illustrated in Fig.~\ref{fig:ejecta-density-temperature}. At this time, the bounce-driven shock-heated ejecta is located far from the central domain. 

We consider a Shakura-Sunyaev-type kinetic viscous coefficient~\citep{Shakura1973a} as
\begin{align}
\nu = \alpha_\mathrm{vis} c_\mathrm{s} \ell_\mathrm{tur}, \label{eq:vis2}
\end{align}
where $c_\mathrm{s}$ is the sound speed and $\alpha_\mathrm{vis}$ is a constant that controls the magnitude of the coefficient. We consider two prescriptions for the length scale of the turbulence $\ell_\mathrm{tur}$. In the first one, we assume a constant value $\ell_\mathrm{tur} = 2GM_0/c^2$ as in \cite{Fujibayashi2020a}. The model with this prescription is denoted by a letter ``M" in its name. Since most of the stellar matter collapses into the black hole, the mass of the formed black hole can be approximated by $M_0$. Therefore, $\ell_\mathrm{tur}$ is approximately the size of the black hole horizon. In the second prescription, we assume
\begin{align}
\ell_\mathrm{tur} = c_\mathrm{s}/\Omega_\mathrm{K}
\end{align}
with the local Keplerian angular velocity $\Omega_\mathrm{K}$, which is approximated with the derivative of the lapse function as
\begin{align}
\Omega_\mathrm{K} = \sqrt{\frac{c^2}{x} \frac{\del \alpha}{\del x}}.
\end{align}
The model with this prescription is denoted by ``SS". For a given cell, we use $\Omega_\mathrm{K}$ evaluated on the equatorial plane at the same cylindrical radius. In the far region from the black hole, we have $\alpha \approx 1-GM_\mathrm{BH}/(\varpi c^2)$ along the equatorial direction, where $\varpi$ is the cylindrical radius. Thus, $\Omega_\mathrm{K}$ is reduced to the usual Newtonian expression $\sqrt{GM_\mathrm{BH}/\varpi^3}$. 

For each viscous prescription, we suppress the viscous coefficient in low-density regions by a factor $(1-e^{-\rho/\rho_\mathrm{crit}})$ with $\rho_\mathrm{crit} = \SI{e-3}{g/cm^3}$ in order not to affect the dynamics of the infalling stellar envelope and expanding ejecta, for which high effective viscosity is not likely to be induced. For each prescription of $\ell_\mathrm{tur}$, we use $\alpha_\mathrm{vis} = 0.03$ and 0.1. In the model name, the imposed values of $\alpha_\mathrm{vis}$ are denoted by the number before the letter denoting the viscosity prescription (e.g., 0.10SS and 0.03M).

% because for such a region the timescale for the development of the magnetorotational instability ($\propto \Omega^{-1}$) would be long. 

Figure~\ref{fig:torus-density-temperature} shows the radial profiles of density and temperature along the $x$-axis for different times. After the viscosity is switched on, the density in the inner region of the torus decreases due to the mass accretion onto the black hole and to outward expansion resulting from the angular momentum transport. Also, the temperature decreases accordingly. Thus, the timescale of the nuclear reactions (in this work CNO cycle and triple-alpha reaction) becomes longer in the later phase of the torus evolution. We also find that at the typical torus radius $x\sim\SI{e12}{cm}$, the viscous timescale is $x^2/\nu\sim\SI{e4}{s}$, which is much shorter than the nuclear reaction timescale $\sim\SI{e12}{s}$ at shortest. This implies that the torus matter is accreted onto the central black hole before it is burnt, and therefore, the nuclear burning in the accreting torus has a negligible effect on its evolution.

Figure~\ref{fig:vis-ejecta} compares the ejecta properties with different viscous parameters and prescriptions for the viscous H4 models. Here we note that the viscosity is switched on at $t-t_\mathrm{BH}\approx \SI{3.4e4}{s}$. We find that an amount of mass that escapes from the computational domain is still bound according to the criterion based on $e_\mathrm{bind}$ (see Eq.~\eqref{eq:ebind}). Such a component may become unbound eventually by being pushed up by the outflow from the torus launched later. The possible range of the ejecta mass taking such a component into account is also indicated by shading in the same figure. For model H4-v0.10SS, the ejecta mass begins to increase at $t-t_\mathrm{BH}\approx\SI{4e4}{s}$ with an approximate saturation at $t-t_\mathrm{BH}\sim \SI{1e5}{s}$. The ejecta mass increases in this period by $\approx \SI{6.5e3}{}M_\odot$. In the same period, the asymptotic kinetic energy of the ejecta increases by $\sim \SI{3e55}{erg}$. This indicates that the viscosity-driven ejecta have an average velocity of $\approx 0.07c$. Considering that the remaining bound mass outside the black hole is $\approx \SI{3e3}{}M_\odot$ at $t-t_\mathrm{BH} = \SI{1e5}{s}$, the viscosity-driven ejecta contributes to the kinetic energy by $\lesssim \SI{5e55}{erg}$, which is smaller than that of the torus-shock-driven ejecta ($\approx \SI{2e56}{erg}$). The contribution of the viscosity-driven ejecta is even smaller for the models with the other prescription of the turbulence length scale or with the smaller viscous parameter. Thus, the viscosity-driven ejecta is a subdominant component of the entire ejecta in this problem. 

\begin{figure}[t]
    \centering
    \includegraphics[width=0.49\textwidth]{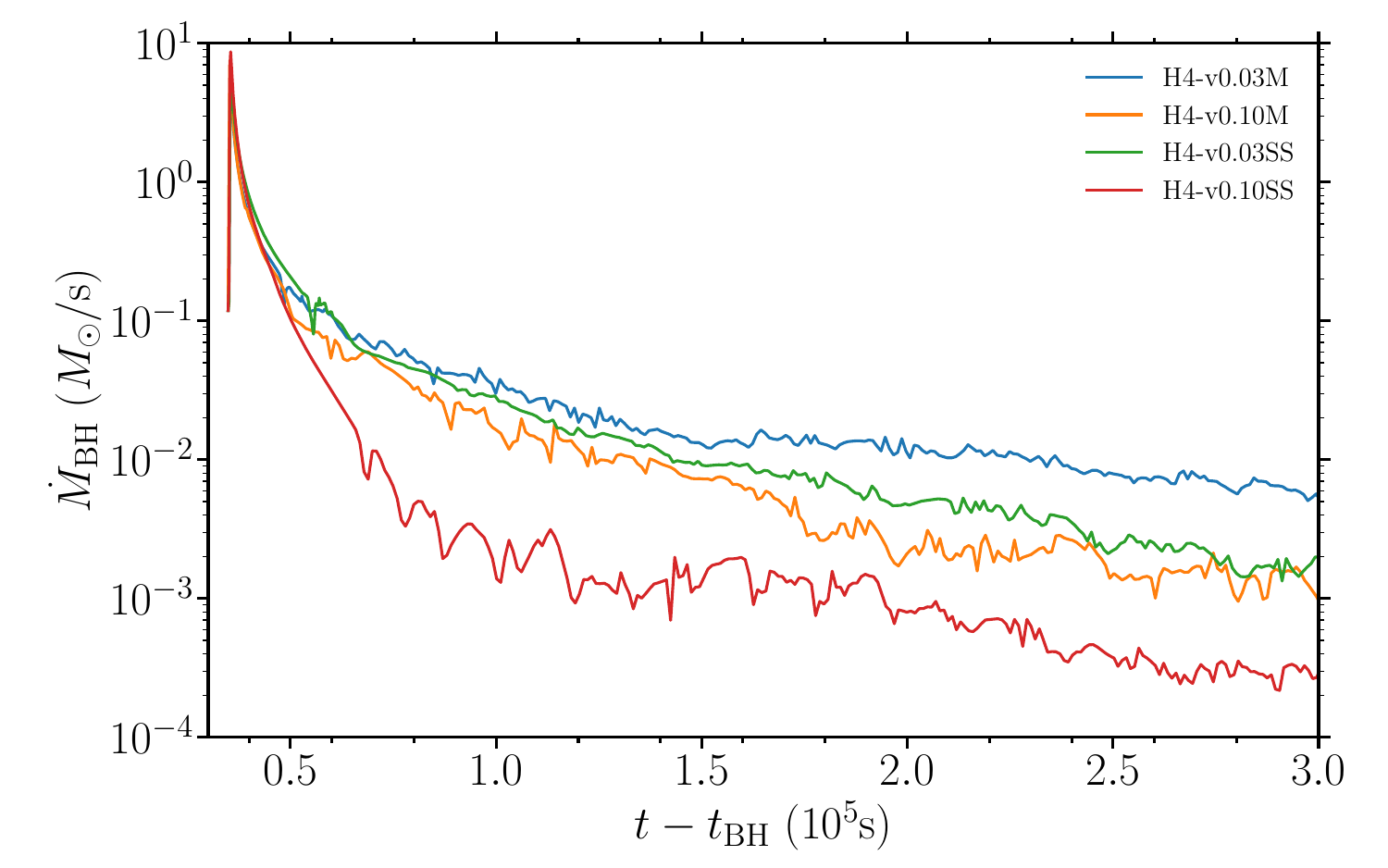}
    \caption{Mass accretion rate onto the black hole for the viscous models after switching on the viscosity. See Sec.~\ref{subsec:vis} for definition.}
    \label{fig:accretion}
\end{figure}

Figure~\ref{fig:accretion} shows the mass accretion rate onto the black hole for the four viscous models. The accretion rate is defined by
\begin{align}
\dot{M}_\mathrm{BH} = \int_{\mathrm{AH}} \rho u^k \sqrt{-g} ds_k,
\end{align}
where $ds_k$ is the area element on the surface of the apparent horizon. The accretion rate decreases with time after an initial steep rise when the viscosity is switched on. The accreted mass amounts to $\approx1.3\times 10^4$--$1.8\times 10^4M_\odot$, which is accreted mainly in the first \SI{3e3}{}--\SI{1e4}{s} depending on the adopted viscosity prescriptions and the values of viscous parameter. For a given value of the viscous parameter, the prescription in Eq.~\eqref{eq:vis2} leads to shorter accretion timescale because Eq.~\eqref{eq:vis2} leads to a longer turbulence length scale than $2GM_0/c^2$ and the larger viscous parameter leads to a shorter mass accretion timescale for a given prescription of the viscosity. 

The mass accretion of the black hole could potentially drive a relativistic jet in the presence of magnetic fields and impact energetic transients. This topic will be discussed in Sec.~\ref{subsec:transients}.

\begin{figure}[t]
    \centering
    \includegraphics[width=0.49\textwidth]{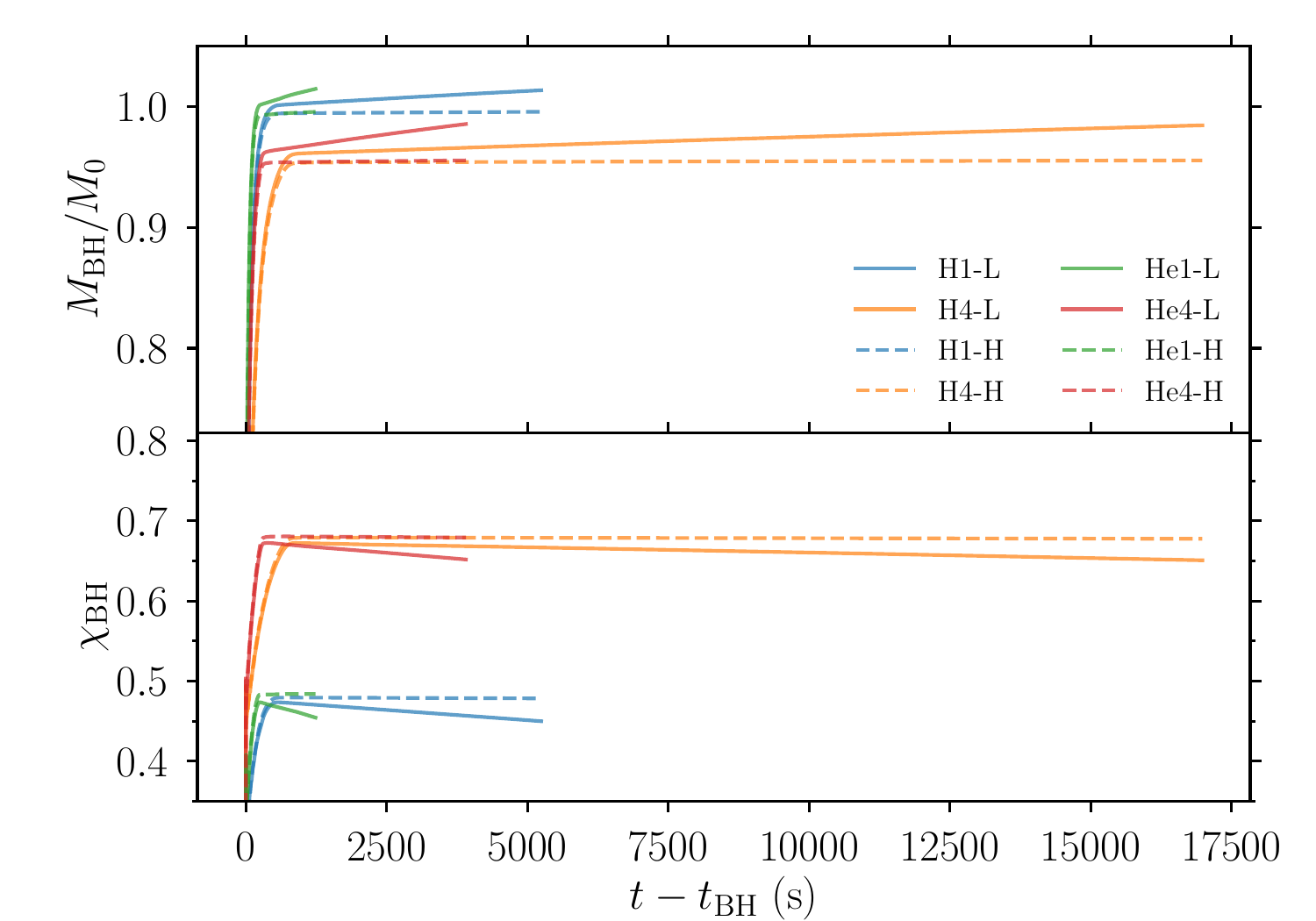}
    \includegraphics[width=0.49\textwidth]{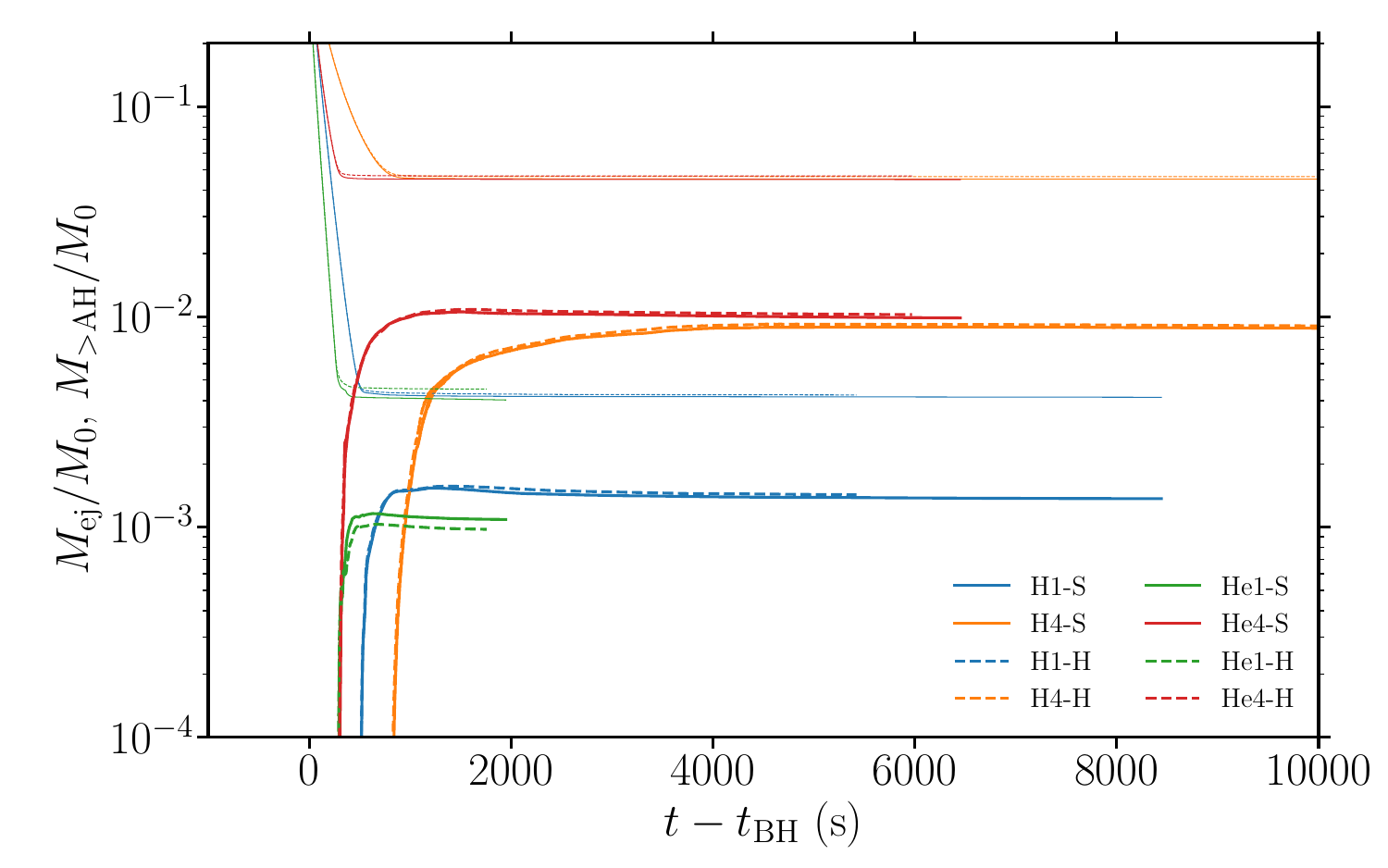}
    \includegraphics[width=0.49\textwidth]{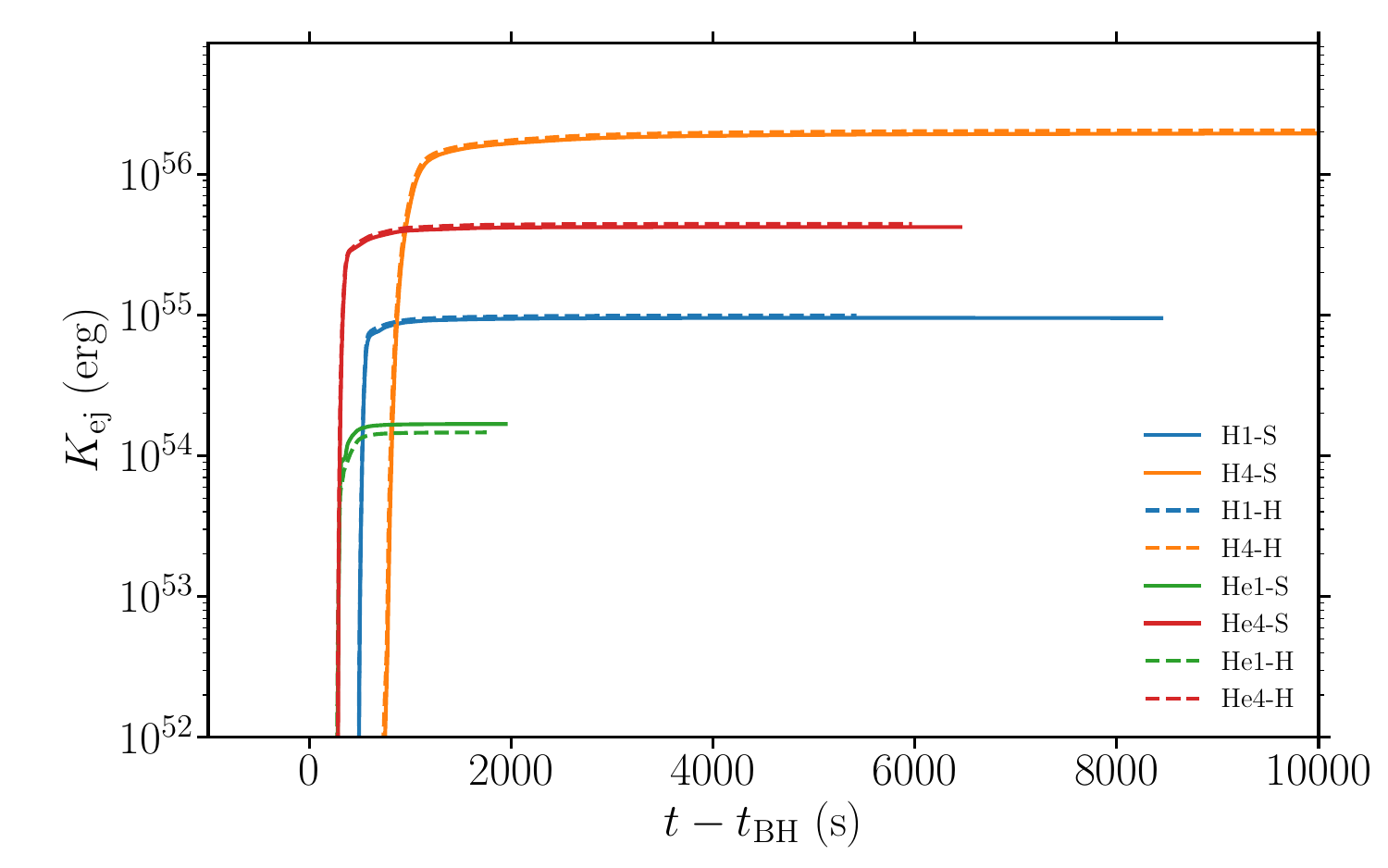}
    \caption{Comparisons of black hole properties (top), ejecta mass relative to the total mass of the star (middle), and ejecta kinetic energy (bottom) between different grid resolutions for models H1, H4, He1, and He4. The standard- and high-resolution results are shown by the solid and dashed curves, respectively. In the middle panel, the rest mass of the matter outside the apparent horizon, $M_\mathrm{>AH}$, is also shown in the thin curves.}
    \label{fig:resolution}
\end{figure}

\subsection{Resolution study} \label{subsec:resolution}
For models H1, H4, He1, and He4, we performed simulations with a higher grid resolution after the last regridding. The top panel of Fig.~\ref{fig:resolution} compares the mass and dimensionless spin of the black hole. The black-hole mass in the high-resolution models is systematically lower than that for the corresponding standard-resolution models, while the dimensionless spin of the black hole is systematically higher than that for standard-resolution simulations. The reason for these trends is that the mass and dimensionless spin increase and decrease spuriously due to numerical errors, which are suppressed in higher grid resolutions approximately at the fourth order (see, e.g., \citealt{Fujibayashi2024jan}). In the present standard-resolution simulations, the black-hole horizon is resolved only by $\approx15$ grid points, and as a result, the black-hole mass spuriously increases $\sim1\%$ during the time $10^3GM_\mathrm{BH}/c^3$ even without mass accretion. For the high-resolution simulations, the black hole horizon is resolved by $\approx 24$ grid points, and thus, the spurious behavior is suppressed by a factor of $\sim (15/24)^4\approx 0.15$. 

The middle and bottom panels of Fig.~\ref{fig:resolution} compare the ejecta mass normalized by $M_0$ and asymptotic kinetic energy of the ejecta. The ejecta mass and kinetic energy tend to be larger for higher-resolution models. This trend is consistent with the smaller mass and the higher dimensionless spin of the formed black hole in higher-resolution models; they result in a smaller radius of the innermost stable circular orbit, and therefore, the torus bounce effect becomes more appreciable.
% } \addsf{The ejecta mass and kinetic energy agree quite well for the standard- and high-resolution simulations. However, a closer look reveals that the high-resolution models have slightly more ejecta mass and kinetic energy.}  

For models with higher values of $T_\mathrm{kin}/|W|$, H4 and He4, the ejecta quantities have no notable dependence on the grid resolution. By contrast, for models H1 and He1, the dependences are clearer in the ejecta mass and kinetic energy simply because the amount of the ejecta mass is relatively small. The differences in the final ejecta mass and kinetic energy are within 30\% even in these cases, and hence, the qualitative picture discussed in this paper is not affected by the grid resolution.

\section{Discussion} \label{sec:discussion}

\subsection{Property of ejecta in realistic environment} \label{subsec:realistic}
Supermassive stars are likely born in an infalling primordial gas cloud which is as massive as or more massive than the supermassive star itself (e.g., \citealt{Johnson2013oct}, \citealt{Whalen2013nov}, and \citealt{Patrick2023jul}). The gas cloud is also likely to be surrounded by a compact halo. The kinetic energy of the ejecta found in the previous section, $10^{55}$--$10^{56}$\,erg, can be large enough to engulf such a surrounding massive cloud and a halo, if these are not extremely massive and compact.\footnote{In \cite{Johnson2013oct} and \cite{Whalen2013nov}, it is shown that the blast wave generated by an explosion with energy of $\sim\SI{e55}{erg}$ is likely to loose energy by several efficient cooling processes. In their work, the matter swept up by the blast wave is likely to recollapse with a time delay of several \SI{10}{Myr}. The delay time depends on how dense the cloud is.}
%Therefore, the mass of the total ejecta of the supermassive stellar collapse in such a realistic environment may also be similar to that of the supermassive star.
Thus, the contribution of the ejecta originated from the supermassive stellar core is significant only in terms of the kinetic energy, because the mass, expected to be $\sim 1$\% of the stellar mass, is minor compared to the mass of the cloud and halo.

Supermassive stars in reality may have an inflated envelope with radius $\sim\SI{e15}{cm}$ because of the high accretion rate, $\gtrsim0.1M_\odot$/yr~\citep{Hosokawa2013dec}. The more diluted structure of the stellar envelope than those studied in this paper may affect the radial distribution of the ejecta properties shown in Fig.~\ref{fig:2d-ejecta}. Nevertheless, the ejecta is formed in the vicinity of the black hole, and the effects of outer structure on the ejecta kinetic energy are likely to be only minor.

\subsection{Electromagnetic transients associated with collapses of supermassive stars} \label{subsec:transients}
The exploded supermassive star inside the primordial (hydrogen-rich) gas cloud may show similar electromagnetic transients to those of type-IIp supernovae, but scaled-up by many orders of magnitude in terms of mass and energy \citep{Uchida2017oct}. Because of the longer diffusion timescale of the ejecta, the transients can last for years in the rest frame of the explosion. 
For example, employing a semi-analytical model of \cite{Matsumoto:2015bjg} (under the assumption of the primodial composition of the ejecta and Thomson scattering dominated opacity), the bolometric luminosity in the plateau phase reaches its peak at 
\begin{align}
    t_{\rm peak}&\approx10\,{\rm yr}\nonumber\\
    &\times\left(\frac{r_0}{10^{15}\,{\rm cm}}\right)^{1/7}\left(\frac{E_{\rm exp}}{10^{56}\,{\rm erg}}\right)^{-5/28}\left(\frac{M}{10^5\,M_\odot}\right)^{15/28}
\end{align}
with the peak value of
\begin{align}
    L_{\rm peak}&\approx3\times10^{45}\,{\rm erg/s}\nonumber\\
    &\times\left(\frac{r_0}{10^{15}\,{\rm cm}}\right)^{4/7}\left(\frac{E_{\rm exp}}{10^{56}\,{\rm erg}}\right)^{11/14}\left(\frac{M}{10^5\,M_\odot}\right)^{-5/14},
\end{align}
where $r_0$, $E_{\rm exp}$, and $M$ denote the supermassive star radius, explosion energy, and ejecta mass (including the contribution from the stellar envelope), respectively.
The emission could be even brighter and longer lasting in the presence of a surrounding optically-thick gas cloud. Hydrodynamics interaction of the ejecta matter with the gas cloud may efficiently convert the ejecta kinetic energy into the internal energy and sustain the opacity of ejecta by keeping the temperature above the recombination temperature. If this is the case, a gas cloud with mass of $M_{\rm c}=10^6\,M_\odot$ and size of $R_{\rm c}=0.3\,{\rm pc}\approx10^{18}\,{\rm cm}$ heated up by the explosion with the energy of $10^{56}\,{\rm erg}$ can be as bright as 
\begin{align}
L&\approx 2\times 10^{46}\,{\rm erg/s}\nonumber\\
&\times\left(\frac{R_{\rm c}}{10^{18}\,{\rm cm}}\right)\left(\frac{E_{\rm exp}}{10^{56}\,{\rm erg}}\right)\left(\frac{M_{\rm c}}{10^6\,M_\odot}\right)^{-1}
\end{align}
with the photon diffusion timescale of 
\begin{align}
t_{\rm diff}\approx  10^2\,{\rm yr}\left(\frac{E_{\rm exp}}{10^{56}\,{\rm erg}}\right)^{-1/4}\left(\frac{M_{\rm c}}{10^6\,M_\odot}\right)^{3/4}.
\end{align}

As supermassive stars possibly form and explode in the high-redshift universe, the duration in the observer's frame can be even longer. At the same time, the typical wavelength of the emission is redshifted. For example, if the explosion in the redshift of $z\approx10$ results in type-IIp-like emission, the emission in the plateau phase, of which spectra in the source frame may be approximated by a black body of the hydrogen recombination temperature ($\approx 6000\,{\rm K}$), will be observed in the Mid-Infrared band of JWST with the duration of $\sim (1+z)t_{\rm peak}\sim 100\,{\rm yr}$ or $\sim (1+z)t_\mathrm{diff} \sim 10^3$\,yr. Hence, we may observe such ``transients" as red quasi-persistent sources. The observational features will be discussed in detail in our follow-up work (Jockel et al., in preparation).
%\addms{COMMENT: In this paragraph, we need to describe at least orders of quantitative numbers, e.g., lumonosity and duration, with an equation as in the following paragraphs. Otherwise it is quite difficult to understand the aim of this paragraph for those who are not familiar with the scenario. Also it would be difficult to understand the purpose why you wrote Sec. 4.1.}

The mass accretion onto the black hole would lead to other activities. If a sufficient magnetic field accretes onto the black hole in association with mass accretion and a magnetically dominated region, supported by the gas pressure of the torus, is subsequently established around the black hole, the Blandford-Znajek (BZ) process~\citep{Blandford1977} can extract the rotational energy of the black hole in the form of a Poynting flux. The extracted energy would then form a relativistic jet towards the more evacuated polar directions~\citep{Matsumoto2015sep}.

\begin{figure}[t]
    \centering
    \includegraphics[width=0.49\textwidth]{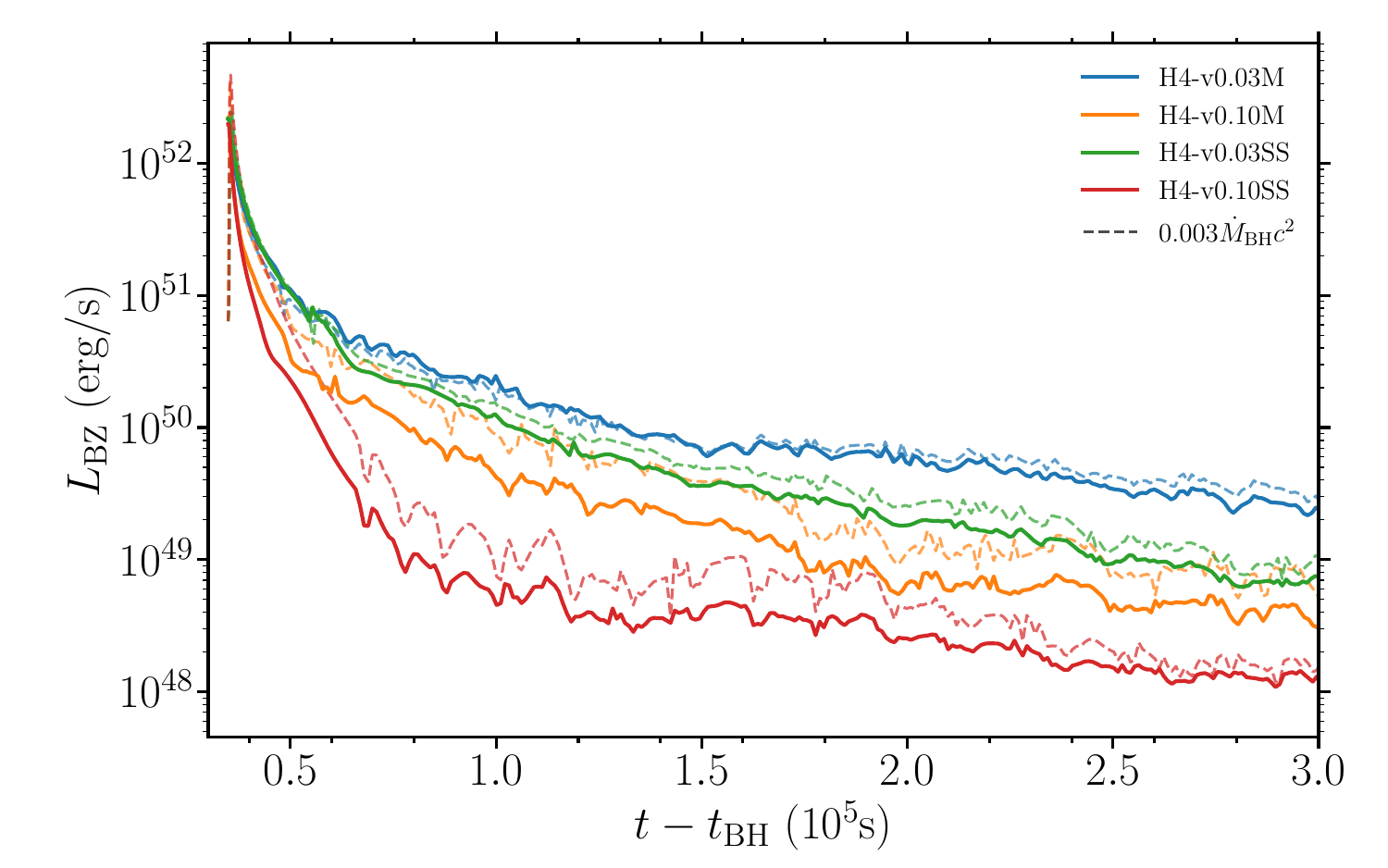}
    \caption{Estimated possible Blandford-Znajek luminosity. The dashed curves denote the rate of the mass accretion onto the black hole multiplied with a constant factor 0.003. See main text in Sec.~\ref{subsec:transients} for the definition.}
    \label{fig:BZ}
\end{figure}

Figure~\ref{fig:BZ} shows the hypothetical BZ luminosity estimated by~\citep{Shibata2024feb}
\begin{align}
L_\mathrm{BZ} \approx \SI{1e49}{erg/s} \bigg(\frac{\chi}{0.7}\bigg)^2 \bigg(\frac{M_\mathrm{BH}}{10^5M_\odot}\bigg)^2 \bigg(\frac{B}{3 \times 10^{9}\mathrm{G}}\bigg)^2,
\end{align}
where $B$ denotes the poloidal magnetic flux penetrating the black hole horizon and is estimated assuming that the magnetic pressure is the same as the matter pressure at an innermost region of the torus around the black hole, i.e., $B^2/8\pi = \xi \rho c_\mathrm{s}^2|_\mathrm{ISCO}$. Here, the value of $\rho c_\mathrm{s}^2|_\mathrm{ISCO}$ is evaluated at $\varpi=6GM_0/c^2$ in the equatorial direction, and the factor $\xi$, which represents the hypothetical saturation level of magnetic energy compared to the internal energy, is set to be 0.1. The shape of the luminosity curve resembles the mass accretion rate (top panel of the same figure), with the efficiency defined by $L_\mathrm{BZ}/\dot{M}c^2$ of $\sim0.003$. If the Poynting flux is assumed to be continuously emitted for the viscous timescale of the torus, the total energy generated by this process is $\sim\SI{e55}{erg}$ for H4 models. %$\chi=0.7$, $M_\mathrm{BH}=10^5M_\odot$, and $B=10^{9.5}$\,G. 
The hypothetical collimated energy injection into the possibly inflated stellar envelope and the gas cloud, which would surround the supermassive star, may form a cocoon, which emits thermal photons when it breaks out from the surface of the star or the cloud~\citep{Kashiyama2013jun,Nakauchi2013nov}. Thus, the jet may also play an important role for the possible electromagnetic signals.

\subsection{Possible qualitative differences from collapses of usual massive stars}
For the collapse of rotating supermassive stars, we observe that the bounce of the torus drives an outflow. Contrary to that, in gravitational collapses of stars of mass $\sim 10$--$100M_\odot$, such a strong bounce and corresponding outflow are not found in the latest numerical simulations (see, e.g., \citealt{Just2022aug,Fujibayashi2024jan,Dean2024apr}).

The difference between the collapses associated with the usually massive and supermassive stars may lie in the unstable mode that triggers their collapses. The collapse of the supermassive stars is triggered by the general relativistic instability, in which the star is unstable with respect to the radial fundamental perturbation of no node \citep{1964ApJ...140..417C}. This indicates that all the stellar matter starts collapsing coherently. As a result, the torus formation proceeds rather coherently after the black-hole formation, enhancing a strong bounce on the torus surface, and furthermore, the density outside the torus becomes very low at their formation; this is preferable for the subsequent prompt shock propagation.

On the other hand, the collapses of usual massive stars are induced by the electron capture and/or photo-dissociation of heavy nuclei, or the thermal production of $e^-e^+$ pairs at their center. Because these processes are active only at the center of the star with high density and temperature, the unstable region is restricted compared to the entire star. As a result, only the central region starts collapsing earlier, and the outer envelope is still in hydrostatic equilibrium at the onset of the central collapse. The matter in the outer region starts collapsing when it looses the pressure support from the inside of the star, typically after the sound-crossing time of the star. When a torus is formed around the black hole, there is still significant matter infalling to the center, which exerts a large ram pressure to prohibit the immediate launch of the bounce-induced outflow (e.g., \citealt{Fujibayashi2024jan}).

For the supermassive star cores with the mass lower than those investigated in this paper, $\lesssim 10^4M_\odot$, the density and temperature of the torus would be higher, and hence, several processes may play an important role in decreasing the pressure of the downstream region of the shock, in the same way for the core-bounce after the proto-neutron star formation (see, e.g., \citealt{Janka2012a} for a review): e.g., the photo-dissociation of heavy nuclei converts the internal energy into the rest mass. In addition, the emitted neutrinos carry the internal energy away from the bounced matter. The importance of such effects in the context of the bounce of the torus is worth investigating for the collapse of low-mass supermassive star cores.

\subsection{Effect of further nuclear burning and prospects of lower-mass stars}
% nuclear burning becomes more important
For the models studied in this paper, the nuclear burning plays a negligible role for the dynamics, because the rate of hydrogen burning is limited by the temperature-independent hot CNO cycle, and the triple-alpha reaction is an inefficient process. %Also, supermassive stars have very high entropy, which makes nuclear reactions inefficient. 
However, after carbon is synthesized, the $\isotope{12}{C}(\alpha,\gamma)\isotope{16}{O}$ reaction will take place and further energy generation may be possible. In \cite{Chen2014aug,Nagele2020aug}, the further nuclear burning indeed synthesizes heavier species up to \isotope{28}{Si}. The feedback due to such energy generation is not taken into account in our present work. In particular for the collapse of helium burning supermassive star cores, such effects may be important.

A supermassive star with a lower final mass will have a lower-entropy core, which has a higher density than that investigated in this work for a given value of temperature. The gravitational collapse (due to the onset of either general relativistic or pair-production instability) of such a lower-mass star will occur in a later evolution stage, i.e., in a later phase of helium burning or after the core helium depletion (e.g., \citealt{Shibata2024prep}). For the collapse of such a star, further nuclear burning and its back-reactions to the dynamics may be more significant. To address such possibilities, the current hydrodynamics has to be coupled with a more sophisticated nuclear reaction network (e.g., the networks in \citealt{Chen2014aug,Nagele2020aug}).

\subsection{Neutrino cooling} \label{subsec:neutrino}
\cite{Uchida2017oct} showed that the neutrino emission plays a negligible role after the black hole formation for models that are essentially the same as ours. In this subsection, we consider the model He4, which is the model resulting in the highest density and temperature torus, and hence the neutrino cooling has the most significant effects among the models considered in this paper. In the following, we will show that the entropy change by the neutrino emission is negligible even for this model. The neutrino luminosity is $\sim\SI{e48}{erg/s}$ after the black hole formation for the model He4 (see figure~7 in \citealt{Uchida2017oct}). The rate of total entropy extraction by neutrino emission can be estimated as
\begin{align}
\dot{S}/k_\mathrm{B} \sim -\frac{L_\nu}{k_\mathrm{B}T} \approx -\SI{7e55}{s^{-1}}\bigg(\frac{L_\nu}{\SI{e48}{erg/s}}\bigg) \bigg(\frac{T}{\SI{e8}{K}}\bigg)^{-1}.
\end{align}
On the other hand, the total entropy of the torus is
\begin{align}
S/k_\mathrm{B} \approx \SI{3.7e63}{}\bigg(\frac{s_\mathrm{torus}/k_\mathrm{B}}{380}\bigg)\bigg(\frac{M_\mathrm{torus}}{8000M_\odot}\bigg),
\end{align}
where the values of entropy per baryon and torus mass are taken from the values for model He4 . The timescale of the change in the entropy is thus estimated as $S/|\dot{S}|\approx \SI{5e7}{s}$ for this model. As the bounce process occurs in a much shorter timescale of $<\SI{e3}{s}$, the neutrino cooling can be safely neglected.

For lower-mass supermassive star cores with masses of $\lesssim10^4M_\odot$, the neutrino cooling may have more significant effects, as the density and temperature become higher. To investigate the outcomes of such stars, we have to include the cooling by various neutrino emission processes (e.g., those mentioned in \citealt{Itoh1996feb}).

\subsection{Possible path to supermassive black holes}
A black hole formed from a supermassive star core of mass $\sim 10^5M_\odot$ may be surrounded by a dense cloud of mass much larger than that of the supermassive star, say $10^7M_\odot$ \citep{Patrick2023jul}. Suppose that the gas cloud is not entirely blown off by the explosion from the torus surrounding the formed black hole. Then, after the black hole formation, a super Eddington mass accretion onto the central black hole may continue because the typical mass accretion rate onto supermassive stars during their growth is $\sim 0.1M_\odot$/yr, which is $\sim 10^2$ times higher than the Eddington accretion rate for the $10^5M_\odot$ black hole~\citep{Johnson2013oct,Whalen2013nov}. Recent numerical simulations (e.g., \citealt{Jiang:2014tpa, Hu:2022qnm}) demonstrate that even for such a very high mass accretion rate, a fraction (an order of $10\%$) of the matter still falls into the black hole although a significant fraction of the infalling matter is outflowed from the system. Hence, a super Eddington accretion growth of the black hole could follow after the formation of a massive black hole from the supermassive stars. This suggests that a black hole formed from a supermassive star with mass $\sim 10^5M_\odot$ may subsequently rapidly grow, leading to a supermassive black hole of mass $\agt 10^6M_\odot$ in $\sim 10^8$\,yrs, which has been observed in the high-redshift universe by JWST. Developing a scenario which connects supermassive star formation, collapse to a seed massive black hole, and subsequent rapid growth of it will be an interesting issue to be explored.

\section{Summary} \label{sec:summary}
In this work, we performed general relativistic hydrodynamics simulations of the collapses of rotating supermassive star cores to investigate the properties of the ejecta as an extension of the previous work~\citep{Uchida2017oct}. We took into account an approximate nuclear burning up to carbon, as in the previous work, and in addition, we incorporated a detailed equation of state, for which ions, photons, electrons, and thermally generated $e^-e^+$ pairs are taken into account.

For all the models we investigated, the energy generation by nuclear burning plays only a minor role, leading to the formation of a black hole without explosion via nuclear burning. However, for rotating models, the stellar explosion sets in from the accreting torus, which forms after the formation of the black hole, with explosion energies up to $10^{-4}$ times the mass energy of the supermassive star cores. We found that, even if we increased the rotation of the progenitor, the ejecta mass saturates at $\sim 1$\% of the total mass of the initial star. The average ejecta velocity also saturates at $\approx 0.2c$. As a result, the ejecta kinetic energy is approximately proportional to the initial mass of the star. 

We further performed viscous hydrodynamics simulations after the black hole and torus formation. We found that, because of the relatively small velocity ($\approx 0.07c$) of the viscosity-driven ejecta, its effect is subdominant in terms of the kinetic energy, although an appreciable fraction of the torus matter can be ejected by this process. 

The collapse of a supermassive star is likely to occur in a dense atomic cooling gas cloud as massive as or more massive than the supermassive star itself according to its formation scenario. As the ejecta mass is minor compared to the mass of the hosting cloud, the explosion plays a role to inject the kinetic energy into the cloud. It may be observed as a very long-duration supernova-like transient. As it likely occurs in a high-redshift ($z\gtrsim10$) universe and the observational duration can be extended by the cosmological redshift effect, we may observe it as a quasi-persistent source. A detail of our analysis on this will be presented in a separate paper (Jockel et al., in preparation). 

The mass accretion of the black hole via the formed torus may also drive a relativistic jet via the BZ process. This outflow may inject energy comparable to that of the ejecta driven by the bounce of the torus. Such an additional energy injection may be important for the electromagnetic signals. We plan to study this process in a future work. 

\acknowledgements
We thank Jan-Torge Schindler, Takashi Hosokawa, Koh Takahashi, Kohei Inayoshi, Shigeo Kimura, and Kazumi Kashiyama for stimulating discussions. This work was in part supported by Grant-in-Aid for Scientific Research (grant Nos.~20H00158 and 23H04900) of Japanese MEXT/JSPS. Numerical computations were performed at Sakura of the Max Planck Computing and Data Facility.

\appendix

\section{Electron equation of state} \label{app:eeos}
In this appendix, we summarize the detailed implementation of the electron contribution to the equation of state. The internal energy density and pressure of electrons and $e^-e^+$ pairs are determined by the net electron number density $n_\mathrm{e}$ and temperature $T$. We prepare a two-dimensional table of the internal energy per electric charge
\begin{align}
\tilde{\varepsilon}_\mathrm{e}(\rho Y_\mathrm{e},T) = \frac{e_\mathrm{e}}{\rho Y_\mathrm{e}},
\end{align}
where we used $\amu n_\mathrm{e} = \rho Y_\mathrm{e}$ instead of $n_\mathrm{e}$ for later convenience. In the same manner, $P_\mathrm{e}$ is tabulated as a function of $\rho Y_\mathrm{e}$ and $T$. In our implementation, we use the equation of state by \cite{Timmes2000jul} to construct the table of $\tilde\varepsilon_\mathrm{e}$ and $P_\mathrm{e}$ as functions of $\rho Y_\mathrm{e}$ and $T$. For given $\rho$, $T$, and $Y_\mathrm{e}$, we first interpolate $\tilde{\varepsilon}_\mathrm{e}$ and $P_\mathrm{e}$ from the table with ($\rho Y_\mathrm{e}$, $T$). Then the specific internal energy of electrons is calculated by
\begin{align}
\varepsilon_\mathrm{e}(\rho, T, Y_\mathrm{e}) = Y_\mathrm{e}\tilde{\varepsilon}_\mathrm{e}(\rho Y_\mathrm{e},T).
\end{align}

\section{Sound speed}
To solve hydrodynamics numerically, we need the sound speed. In this appendix, we present a way to calculate the sound speed if there are several contributions, a part of which is obtained by interpolating equation of state tables. Suppose that there are no changes of $\langle A \rangle$, $\langle \Delta m \rangle$, and $Y_\mathrm{e}$: we may then write the variation of the pressure in terms of the variations of $\rho$ and $\varepsilon$ as
\begin{align}
dP &= \biggl(\frac{\del P}{\del \rho}\biggr)_\varepsilon d\rho + \biggl(\frac{\del P}{\del \varepsilon}\biggr)_\rho d\varepsilon\notag\\
&=\biggl(\frac{\del P}{\del \rho}\biggr)_\varepsilon d\rho + \biggl(\frac{\del P}{\del \varepsilon}\biggr)_\rho \biggl(Tds + \frac{P}{\rho^2}d\rho\biggr),
\end{align}
where the first law of thermodynamics was used in the second line. In the equation above and in the rest of this appendix, we fix $Y_\mathrm{e}$ in the partial derivatives without writing it explicitly. In a similar way, from this expression, the sound speed is then written as
\begin{align}
h{c_\mathrm{s}}^2 = \biggl(\frac{\del P}{\del \rho}\biggr)_s = \frac{P}{\rho^2}\biggl(\frac{\del P}{\del \varepsilon}\biggr)_\rho + \biggl(\frac{\del P}{\del \rho}\biggr)_\varepsilon,
\end{align}
where $h=1+\varepsilon/c^2 + P/\rho c^2$ is the specific enthalpy. It is beneficial if there is a way to describe the sound speed using the derivatives of thermodynamical quantities with respect to $T$ and $\rho$, which are the usual inputs of equations of state. From an expression of $\varepsilon$ as a function of $\rho$ and $T$, and the first law of thermodynamics, we have
\begin{align}
d\varepsilon = \biggl(\frac{\del \varepsilon}{\del \rho}\biggr)_T d\rho + \biggl(\frac{\del \varepsilon}{\del T}\biggr)_\rho dT = Tds + \frac{P}{\rho^2}d\rho.
\end{align}
In the above, the temperature $T$ is expressed as a function of $s$ and $\rho$. Then
\begin{align}
dP &= \biggl(\frac{\del P}{\del \rho}\biggr)_T d\rho + \biggl(\frac{\del P}{\del T}\biggr)_\rho dT \notag\\
&= \biggl(\frac{\del P}{\del \rho}\biggr)_T d\rho \notag
+ \biggl(\frac{\del P}{\del T}\biggr)_\rho \biggl(\frac{\del \varepsilon}{\del T}\biggr)_\rho^{-1} \biggl[Tds + \biggl(\frac{P}{\rho^2} - \biggl(\frac{\del \varepsilon}{\del \rho}\biggr)_T \biggr) d\rho \biggr].
\end{align}
The last expression indicates the sound speed as
\begin{align}
h{c_\mathrm{s}}^2 = \biggl(\frac{\del P}{\del \rho}\biggr)_T + \biggl(\frac{\del P}{\del T}\biggr)_\rho \biggl(\frac{\del \varepsilon}{\del T}\biggr)_\rho^{-1} \biggl[\frac{P}{\rho^2} - \biggl(\frac{\del \varepsilon}{\del \rho}\biggr)_T \biggr]. \label{eq:cs}
\end{align}
For our physical ingredients, we have
\begin{alignat}{3}
\biggl(\frac{\del P}{\del \rho}\biggr)_T &= &&\frac{k_\mathrm{B}T}{\langle A\rangle \amu} &&+ \biggl(\frac{\del P_\mathrm{e}}{\del \rho}\biggr)_T,\\
\biggl(\frac{\del P}{\del T}\biggr)_\rho &= \frac{4}{3}a_\mathrm{rad}T^3 + &&\frac{\rho k_\mathrm{B}}{\langle A\rangle \amu} &&+ \biggl(\frac{\del P_\mathrm{e}}{\del T}\biggr)_\rho,\\
\biggl(\frac{\del \varepsilon}{\del \rho}\biggr)_T &= -\frac{a_\mathrm{rad}T^4}{\rho^2}&& &&+ \biggl(\frac{\del \varepsilon_\mathrm{e}}{\del \rho}\biggr)_T,\\
\biggl(\frac{\del \varepsilon}{\del T}\biggr)_\rho &= \frac{4a_\mathrm{rad}T^3}{\rho} +&& \frac{3}{2}\frac{k_\mathrm{B}}{\langle A\rangle \amu} &&+ \biggl(\frac{\del \varepsilon_\mathrm{e}}{\del T}\biggr)_\rho.
\end{alignat}
Again, $P_\mathrm{e}$ and $\varepsilon_\mathrm{e} = Y_\mathrm{e} \tilde \varepsilon_\mathrm{e}$ are functions of $\rho Y_\mathrm{e}$ and $T$. Their partial derivatives with respect to $T$ are calculated trivially as
\begin{align}
\biggl(\frac{\del P_\mathrm{e}}{\del T}\biggr)_{\rho} &= \frac{\del P_\mathrm{e}(\rho Y_\mathrm{e},T)}{\del T}\bigg|_{\rho Y_\mathrm{e}},\\
\biggl(\frac{\del \varepsilon_\mathrm{e}}{\del T}\biggr)_{\rho} &= \frac{\del}{\del T} \bigg(Y_\mathrm{e} \tilde{\varepsilon}_\mathrm{e}(\rho Y_\mathrm{e},T)\bigg)\bigg|_{\rho Y_\mathrm{e}} = Y_\mathrm{e} \frac{\del \tilde{\varepsilon}_\mathrm{e}}{\del T}\bigg|_{\rho Y_\mathrm{e}}.
\end{align}
The derivatives with respect to $\rho$ (fixing $Y_\mathrm{e}$) are more complicated as
\begin{align}
\biggl(\frac{\del P_\mathrm{e}}{\del \rho}\biggr)_{T} &= \frac{\del (\rho Y_\mathrm{e})}{\del \rho}\frac{\del P_\mathrm{e}(\rho Y_\mathrm{e},T)}{\del (\rho Y_\mathrm{e})}\bigg|_T =  Y_\mathrm{e}\frac{\del P_\mathrm{e}}{\del (\rho Y_\mathrm{e})}\bigg|_T,\\
\biggl(\frac{\del \varepsilon_\mathrm{e}}{\del \rho}\biggr)_{T} &= Y_\mathrm{e} \frac{\del (\rho Y_\mathrm{e})}{\del \rho} \frac{\del \tilde{\varepsilon}_\mathrm{e}(\rho Y_\mathrm{e}, T)}{\del (\rho Y_\mathrm{e})} \bigg|_T =  {Y_\mathrm{e}}^2\frac{\del \tilde{\varepsilon}_\mathrm{e}}{\del (\rho Y_\mathrm{e})}\bigg|_T.
\end{align}
In our implementation, we also use Timmes equation of state to tabulate $\del P_\mathrm{e}/\del (\rho Y_\mathrm{e})|_T$, $\del P_\mathrm{e}/\del T|_{\rho Y_\mathrm{e}}$, $\del \tilde \varepsilon_\mathrm{e}/\del (\rho Y_\mathrm{e})|_T$ and $\del \tilde \varepsilon_\mathrm{e}/\del T|_{\rho Y_\mathrm{e}}$ as functions of $(T, \rho Y_\mathrm{e})$, and interpolate them to a given set of $(\rho, T,  Y_\mathrm{e})$ to calculate sound speed with Eq.~\eqref{eq:cs}.

\bibliography{reference}

\begin{thebibliography}{}
\expandafter\ifx\csname natexlab\endcsname\relax\def\natexlab#1{#1}\fi
\providecommand{\url}[1]{\href{#1}{#1}}
\providecommand{\dodoi}[1]{doi:~\href{http://doi.org/#1}{\nolinkurl{#1}}}
\providecommand{\doeprint}[1]{\href{http://ascl.net/#1}{\nolinkurl{http://ascl.net/#1}}}
\providecommand{\doarXiv}[1]{\href{https://arxiv.org/abs/#1}{\nolinkurl{https://arxiv.org/abs/#1}}}

\bibitem[{{Alcubierre} {et~al.}(2001){Alcubierre}, {Br{\"u}gmann}, {Holz},
  {Takahashi}, {Brandt}, {Seidel}, {Thornburg}, \&
  {Ashtekar}}]{Alcubierre2001a}
{Alcubierre}, M., {Br{\"u}gmann}, B., {Holz}, D., {et~al.} 2001, International
  Journal of Modern Physics D, 10, 273, \dodoi{10.1142/S0218271801000834}

\bibitem[{{Arnowitt} {et~al.}(1960){Arnowitt}, {Deser}, \&
  {Misner}}]{arnowitt1960a}
{Arnowitt}, R., {Deser}, S., \& {Misner}, C.~W. 1960, Physical Review, 118,
  1100, \dodoi{10.1103/PhysRev.118.1100}

\bibitem[{{Balbus} \& {Hawley}(1991)}]{Balbus1991a}
{Balbus}, S.~A., \& {Hawley}, J.~F. 1991, \apj, 376, 214,
  \dodoi{10.1086/170270}

\bibitem[{Balbus \& Hawley(1998)}]{Balbus:1998ja}
Balbus, S.~A., \& Hawley, J.~F. 1998, Rev. Mod. Phys., 70, 1,
  \dodoi{10.1103/RevModPhys.70.1}

\bibitem[{{Baumgarte} \& {Shapiro}(1998)}]{baumgarte1998a}
{Baumgarte}, T.~W., \& {Shapiro}, S.~L. 1998, \apj, 504, 431,
  \dodoi{10.1086/306067}

\bibitem[{Baumgarte {et~al.}(2000)Baumgarte, Shapiro, \&
  Shibata}]{Baumgarte:1999cq}
Baumgarte, T.~W., Shapiro, S.~L., \& Shibata, M. 2000, Astrophys. J. Lett.,
  528, L29, \dodoi{10.1086/312425}

\bibitem[{{Blandford} \& {Znajek}(1977)}]{Blandford1977}
{Blandford}, R.~D., \& {Znajek}, R.~L. 1977, \mnras, 179, 433,
  \dodoi{10.1093/mnras/179.3.433}

\bibitem[{{Bogd{\'a}n} {et~al.}(2024){Bogd{\'a}n}, {Goulding}, {Natarajan},
  {Kov{\'a}cs}, {Tremblay}, {Chadayammuri}, {Volonteri}, {Kraft}, {Forman},
  {Jones}, {Churazov}, \& {Zhuravleva}}]{Bogdan2024jan}
{Bogd{\'a}n}, {\'A}., {Goulding}, A.~D., {Natarajan}, P., {et~al.} 2024, Nature
  Astronomy, 8, 126, \dodoi{10.1038/s41550-023-02111-9}

\bibitem[{Bond {et~al.}(1984)Bond, Arnett, \& Carr}]{Bond:1984sn}
Bond, J.~R., Arnett, W.~D., \& Carr, B.~J. 1984, Astrophys. J., 280, 825,
  \dodoi{10.1086/162057}

\bibitem[{{Bromm} \& {Loeb}(2003)}]{Bromm2003oct}
{Bromm}, V., \& {Loeb}, A. 2003, \apj, 596, 34, \dodoi{10.1086/377529}

\bibitem[{Campanelli {et~al.}(2006)Campanelli, Lousto, Marronetti, \&
  Zlochower}]{Campanelli:2005dd}
Campanelli, M., Lousto, C.~O., Marronetti, P., \& Zlochower, Y. 2006, Phys.
  Rev. Lett., 96, 111101, \dodoi{10.1103/PhysRevLett.96.111101}

\bibitem[{{Chandrasekhar}(1964)}]{1964ApJ...140..417C}
{Chandrasekhar}, S. 1964, \apj, 140, 417, \dodoi{10.1086/147938}

\bibitem[{{Chen} {et~al.}(2014{\natexlab{a}}){Chen}, {Heger}, {Woosley},
  {Almgren}, \& {Whalen}}]{2014ApJ...792...44C}
{Chen}, K.-J., {Heger}, A., {Woosley}, S., {Almgren}, A., \& {Whalen}, D.~J.
  2014{\natexlab{a}}, \apj, 792, 44, \dodoi{10.1088/0004-637X/792/1/44}

\bibitem[{{Chen} {et~al.}(2014{\natexlab{b}}){Chen}, {Heger}, {Woosley},
  {Almgren}, {Whalen}, \& {Johnson}}]{Chen2014aug}
{Chen}, K.-J., {Heger}, A., {Woosley}, S., {et~al.} 2014{\natexlab{b}}, \apj,
  790, 162, \dodoi{10.1088/0004-637X/790/2/162}

\bibitem[{{Chon} {et~al.}(2016){Chon}, {Hirano}, {Hosokawa}, \&
  {Yoshida}}]{Chon2016dec}
{Chon}, S., {Hirano}, S., {Hosokawa}, T., \& {Yoshida}, N. 2016, \apj, 832,
  134, \dodoi{10.3847/0004-637X/832/2/134}

\bibitem[{{Dean} \& {Fern{\'a}ndez}(2024)}]{Dean2024apr}
{Dean}, C., \& {Fern{\'a}ndez}, R. 2024, \prd, 109, 083010,
  \dodoi{10.1103/PhysRevD.109.083010}

\bibitem[{{Fan} {et~al.}(2023){Fan}, {Ba{\~n}ados}, \& {Simcoe}}]{Fan2023aug}
{Fan}, X., {Ba{\~n}ados}, E., \& {Simcoe}, R.~A. 2023, \araa, 61, 373,
  \dodoi{10.1146/annurev-astro-052920-102455}

\bibitem[{{Fricke}(1973)}]{1973ApJ...183..941F}
{Fricke}, K.~J. 1973, \apj, 183, 941, \dodoi{10.1086/152280}

\bibitem[{{Fujibayashi} {et~al.}(2024){Fujibayashi}, {Lam}, {Shibata}, \&
  {Sekiguchi}}]{Fujibayashi2024jan}
{Fujibayashi}, S., {Lam}, A. T.-L., {Shibata}, M., \& {Sekiguchi}, Y. 2024,
  \prd, 109, 023031, \dodoi{10.1103/PhysRevD.109.023031}

\bibitem[{{Fujibayashi} {et~al.}(2017){Fujibayashi}, {Sekiguchi}, {Kiuchi}, \&
  {Shibata}}]{fujibayashi2017a}
{Fujibayashi}, S., {Sekiguchi}, Y., {Kiuchi}, K., \& {Shibata}, M. 2017, \apj,
  846, 114, \dodoi{10.3847/1538-4357/aa8039}

\bibitem[{{Fujibayashi} {et~al.}(2020{\natexlab{a}}){Fujibayashi}, {Shibata},
  {Wanajo}, {Kiuchi}, {Kyutoku}, \& {Sekiguchi}}]{Fujibayashi2020a}
{Fujibayashi}, S., {Shibata}, M., {Wanajo}, S., {et~al.} 2020{\natexlab{a}},
  \prd, 101, 083029, \dodoi{10.1103/PhysRevD.101.083029}

\bibitem[{{Fujibayashi} {et~al.}(2020{\natexlab{b}}){Fujibayashi}, {Shibata},
  {Wanajo}, {Kiuchi}, {Kyutoku}, \& {Sekiguchi}}]{Fujibayashi2020b}
---. 2020{\natexlab{b}}, \prd, 102, 123014, \dodoi{10.1103/PhysRevD.102.123014}

\bibitem[{{Fujibayashi} {et~al.}(2021){Fujibayashi}, {Takahashi}, {Sekiguchi},
  \& {Shibata}}]{fujibayashi2021oct}
{Fujibayashi}, S., {Takahashi}, K., {Sekiguchi}, Y., \& {Shibata}, M. 2021,
  \apj, 919, 80, \dodoi{10.3847/1538-4357/ac10cb}

\bibitem[{{Fujibayashi} {et~al.}(2020{\natexlab{c}}){Fujibayashi}, {Wanajo},
  {Kiuchi}, {Kyutoku}, {Sekiguchi}, \& {Shibata}}]{Fujibayashi2020c}
{Fujibayashi}, S., {Wanajo}, S., {Kiuchi}, K., {et~al.} 2020{\natexlab{c}},
  \apj, 901, 122, \dodoi{10.3847/1538-4357/abafc2}

\bibitem[{{Fuller} {et~al.}(1986){Fuller}, {Woosley}, \&
  {Weaver}}]{1986ApJ...307..675F}
{Fuller}, G.~M., {Woosley}, S.~E., \& {Weaver}, T.~A. 1986, \apj, 307, 675,
  \dodoi{10.1086/164452}

\bibitem[{{Goulding} {et~al.}(2023){Goulding}, {Greene}, {Setton}, {Labbe},
  {Bezanson}, {Miller}, {Atek}, {Bogd{\'a}n}, {Brammer}, {Chemerynska},
  {Cutler}, {Dayal}, {Fudamoto}, {Fujimoto}, {Furtak}, {Kokorev}, {Khullar},
  {Leja}, {Marchesini}, {Natarajan}, {Nelson}, {Oesch}, {Pan}, {Papovich},
  {Price}, {van Dokkum}, {Wang}, {Weaver}, {Whitaker}, \&
  {Zitrin}}]{Goulding2023sep}
{Goulding}, A.~D., {Greene}, J.~E., {Setton}, D.~J., {et~al.} 2023, \apjl, 955,
  L24, \dodoi{10.3847/2041-8213/acf7c5}

\bibitem[{{Haemmerl{\'e}} {et~al.}(2018){Haemmerl{\'e}}, {Woods}, {Klessen},
  {Heger}, \& {Whalen}}]{2018ApJ...853L...3H}
{Haemmerl{\'e}}, L., {Woods}, T.~E., {Klessen}, R.~S., {Heger}, A., \&
  {Whalen}, D.~J. 2018, \apjl, 853, L3, \dodoi{10.3847/2041-8213/aaa462}

\bibitem[{{Hilditch} {et~al.}(2013){Hilditch}, {Bernuzzi}, {Thierfelder},
  {Cao}, {Tichy}, \& {Br{\"u}gmann}}]{Hilditch2013a}
{Hilditch}, D., {Bernuzzi}, S., {Thierfelder}, M., {et~al.} 2013, \prd, 88,
  084057, \dodoi{10.1103/PhysRevD.88.084057}

\bibitem[{{Hirano} {et~al.}(2017){Hirano}, {Hosokawa}, {Yoshida}, \&
  {Kuiper}}]{Hirano2017sep}
{Hirano}, S., {Hosokawa}, T., {Yoshida}, N., \& {Kuiper}, R. 2017, Science,
  357, 1375, \dodoi{10.1126/science.aai9119}

\bibitem[{{Hirano} {et~al.}(2014){Hirano}, {Hosokawa}, {Yoshida}, {Umeda},
  {Omukai}, {Chiaki}, \& {Yorke}}]{Hirano2014feb}
{Hirano}, S., {Hosokawa}, T., {Yoshida}, N., {et~al.} 2014, \apj, 781, 60,
  \dodoi{10.1088/0004-637X/781/2/60}

\bibitem[{{Hosokawa} {et~al.}(2013){Hosokawa}, {Yorke}, {Inayoshi}, {Omukai},
  \& {Yoshida}}]{Hosokawa2013dec}
{Hosokawa}, T., {Yorke}, H.~W., {Inayoshi}, K., {Omukai}, K., \& {Yoshida}, N.
  2013, \apj, 778, 178, \dodoi{10.1088/0004-637X/778/2/178}

\bibitem[{Hu {et~al.}(2022)Hu, Inayoshi, Haiman, Quataert, \&
  Kuiper}]{Hu:2022qnm}
Hu, H., Inayoshi, K., Haiman, Z., Quataert, E., \& Kuiper, R. 2022, Astrophys.
  J., 934, 132, \dodoi{10.3847/1538-4357/ac75d8}

\bibitem[{{Inayoshi} \& {Omukai}(2012)}]{Inayoshi2012may}
{Inayoshi}, K., \& {Omukai}, K. 2012, \mnras, 422, 2539,
  \dodoi{10.1111/j.1365-2966.2012.20812.x}

\bibitem[{{Inayoshi} {et~al.}(2020){Inayoshi}, {Visbal}, \&
  {Haiman}}]{Inayoshi2020aug}
{Inayoshi}, K., {Visbal}, E., \& {Haiman}, Z. 2020, \araa, 58, 27,
  \dodoi{10.1146/annurev-astro-120419-014455}

\bibitem[{{Itoh} {et~al.}(1996){Itoh}, {Hayashi}, {Nishikawa}, \&
  {Kohyama}}]{Itoh1996feb}
{Itoh}, N., {Hayashi}, H., {Nishikawa}, A., \& {Kohyama}, Y. 1996, \apjs, 102,
  411, \dodoi{10.1086/192264}

\bibitem[{{Janka}(2012)}]{Janka2012a}
{Janka}, H.-T. 2012, Annual Review of Nuclear and Particle Science, 62, 407,
  \dodoi{10.1146/annurev-nucl-102711-094901}

\bibitem[{Jiang {et~al.}(2014)Jiang, Stone, \& Davis}]{Jiang:2014tpa}
Jiang, Y.-F., Stone, J.~M., \& Davis, S.~W. 2014, Astrophys. J., 796, 106,
  \dodoi{10.1088/0004-637X/796/2/106}

\bibitem[{{Johnson} {et~al.}(2013){Johnson}, {Whalen}, {Even}, {Fryer},
  {Heger}, {Smidt}, \& {Chen}}]{Johnson2013oct}
{Johnson}, J.~L., {Whalen}, D.~J., {Even}, W., {et~al.} 2013, \apj, 775, 107,
  \dodoi{10.1088/0004-637X/775/2/107}

\bibitem[{{Just} {et~al.}(2022){Just}, {Aloy}, {Obergaulinger}, \&
  {Nagataki}}]{Just2022aug}
{Just}, O., {Aloy}, M.~A., {Obergaulinger}, M., \& {Nagataki}, S. 2022,
  Astrophys. J. Lett., 934, L30, \dodoi{10.3847/2041-8213/ac83a1}

\bibitem[{{Kashiyama} {et~al.}(2013){Kashiyama}, {Nakauchi}, {Suwa}, {Yajima},
  \& {Nakamura}}]{Kashiyama2013jun}
{Kashiyama}, K., {Nakauchi}, D., {Suwa}, Y., {Yajima}, H., \& {Nakamura}, T.
  2013, \apj, 770, 8, \dodoi{10.1088/0004-637X/770/1/8}

\bibitem[{{Kippenhahn} \& {Weigert}(1990)}]{1990sse..book.....K}
{Kippenhahn}, R., \& {Weigert}, A. 1990, {Stellar Structure and Evolution}

\bibitem[{{Kiuchi} {et~al.}(2009){Kiuchi}, {Sekiguchi}, {Shibata}, \&
  {Taniguchi}}]{Kiuchi2009sep}
{Kiuchi}, K., {Sekiguchi}, Y., {Shibata}, M., \& {Taniguchi}, K. 2009, \prd,
  80, 064037, \dodoi{10.1103/PhysRevD.80.064037}

\bibitem[{{Kov{\'a}cs} {et~al.}(2024){Kov{\'a}cs}, {Bogd{\'a}n}, {Natarajan},
  {Werner}, {Azadi}, {Volonteri}, {Tremblay}, {Chadayammuri}, {Forman},
  {Jones}, \& {Kraft}}]{Kovacs2024apr}
{Kov{\'a}cs}, O.~E., {Bogd{\'a}n}, {\'A}., {Natarajan}, P., {et~al.} 2024,
  \apjl, 965, L21, \dodoi{10.3847/2041-8213/ad391f}

\bibitem[{Lee \& Yoon(2016)}]{Lee_2016}
Lee, H., \& Yoon, S.-C. 2016, The Astrophysical Journal, 820, 135,
  \dodoi{10.3847/0004-637X/820/2/135}

\bibitem[{{Lee} \& {Ramirez-Ruiz}(2006)}]{Lee2006apr}
{Lee}, W.~H., \& {Ramirez-Ruiz}, E. 2006, \apj, 641, 961,
  \dodoi{10.1086/500533}

\bibitem[{{Liu} {et~al.}(2007{\natexlab{a}}){Liu}, {Shapiro}, \&
  {Stephens}}]{2007PhRvD..76h4017L}
{Liu}, Y.~T., {Shapiro}, S.~L., \& {Stephens}, B.~C. 2007{\natexlab{a}}, \prd,
  76, 084017, \dodoi{10.1103/PhysRevD.76.084017}

\bibitem[{{Liu} {et~al.}(2007{\natexlab{b}}){Liu}, {Shapiro}, \&
  {Stephens}}]{Liu2007oct}
---. 2007{\natexlab{b}}, \prd, 76, 084017, \dodoi{10.1103/PhysRevD.76.084017}

\bibitem[{{Matsumoto} {et~al.}(2015){Matsumoto}, {Nakauchi}, {Ioka}, {Heger},
  \& {Nakamura}}]{Matsumoto2015sep}
{Matsumoto}, T., {Nakauchi}, D., {Ioka}, K., {Heger}, A., \& {Nakamura}, T.
  2015, \apj, 810, 64, \dodoi{10.1088/0004-637X/810/1/64}

\bibitem[{Matsumoto {et~al.}(2016)Matsumoto, Nakauchi, Ioka, \&
  Nakamura}]{Matsumoto:2015bjg}
Matsumoto, T., Nakauchi, D., Ioka, K., \& Nakamura, T. 2016, Astrophys. J.,
  823, 83, \dodoi{10.3847/0004-637X/823/2/83}

\bibitem[{{Montero} {et~al.}(2012){Montero}, {Janka}, \&
  {M{\"u}ller}}]{Montero2012apr}
{Montero}, P.~J., {Janka}, H.-T., \& {M{\"u}ller}, E. 2012, \apj, 749, 37,
  \dodoi{10.1088/0004-637X/749/1/37}

\bibitem[{Nagele \& Umeda(2024)}]{nagele2024arxiv}
Nagele, C., \& Umeda, H. 2024, The formation of black holes from rapidly
  accreting supermassive stars is not trivial: Simulations of thermonuclear
  pulsations and explosions.
\newblock \doarXiv{2408.08352}

\bibitem[{{Nagele} {et~al.}(2023){Nagele}, {Umeda}, \&
  {Takahashi}}]{Nagele2023aug}
{Nagele}, C., {Umeda}, H., \& {Takahashi}, K. 2023, \mnras, 523, 1629,
  \dodoi{10.1093/mnras/stad1522}

\bibitem[{{Nagele} {et~al.}(2020){Nagele}, {Umeda}, {Takahashi}, {Yoshida}, \&
  {Sumiyoshi}}]{Nagele2020aug}
{Nagele}, C., {Umeda}, H., {Takahashi}, K., {Yoshida}, T., \& {Sumiyoshi}, K.
  2020, \mnras, 496, 1224, \dodoi{10.1093/mnras/staa1636}

\bibitem[{{Nagele} {et~al.}(2022){Nagele}, {Umeda}, {Takahashi}, {Yoshida}, \&
  {Sumiyoshi}}]{Nagele2022dec}
---. 2022, \mnras, 517, 1584, \dodoi{10.1093/mnras/stac2495}

\bibitem[{{Nakauchi} {et~al.}(2013){Nakauchi}, {Kashiyama}, {Suwa}, \&
  {Nakamura}}]{Nakauchi2013nov}
{Nakauchi}, D., {Kashiyama}, K., {Suwa}, Y., \& {Nakamura}, T. 2013, \apj, 778,
  67, \dodoi{10.1088/0004-637X/778/1/67}

\bibitem[{{Omukai}(2001)}]{Omukai2001jan}
{Omukai}, K. 2001, \apj, 546, 635, \dodoi{10.1086/318296}

\bibitem[{{Patrick} {et~al.}(2023){Patrick}, {Whalen}, {Latif}, \&
  {Elford}}]{Patrick2023jul}
{Patrick}, S.~J., {Whalen}, D.~J., {Latif}, M.~A., \& {Elford}, J.~S. 2023,
  \mnras, 522, 3795, \dodoi{10.1093/mnras/stad1179}

\bibitem[{{Rees}(1978)}]{Rees1978oct}
{Rees}, M.~J. 1978, The Observatory, 98, 210

\bibitem[{{Saio} {et~al.}(2024){Saio}, {Nandal}, {Ekstroem}, \&
  {Meynet}}]{Saio2024arxiv}
{Saio}, H., {Nandal}, D., {Ekstroem}, S., \& {Meynet}, G. 2024, arXiv e-prints,
  arXiv:2406.18040, \dodoi{10.48550/arXiv.2406.18040}

\bibitem[{{Sekiguchi}(2010)}]{Sekiguchi2010a}
{Sekiguchi}, Y. 2010, Progress of Theoretical Physics, 124, 331,
  \dodoi{10.1143/PTP.124.331}

\bibitem[{{Shakura} \& {Sunyaev}(1973)}]{Shakura1973a}
{Shakura}, N.~I., \& {Sunyaev}, R.~A. 1973, \aap, 500, 33

\bibitem[{{Shen} \& {Bildsten}(2007)}]{Shen2007may}
{Shen}, K.~J., \& {Bildsten}, L. 2007, \apj, 660, 1444, \dodoi{10.1086/513457}

\bibitem[{{Shibata}(2000)}]{Shibata2000a}
{Shibata}, M. 2000, Progress of Theoretical Physics, 104, 325,
  \dodoi{10.1143/PTP.104.325}

\bibitem[{{Shibata}(2016)}]{Shibata2016a}
---. 2016, {Numerical Relativity} ({World Scientific Publishing Company}),
  \dodoi{10.1142/9692}

\bibitem[{{Shibata} {et~al.}(2024{\natexlab{a}}){Shibata}, {Fujibayashi},
  {Jockel}, \& {Kawaguchi}}]{Shibata2024prep}
{Shibata}, M., {Fujibayashi}, S., {Jockel}, C., \& {Kawaguchi}, K.
  2024{\natexlab{a}}, \apj

\bibitem[{{Shibata} {et~al.}(2024{\natexlab{b}}){Shibata}, {Fujibayashi},
  {Lam}, {Ioka}, \& {Sekiguchi}}]{Shibata2024feb}
{Shibata}, M., {Fujibayashi}, S., {Lam}, A. T.-L., {Ioka}, K., \& {Sekiguchi},
  Y. 2024{\natexlab{b}}, \prd, 109, 043051, \dodoi{10.1103/PhysRevD.109.043051}

\bibitem[{{Shibata} {et~al.}(2011){Shibata}, {Kiuchi}, {Sekiguchi}, \&
  {Suwa}}]{shibata2011a}
{Shibata}, M., {Kiuchi}, K., {Sekiguchi}, Y., \& {Suwa}, Y. 2011, Progress of
  Theoretical Physics, 125, 1255, \dodoi{10.1143/PTP.125.1255}

\bibitem[{{Shibata} {et~al.}(2017){Shibata}, {Kiuchi}, \&
  {Sekiguchi}}]{shibata2017apr}
{Shibata}, M., {Kiuchi}, K., \& {Sekiguchi}, Y.-i. 2017, \prd, 95, 083005,
  \dodoi{10.1103/PhysRevD.95.083005}

\bibitem[{{Shibata} \& {Nakamura}(1995)}]{shibata1995a}
{Shibata}, M., \& {Nakamura}, T. 1995, \prd, 52, 5428,
  \dodoi{10.1103/PhysRevD.52.5428}

\bibitem[{{Shibata} \& {Sekiguchi}(2012)}]{Shibata2012mar}
{Shibata}, M., \& {Sekiguchi}, Y. 2012, Progress of Theoretical Physics, 127,
  535, \dodoi{10.1143/PTP.127.535}

\bibitem[{Shibata \& Shapiro(2002)}]{Shibata:2002br}
Shibata, M., \& Shapiro, S.~L. 2002, Astrophys. J. Lett., 572, L39,
  \dodoi{10.1086/341516}

\bibitem[{{Shibata} {et~al.}(2016){Shibata}, {Uchida}, \&
  {Sekiguchi}}]{Shibata2016feb}
{Shibata}, M., {Uchida}, H., \& {Sekiguchi}, Y.-i. 2016, \apj, 818, 157,
  \dodoi{10.3847/0004-637X/818/2/157}

\bibitem[{{Tanaka} \& {Haiman}(2009)}]{Tanaka2009may}
{Tanaka}, T., \& {Haiman}, Z. 2009, \apj, 696, 1798,
  \dodoi{10.1088/0004-637X/696/2/1798}

\bibitem[{{Thorne}(1981)}]{Thorne1981a}
{Thorne}, K.~S. 1981, \mnras, 194, 439, \dodoi{10.1093/mnras/194.2.439}

\bibitem[{{Timmes} {et~al.}(2000){Timmes}, {Hoffman}, \&
  {Woosley}}]{Timmes2000jul}
{Timmes}, F.~X., {Hoffman}, R.~D., \& {Woosley}, S.~E. 2000, \apjs, 129, 377,
  \dodoi{10.1086/313407}

\bibitem[{{Timmes} \& {Swesty}(2000)}]{timmes2000a}
{Timmes}, F.~X., \& {Swesty}, F.~D. 2000, \apjs, 126, 501,
  \dodoi{10.1086/313304}

\bibitem[{{Uchida} {et~al.}(2017){Uchida}, {Shibata}, {Yoshida}, {Sekiguchi},
  \& {Umeda}}]{Uchida2017oct}
{Uchida}, H., {Shibata}, M., {Yoshida}, T., {Sekiguchi}, Y., \& {Umeda}, H.
  2017, \prd, 96, 083016, \dodoi{10.1103/PhysRevD.96.083016}

\bibitem[{{Umeda} {et~al.}(2016){Umeda}, {Hosokawa}, {Omukai}, \&
  {Yoshida}}]{Umeda2016oct}
{Umeda}, H., {Hosokawa}, T., {Omukai}, K., \& {Yoshida}, N. 2016, \apjl, 830,
  L34, \dodoi{10.3847/2041-8205/830/2/L34}

\bibitem[{{Volonteri} {et~al.}(2021){Volonteri}, {Habouzit}, \&
  {Colpi}}]{Volonteri2021sep}
{Volonteri}, M., {Habouzit}, M., \& {Colpi}, M. 2021, Nature Reviews Physics,
  3, 732, \dodoi{10.1038/s42254-021-00364-9}

\bibitem[{{Waxman} \& {Shvarts}(1993)}]{Waxman1993apr}
{Waxman}, E., \& {Shvarts}, D. 1993, Physics of Fluids A, 5, 1035,
  \dodoi{10.1063/1.858668}

\bibitem[{{Whalen} {et~al.}(2013){Whalen}, {Johnson}, {Smidt}, {Heger}, {Even},
  \& {Fryer}}]{Whalen2013nov}
{Whalen}, D.~J., {Johnson}, J.~L., {Smidt}, J., {et~al.} 2013, \apj, 777, 99,
  \dodoi{10.1088/0004-637X/777/2/99}

\bibitem[{{Wiescher} {et~al.}(1999){Wiescher}, {G{\"o}rres}, \&
  {Schatz}}]{Wiescher1999jun}
{Wiescher}, M., {G{\"o}rres}, J., \& {Schatz}, H. 1999, Journal of Physics G
  Nuclear Physics, 25, R133, \dodoi{10.1088/0954-3899/25/6/201}

\bibitem[{{Wise} {et~al.}(2019){Wise}, {Regan}, {O'Shea}, {Norman}, {Downes},
  \& {Xu}}]{Wise2019jan}
{Wise}, J.~H., {Regan}, J.~A., {O'Shea}, B.~W., {et~al.} 2019, \nat, 566, 85,
  \dodoi{10.1038/s41586-019-0873-4}

\end{thebibliography}

\end{document}